%% file: 19561.tex
\documentclass[traditabstract]{aa}
\usepackage{txfonts}
\usepackage{graphicx,longtable,lscape,stfloats}

\usepackage{natbib}
\bibpunct{(}{)}{;}{a}{}{,} 

\newcommand{\ltsima} {$\; \buildrel < \over \sim \;$}  
\newcommand{\gtsima} {$\; \buildrel > \over \sim \;$}  
\newcommand{\lta} {\lower.5ex\hbox{\ltsima}}  
\newcommand{\gta} {\lower.5ex\hbox{\gtsima}}  
\newcommand{\Ha} {H$\alpha$}  
\newcommand{\Hb} {H$\beta$}

\newcommand{\mekal}{\textsc{Mekal}}
\newcommand{\xstar}{\textsc{Xstar}}
\newcommand{\xspec}{\textsc{Xspec}}

\begin{document}
\title{Extended soft X-ray emission in 3CR radio galaxies at $z<0.3$:\,\,\,
High Excitation  and  Broad  Line  Galaxies.}
\subtitle{} \titlerunning{Extended soft X-ray emission in 3CR radio galaxies}
\authorrunning{Balmaverde et al.}

\author{
B.~Balmaverde\inst{1}
\and A.~Capetti\inst{1}
\and P.~Grandi\inst{2}
\and E.~Torresi\inst{2}
\and M.~ Chiaberge\inst{3,4,5}
\and J.~ Rodriguez Zaurin\inst{6}
\and G.~R.~Tremblay\inst{7}
\and D.~J.~Axon\inst{8,9}
\and S.~A.~Baum\inst{10,11}
\and G.~Giovannini\inst{3,12}
\and P.~Kharb\inst{8}
\and F.~D.~Macchetto\inst{4}
\and C.~P.~O'Dea\inst{8,13}
\and W.~Sparks\inst{4}}

\institute {INAF - Osservatorio Astrofisico di Torino, Strada Osservatorio 20,
  I-10025 Pino Torinese, Italy 
\and INAF-IASF - Istituto di Astrofisica e Fisica cosmica di Bologna, Via P. Gobetti 101, 40129, Bologna,
  Italy 
\and INAF - Istituto di Radioastronomia di Bologna, via
  Gobetti 101 40129 Bologna, Italy 
\and
  Space Telescope Science Institute, 3700 San Martin Drive, Baltimore, MD
  21218 
\and Center for Astrophysical Sciences, Johns Hopkins University, 3400 N. Charles Street Baltimore, MD 21218 
\and Instituto de Astrof\'isica de Canarias, Calle V\'ia L\'actea s/n, E-38205 La Laguna, Tenerife, Spain 
\and European Southern Observatory, Karl-Schwarzschild-Str. 2, 
85748 Garching bei Muenchen, Germany 
\and Dept of Physics,
  Rochester Institute of Technology, 84 Lomb Memorial Dr., Rochester, NY 14623  
\and School of Mathematical \&
  Physical Sciences, University of Sussex, Falmer, Brighton, BN2 9BH, UK
\and Carlson Center for
  Imaging Science, 84 Lomb Memorial Dr., Rochester, NY 14623
\and Radcliffe Institute for Advanced Study, 10 Garden St. Cambridge, 
MA 02138
\and Dipartimento di Astronomia, Universit\'a di
  Bologna, via Ranzani 1, 40127 Bologna, Italy 
\and Harvard Smithsonian Center for Astrophysics, 60 Garden St. 
Cambridge, MA 02138}

\offprints{balmaverde@oato.inaf.it} 

\abstract  {We  analyze Chandra  observations  of  diffuse soft  X-ray
  emission associated with a complete  sample of 3CR radio galaxies at
  z $<$ 0.3. In this paper we focus on the properties of the spectroscopic
  sub-classes  of  high  excitation  galaxies (HEGs)  and  broad  line
  objects (BLOs).  Among the 33 HEGs we detect extended (or possibly
  extended) emission  in about  40\% of the  sources; the  fraction is
  even  higher (8/10)  restricting the  analysis to  the  objects with
  exposure times larger than 10  ks. In the 18 BLOs, extended emission
  is seen only in 2 objects; this lower detection rate can be ascribed
  to the  presence of their  bright X-ray nuclei that  easily outshine
  any genuine diffuse emission.

  A very close correspondence between  the soft X-ray and optical line
  morphology emerges. We also find  that the ratio between [O~III] and
  extended soft X-ray  luminosity  is confined within  a factor  of 2
  around a median value of 5.  Both results are similar to what is
  seen in Seyfert galaxies.

We  discuss different  processes  that could  explain  the soft  X-ray
emission  and  conclude  that  the photoionization  of  extended  gas,
coincident with the narrow line region, is the favored mechanism.}

\maketitle

\section{Introduction}

\begin{figure*}
\centering
\includegraphics[width=4.5cm]{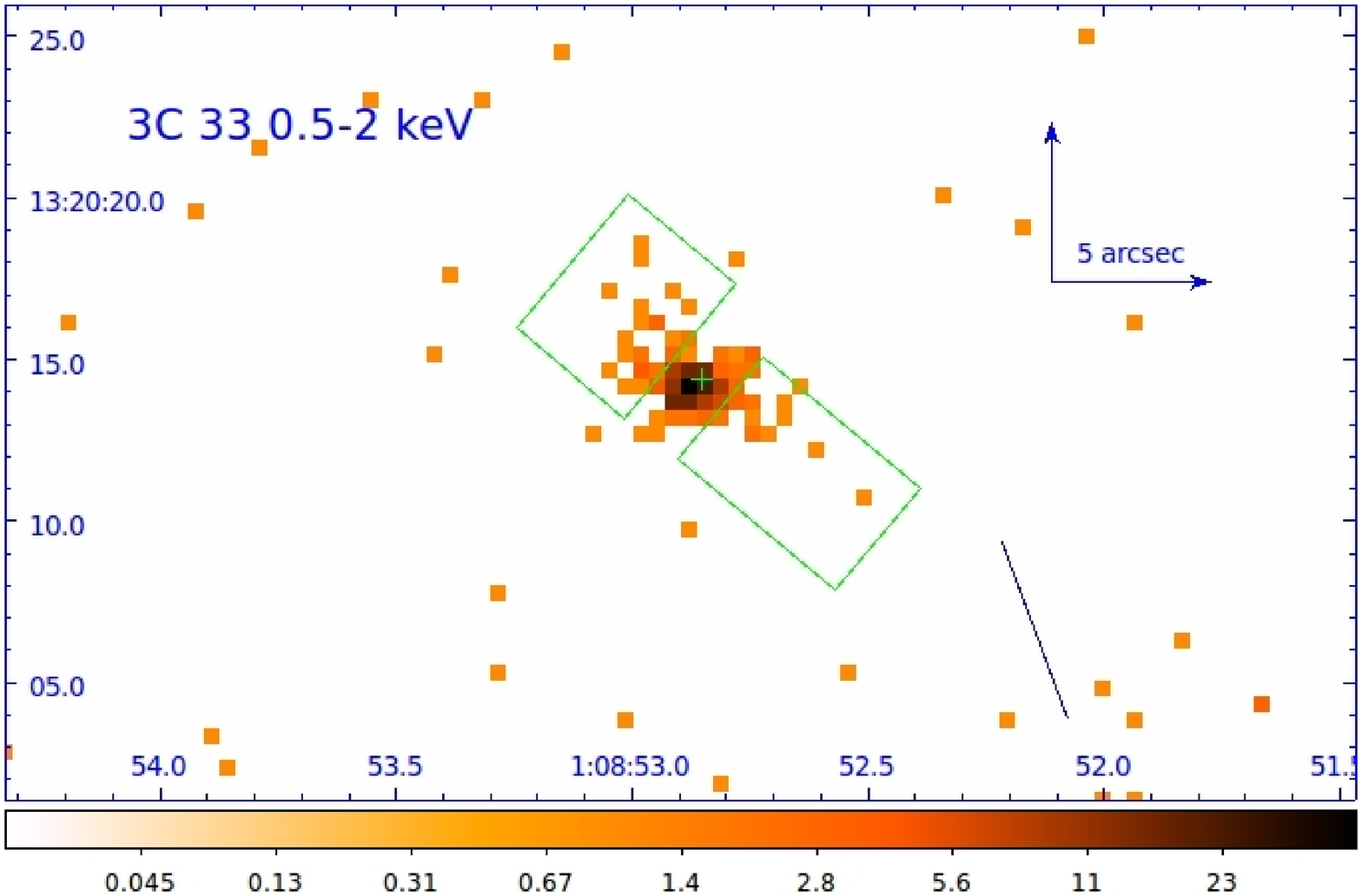} 
\includegraphics[width=4.5cm]{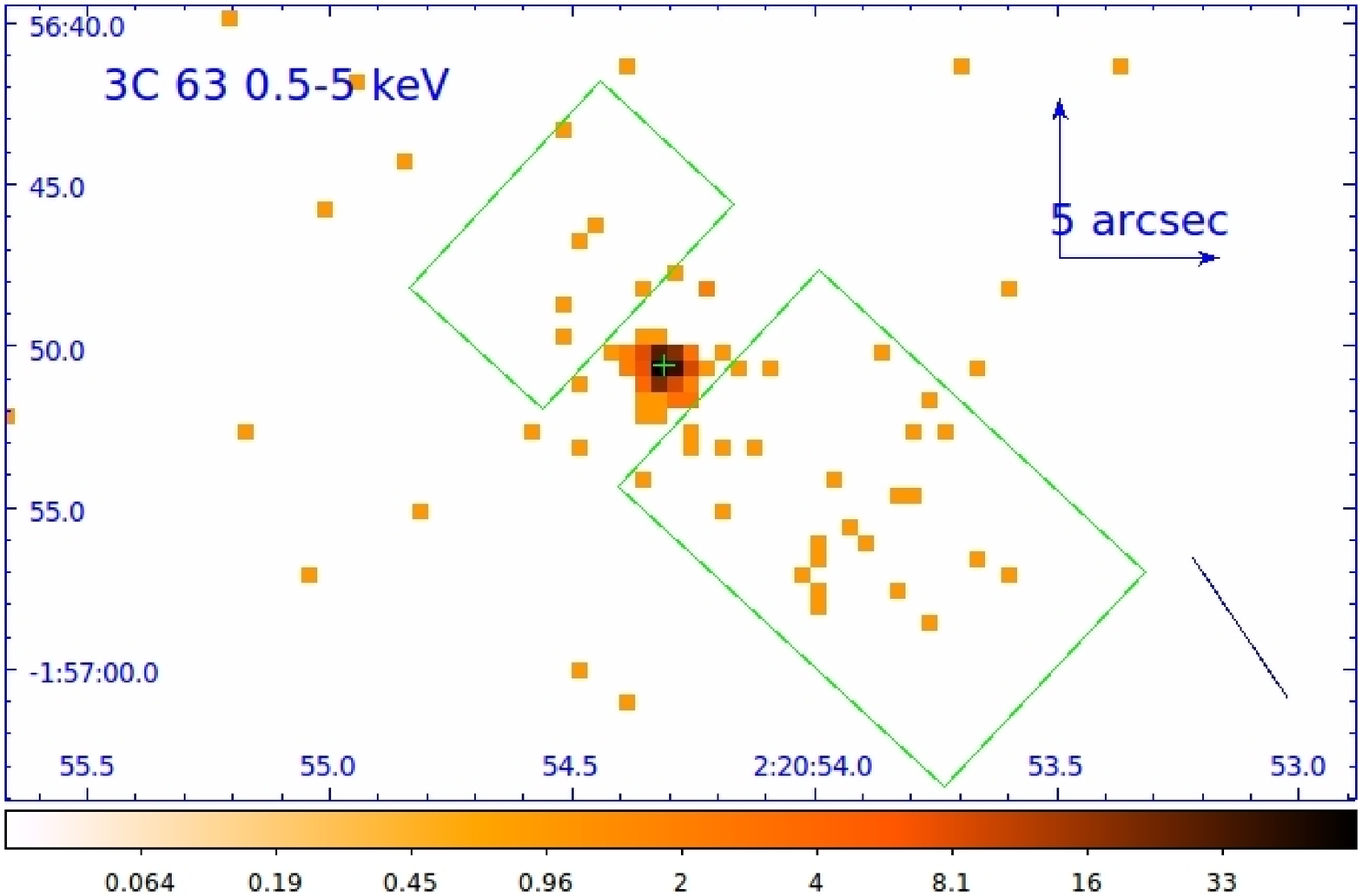} 
\includegraphics[width=4.5cm]{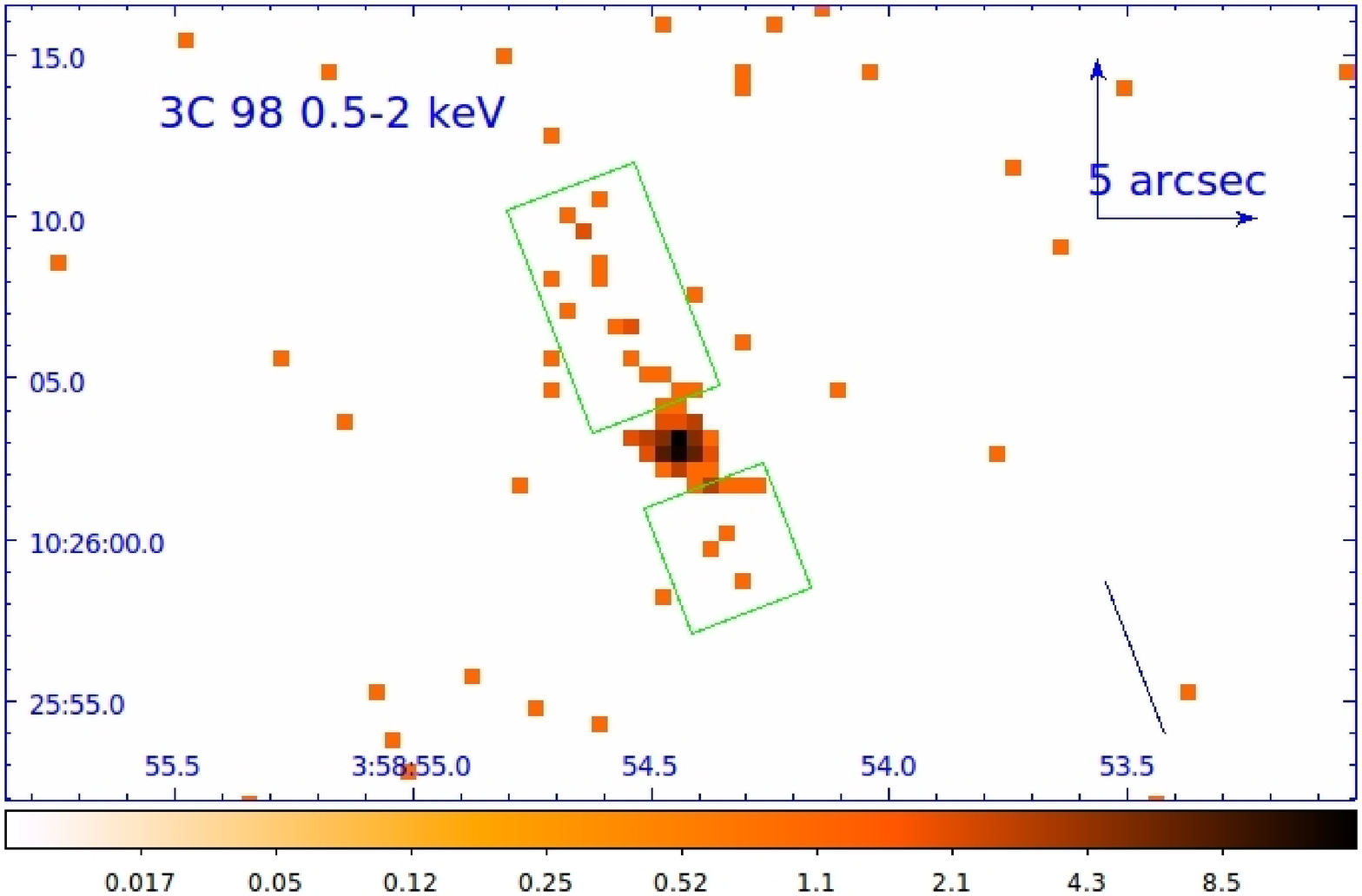} 
\includegraphics[width=4.5cm]{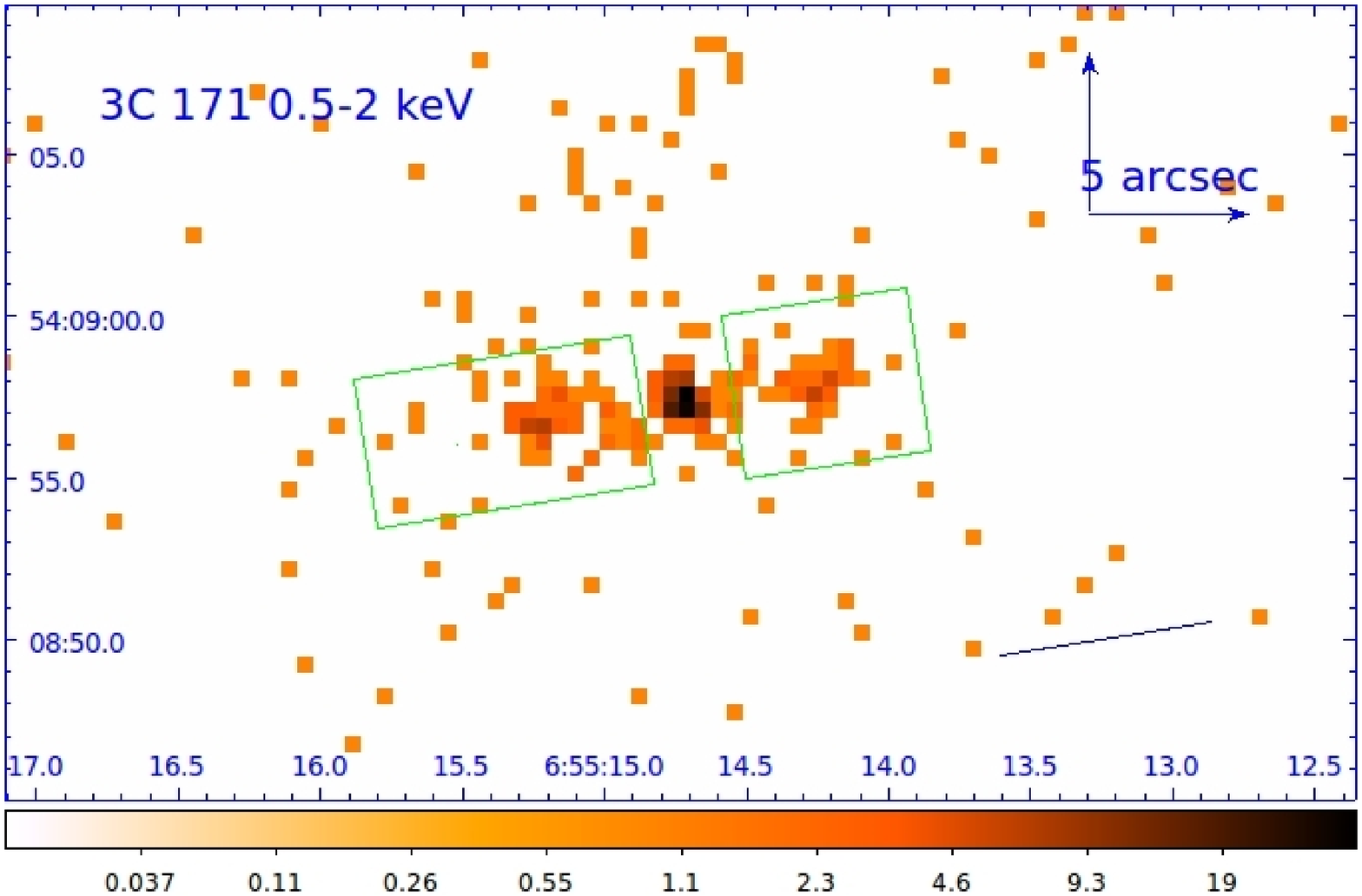} 
\includegraphics[width=4.5cm]{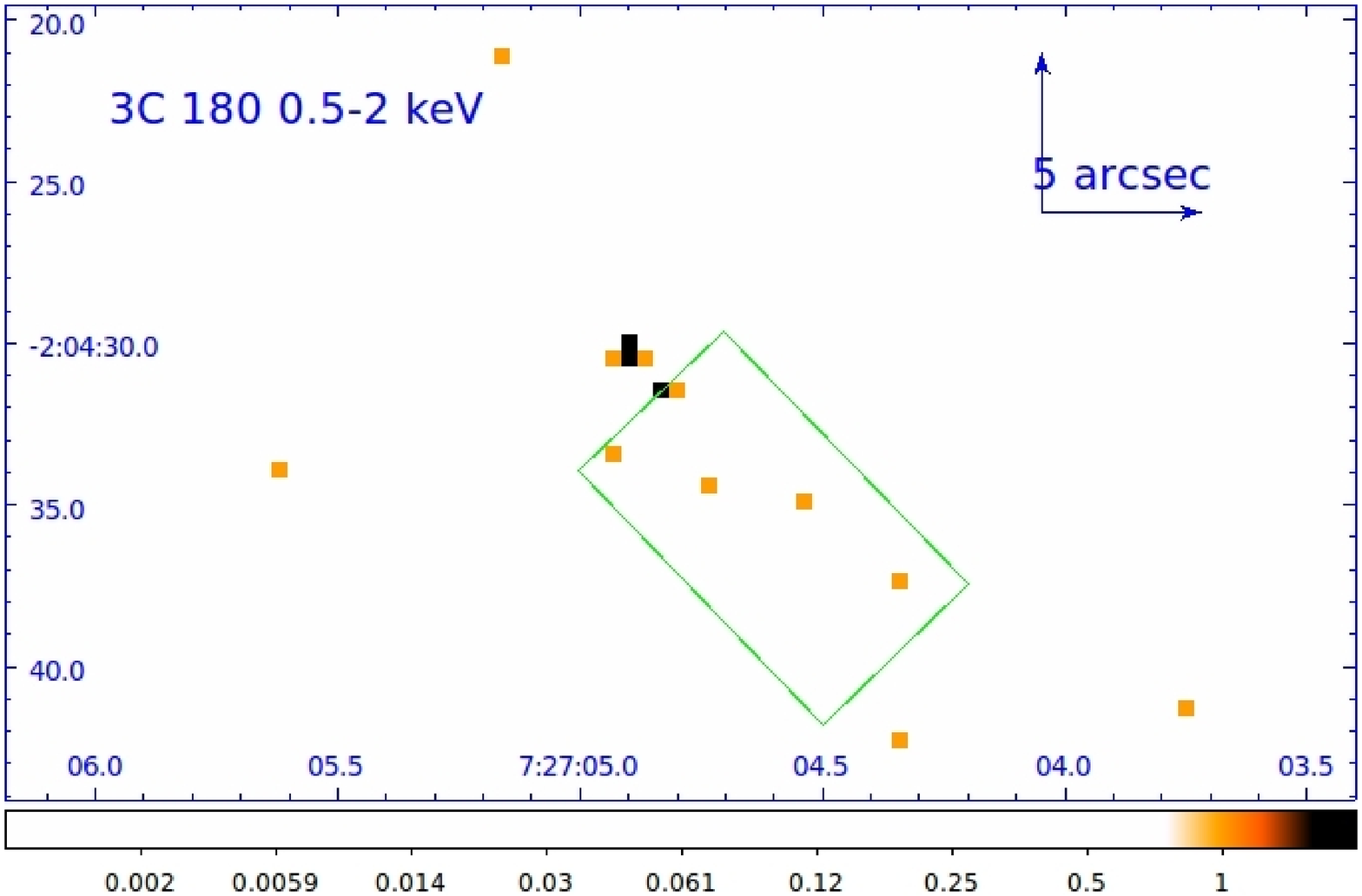} 
\includegraphics[width=4.5cm]{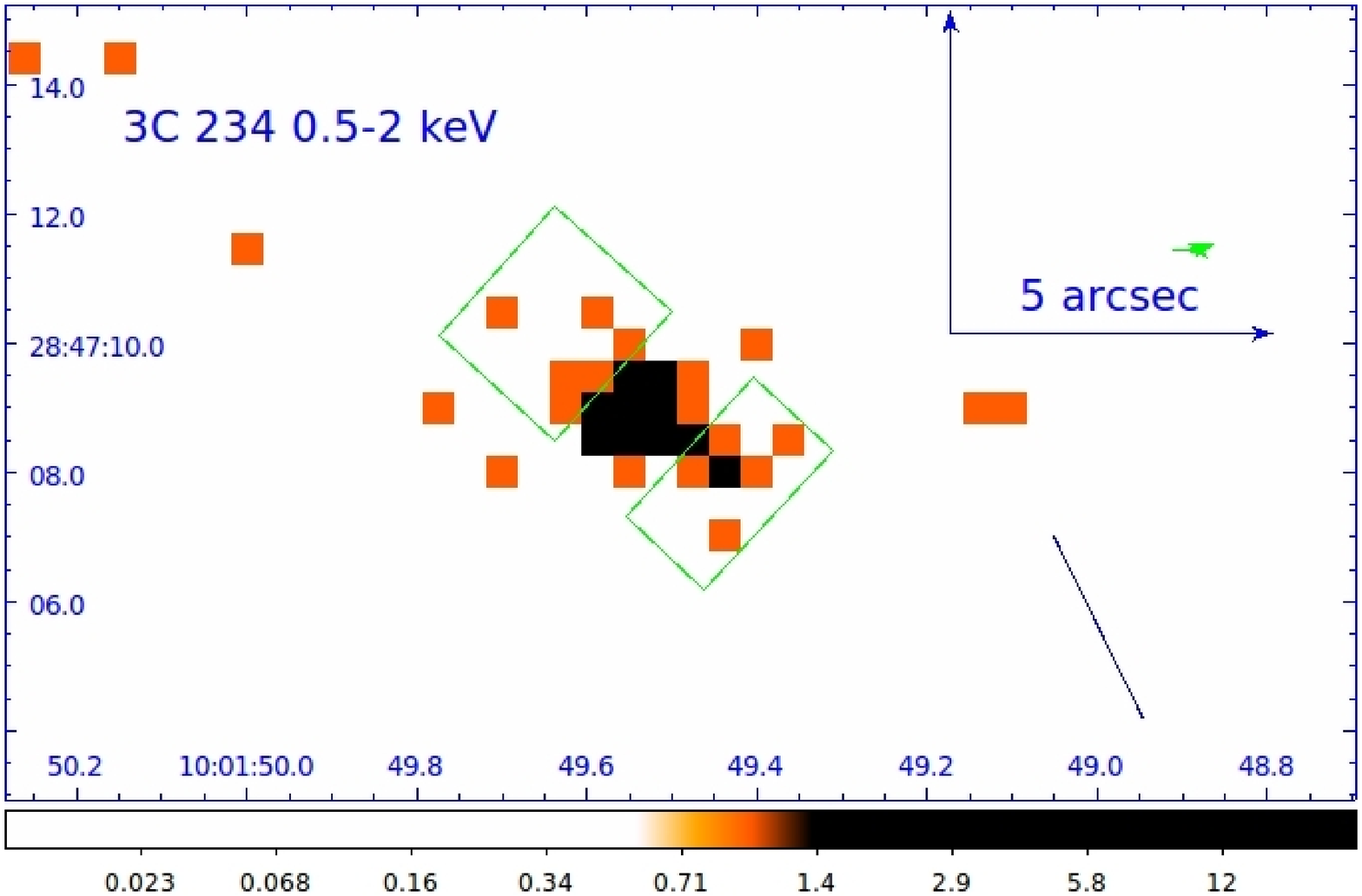} 
\includegraphics[width=4.5cm]{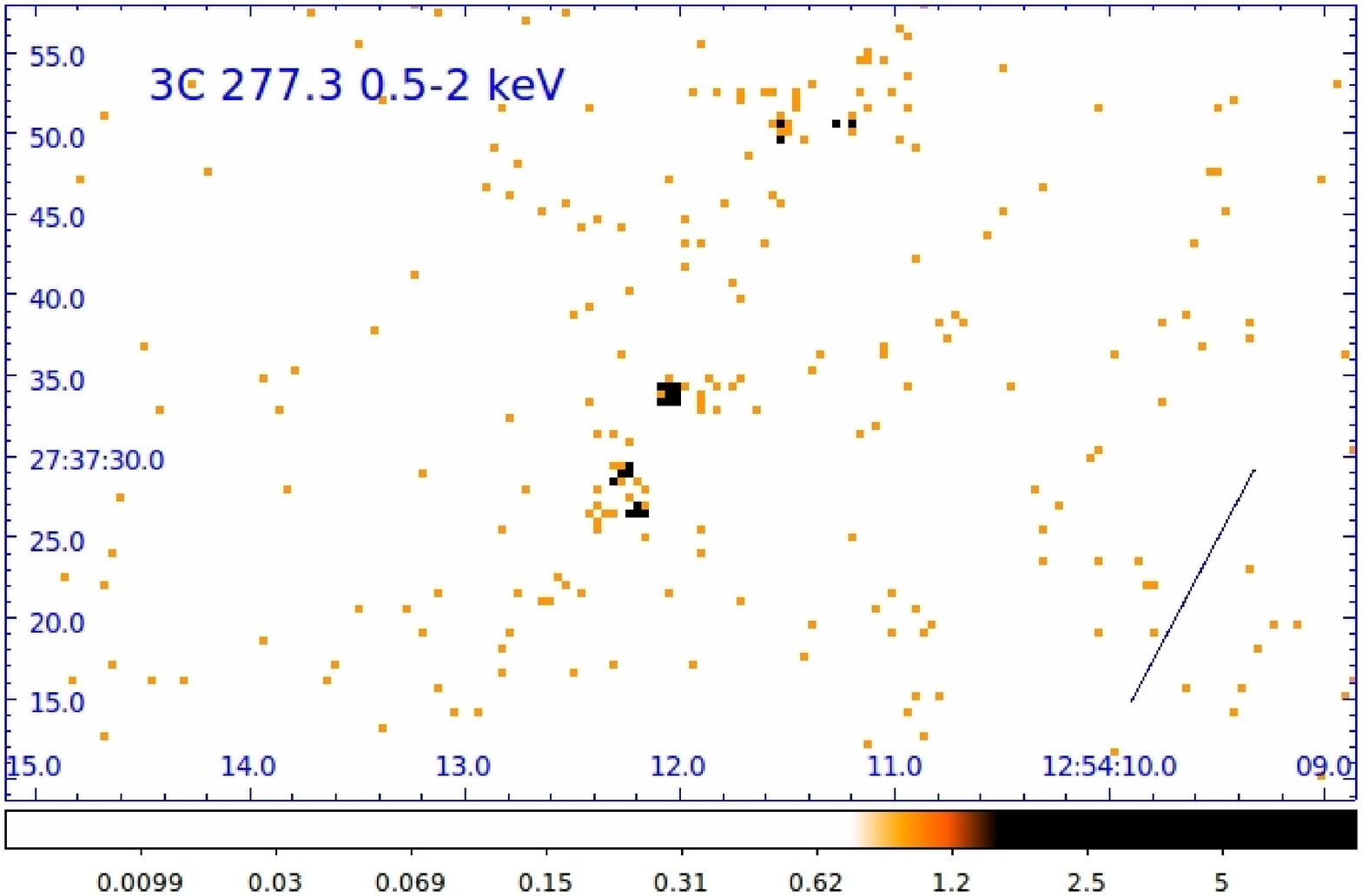} 
\includegraphics[width=4.5cm]{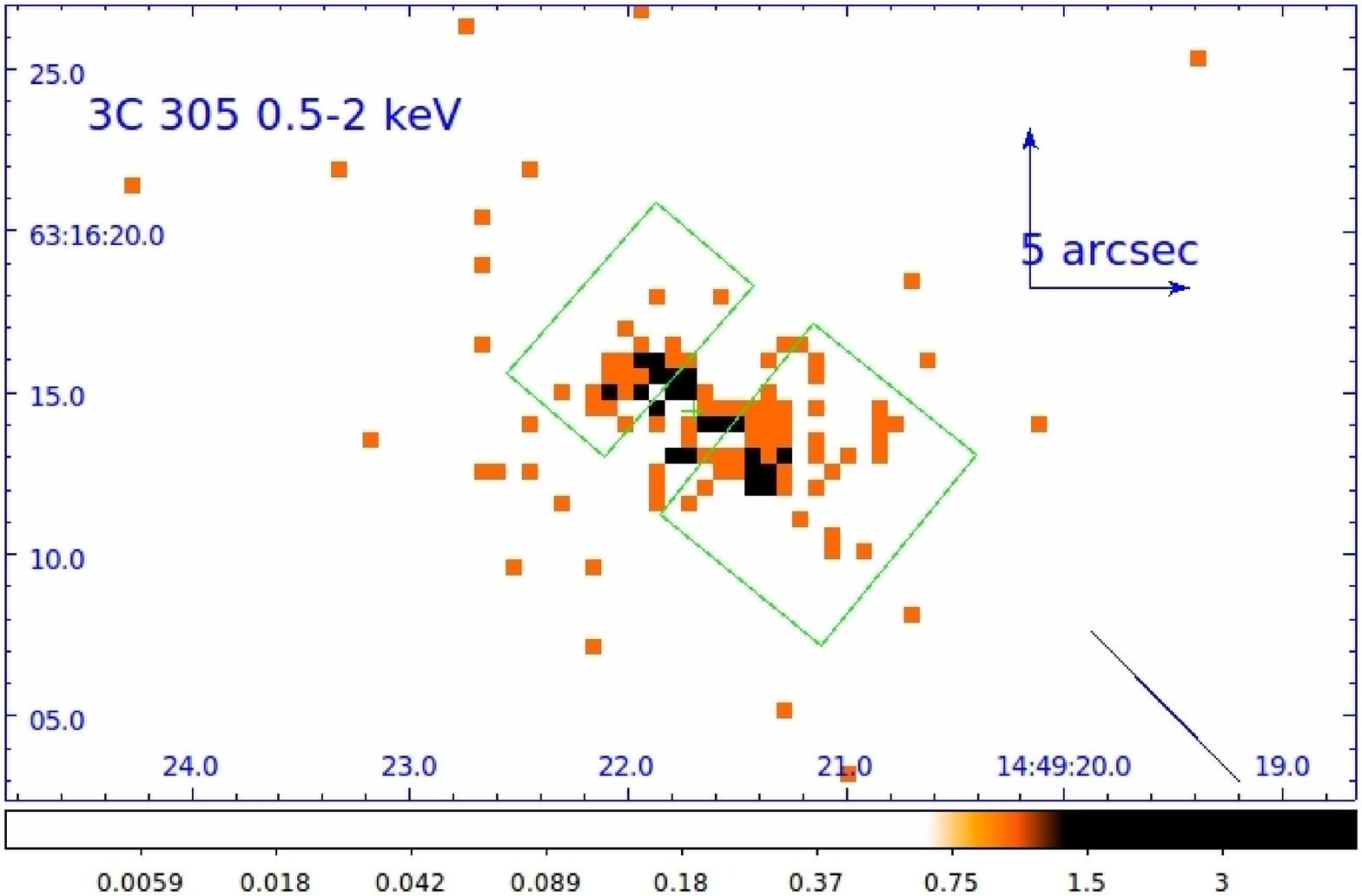} 
\includegraphics[width=4.5cm]{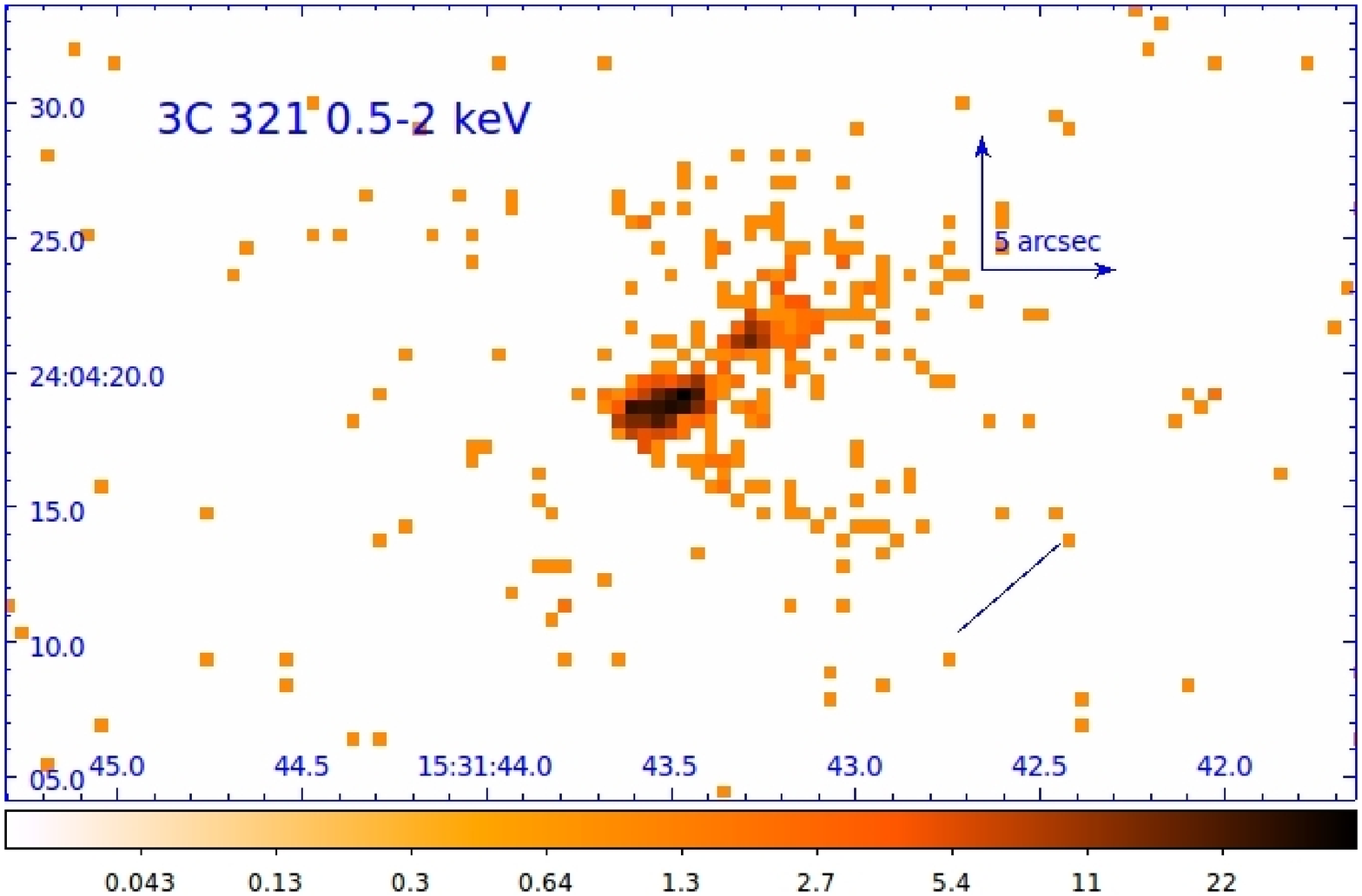} 
\includegraphics[width=4.5cm]{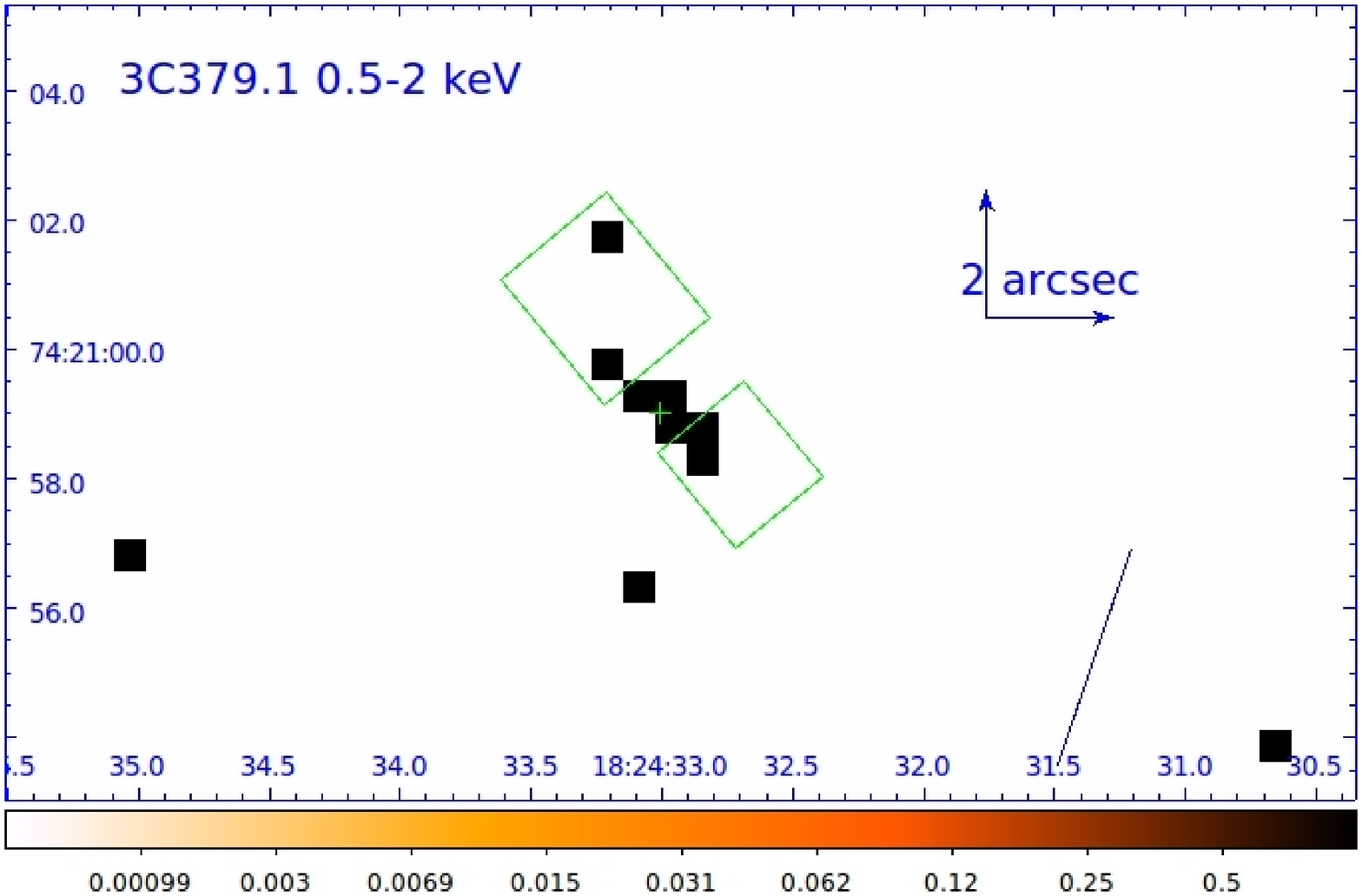}
\includegraphics[width=4.5cm]{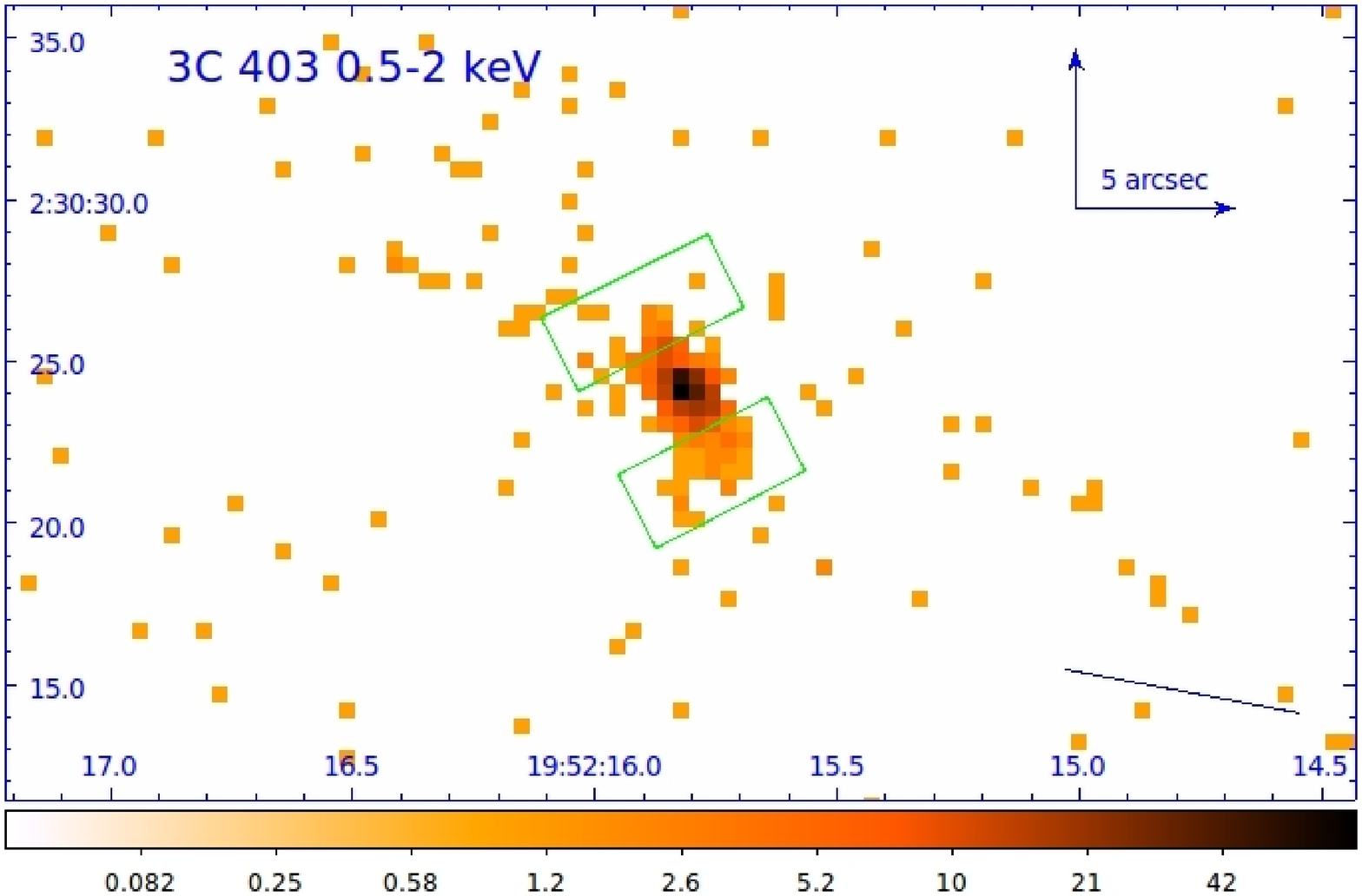}
\includegraphics[width=4.5cm]{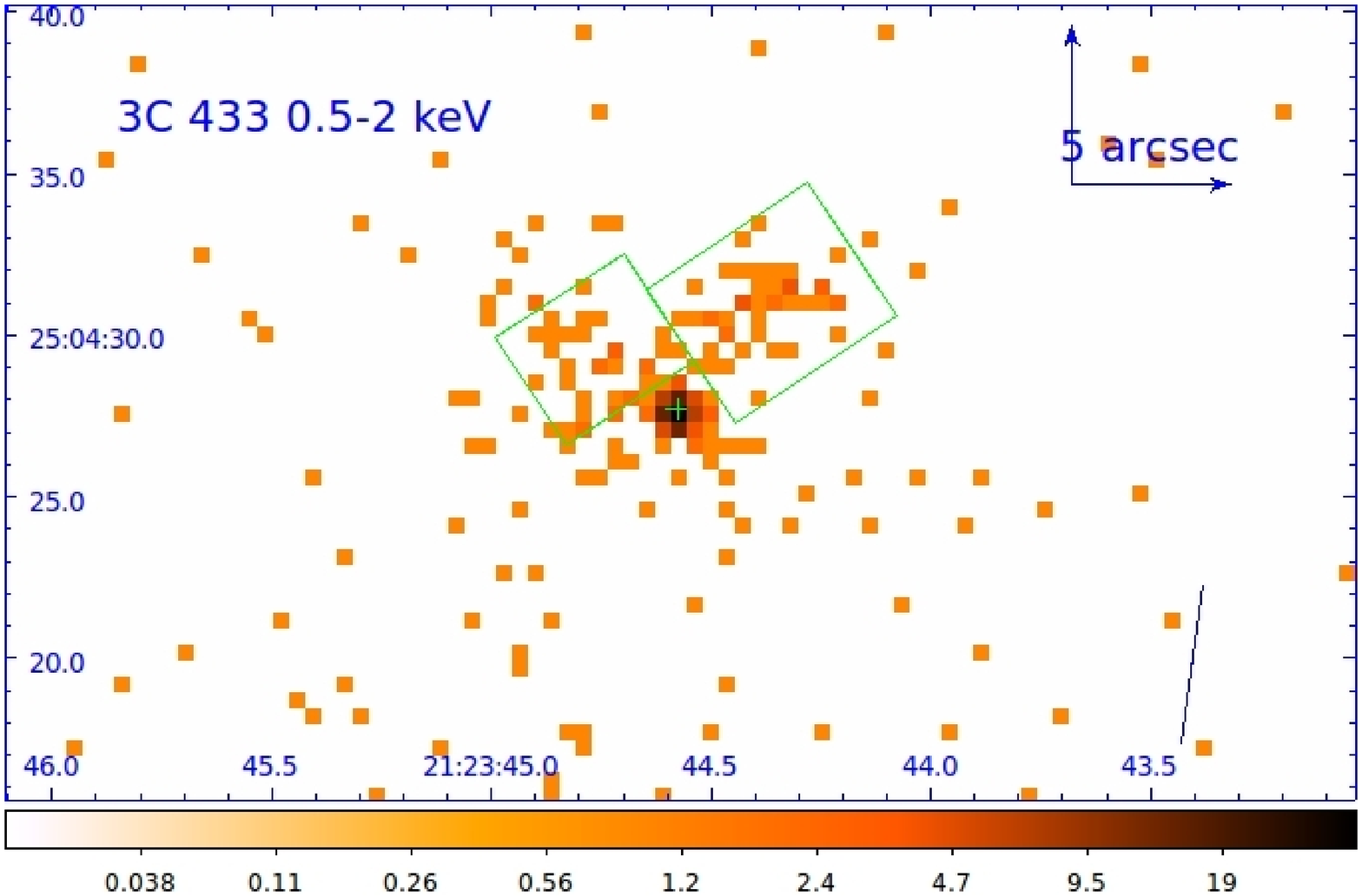}
\includegraphics[width=4.5cm]{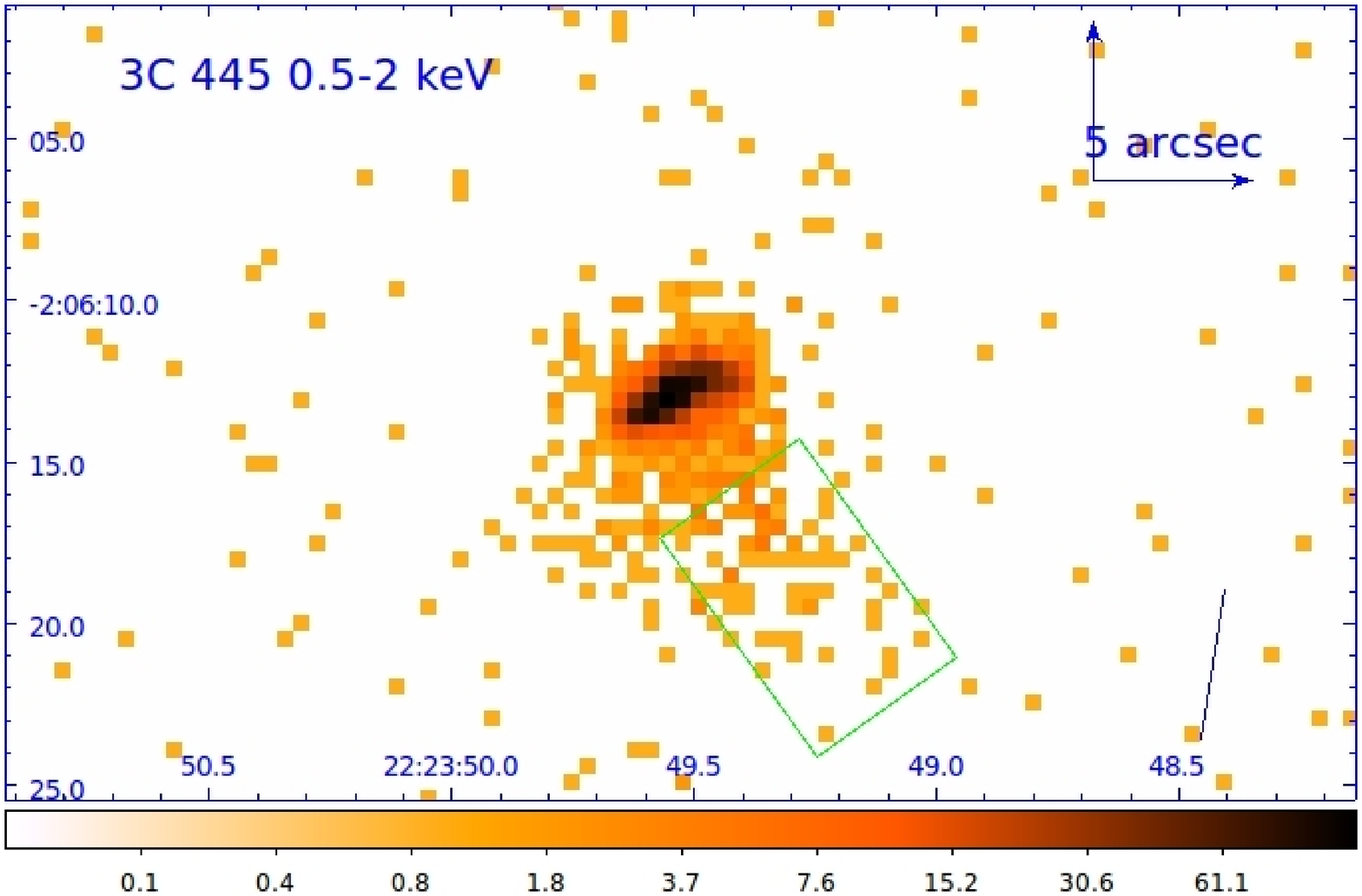}
\includegraphics[width=4.5cm]{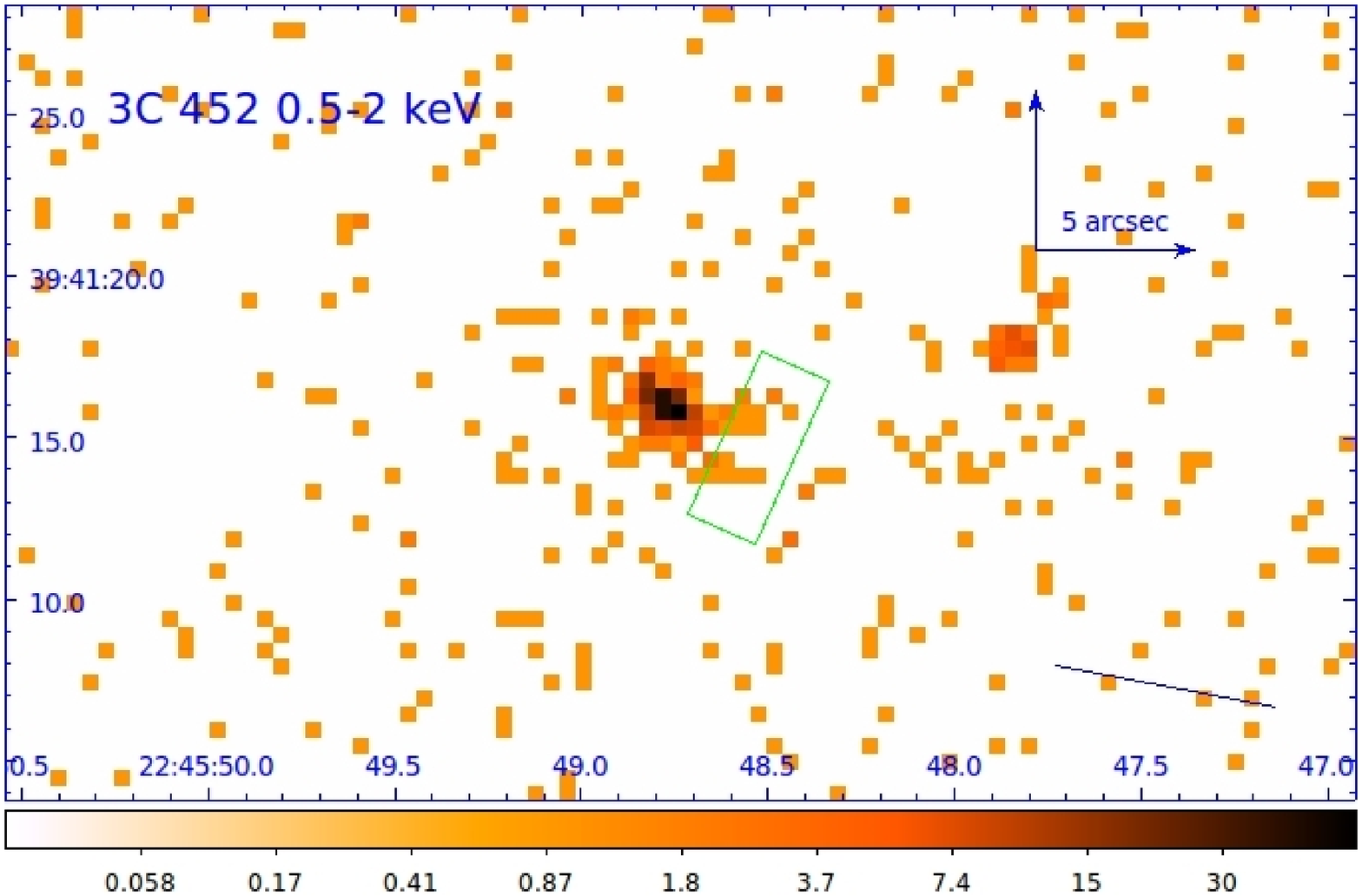}
\includegraphics[width=4.5cm]{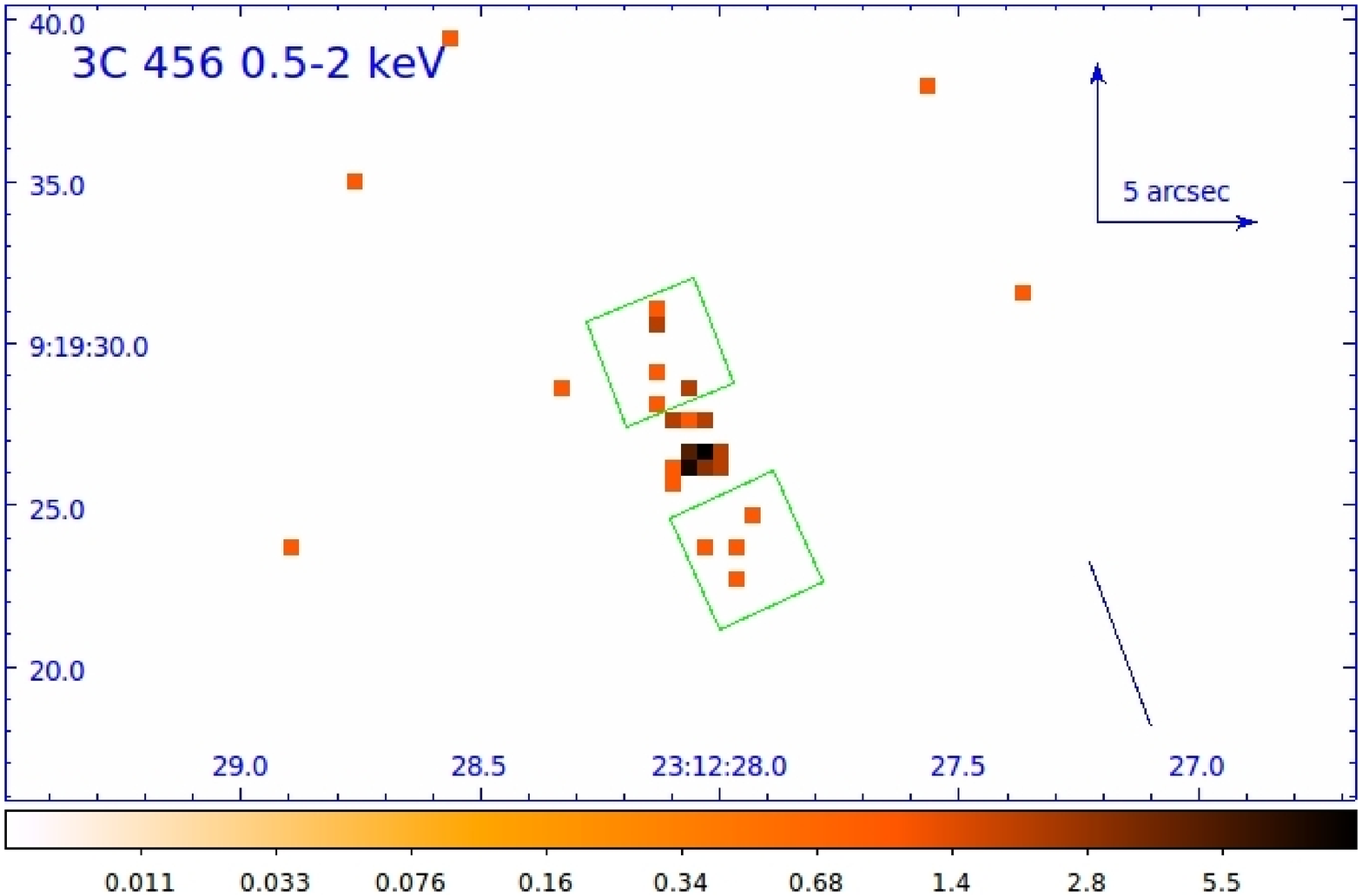}
\includegraphics[width=4.5cm]{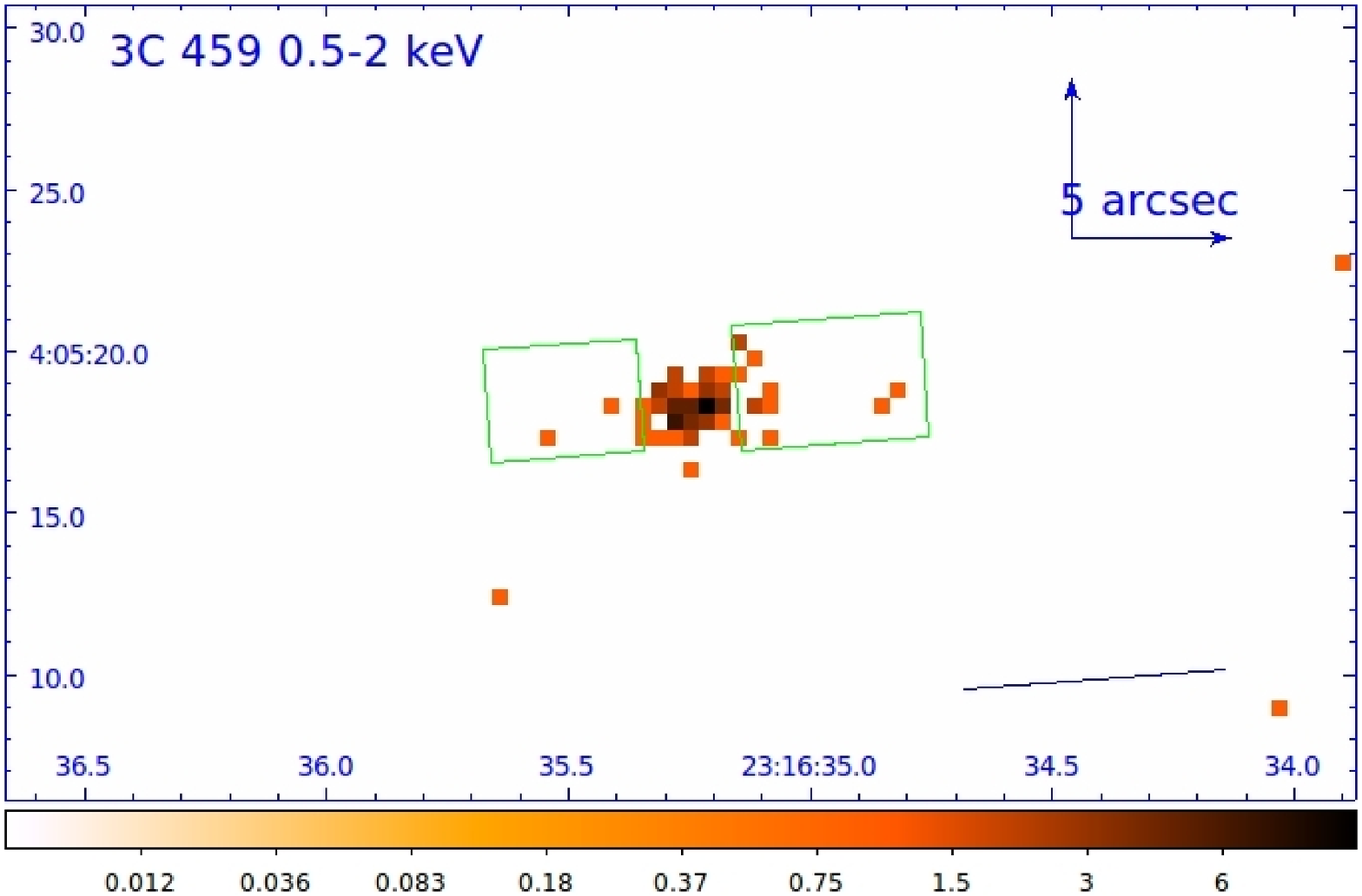}
\caption{Images in the  soft X-ray band for the  16 sources classified
  as extended  or possibly extended  in Tab. \ref{bigtable}  (plus the
  complex source 3C~321 and the "blobby" source 3C~277.3).  The images
  are unbinned  (the pixel  size is 0.49\arcsec)  and we  indicate the
  orientation of  the radio axis.  When extended  emission is present,
  we superposed the region used for the spectral extraction.}
\label{images}
\end{figure*}

Supermassive  black  holes  (SMBHs)  have  a profound  effect  on  the
evolution of galaxies but the nature of the relationship between these
two entities is still an open problem. This depends on how much of the
released  energy interacts with  the surrounding  matter and  how this
accretes onto the SMBHs. This feedback process can be explored through
a  multi-wavelength   analysis  of   the  emission  observed   in  the
circumnuclear regions. The vast  collection of emission lines, ranging
from  the mm  to the  X-ray bands,  reveal the  presence of  a complex
multiphase  medium surrounding the  AGN. Of  special interest  for our
purposes  is the  so-called  Narrow Line  Region  (NLR) where  optical
emission  lines with  widths of  several hundreds  of km  s$^{-1}$ are
produced.  The NLR  is just outside (or even  within) the SMBHs sphere
of influence  and its physical  and dynamical properties  are strongly
affected  by the  central  engine.  It  exists  at the  interface
between the  active nucleus and the  galaxy, and is  thus a convenient
laboratory  in  which  to  explore  the energy  exchange  between  the
two.  Moreover,  the NLR  is  mostly  free  from the  effects  of
obscuration  and it is  resolved in  most nearby  AGN, allowing  us to
perform  spatially resolved studies  and morphological  comparisons in
different energy bands. Indeed, it offers a wide variety of diagnostic
tools to probe the gas physical conditions.

The  NLR and  the processes  that occur  in it  have  been extensively
studied  in many  bands.  In this  paper  we focus  our  study on  the
properties  of the  circumnuclear emission  seen in  the  Chandra soft
X-ray images in  the complete sub-sample of nearby  ($z<0.3$) 3CR radio galaxies
with High Excitation (HEGs) or Broad (BLO) emission lines.

The first case reported in  the literature of extended soft X-ray emission
(below  $\sim$2 keV)  associated with  an  AGN is  the Seyfert  galaxy
NGC4151 (\citealt{elvis83}) observed  with {\it Einstein}. Since then,
it has  been recognized that a  few bright, nearby  Seyfert 2 galaxies are associated with
soft X-ray emission  extending over $\sim$1 kpc and matching
very   closely   the   morphology    of   the   optical   NLR   (e.g.,
\citealt{elvis90}). This result was strengthened by later observations by the
{\it   Einstein}  and   ROSAT   satellites  (e.g.,   NGC~1068:
\citealt{wilson92},   NGC~2992:  \citealt{elvis90},  NGC~2110:
\citealt{weaver95}). This was not a totally unexpected result,
because a hot  medium in pressure equilibrium with  the NLR is
thought to  be necessary to prevent the  optical emitting line
clouds  from evaporating  \citep{krolik84}. In  fact initially
the preferred explanation of these extended soft X-ray regions
was an outflow of hot, collisionally ionized gas, that confines
the narrow-line clouds (\citealt{wilson92}, \citealt{elvis90},
\citealt{weaver95}), although  other emission mechanisms, such
as scattering of nuclear light, could not be ruled out.

Recently, with the advent of a new generation of higher resolution and
sensitivity  X-ray telescopes,  such as  Chandra and  XMM-Newton, more
detailed comparison  on sub-arcsec  scales between the  X-ray, optical,
and  radio  emission  has   been  possible.  Chandra  images  at  high
resolution and  XMM spectroscopic observations, combined  with HST and
VLA images, have  been used to investigate a  larger number of bright,
nearby    AGN,    mostly     Seyfert    galaxies    (e.g.,    NGC~2110
\citealt{evans06}). \citealt{bianchi06} studied a sample of 8 Seyferts
and found extended soft X-ray emission co-spatial with the NLR for all
of their  sources. High  resolution spectroscopic observations  of the
extended X-ray emitting regions performed with Chandra/HETGS, possible
in   a   few   cases   (e.g.   NGC~1068   \citealt{ogle03},   NGC~4151
\citealt{wang11}) reveals that the soft  X-ray part of the spectrum is
dominated by emission  lines mainly from He- and  H-like K transitions
of light metals similar to those observed in their nuclei (e.g. MRK 3:
\citealt{sako00}, Circinus:  \citealt{sambruna01}). X-ray observations
made   with  the  Reflection   Grating  Spectrometers   (RGS)  onboard
XMM-Newton have lower resolution (15$\arcsec$), thus encompassing both
the nucleus and the soft  X-ray extended region. The resulting spectra
therefore represent intensity-weighted conditions over kpc scales,
but they  are consistent with  an ensemble of several  narrow emission
lines (e.g.  \citealt{kinkhabwala02}, \citealt{guainazzi07}, NGC~5252:
\citealt{dadina10}).

Nevertheless,  although  in  Seyfert  2  galaxies  the  most  probable
explanation is  that the gas is  photoionized by the AGN,  there is no
general consensus  about the dominant  process that produces  the soft
X-ray  extended emission  in other  classes of  galaxies.  In  fact, a
variety of  different mechanisms could  be considered, such as:  a hot
component  of a multiphase  interstellar medium,  hot gas  shocked and
evacuated  by an  outflow or  a jet,  or an  outflow  of collisionally
ionized  gas escaping  from  the nucleus,  driven  by radiation,  that
interacts  with  the  NLR  clouds.  In an  ”outflow”  model,  the  NLR
kinematics  are dominated  by radiation  and/or wind  pressure driving
clouds outwards  from the  nucleus \citep{crenshaw00}  and indeed
complex motions in the NLR are often inferred.

The literature on the X-ray properties of the NLR in radio galaxies is
more limited  than for  Seyferts. At high  energies (between 2  and 10
keV)  radio galaxies  show a  compact nuclear  component and,  in some
cases, collimated structures co-spatial  with the radio emission. Only
recently extended regions in soft  X-rays, similar to that observed in
Seyfert   galaxies,   have   been   discovered   (e.g.,   in   3C~171:
\citealt{hardcastle2010},    3C~33:    \citealt{torresi09},    3C~305:
\citealt{massaro09}  and \citealt{hardcastle12},  and  PKS 1138\--262:
\citealt{carilli02}). In all cases the soft X-ray emitting regions are
closely spatially  related to the  optical emission lines,  similar to what 
is observed for radio quiet AGN. Also, analysis of the nuclei
performed with  high or  medium resolution spectroscopic  data reveals
many emission lines  in the soft band, as  observed in the radio-quiet
AGN (e.g.  3C~445: \citealt{sambruna2007} et  \citealt{grandi2007}, 3C~33:
\citealt{torresi09}, 3C~234: \citealt{piconcelli2008}).

In this work we perform a complete analysis of the properties of X-ray
emission in  the soft band (0.5-2  keV) for 113 3CR radio
galaxies at $z<0.3$  (from \citealt{spinrad1985}).  A Chandra snapshot
survey has been recently completed  to provide us with observations of
all 3CR sources not already covered by other programs. Now we can take
advantage   of  a   complete   database  of   high  resolution   X-ray
observations. In  this paper we  focus on  the extended
circumnuclear emission of the galaxies  classified as HEG and BLO. Our
aim  is to  establish whether  extended  emission is  present, and  to
explore the relationship with the overall multi-band properties of the
galaxy.

The multi-wavelength data we will use are mainly from the ground based
spectroscopic              survey             presented             by
\citet{buttiglione09,buttiglione10,buttiglione11},    leading   to   a
classification into the three  main classes: broad line objects (BLO),
low and high excitation galaxies (LEG and HEG).  Furthermore, emission
line imaging  surveys of the 3CR  sources have been  carried out using
both ground-based telescopes  \citep{mccarthy95,baum88} and the Hubble Space
Telescope \citep{privon08,tremblay09}.

We adopted the following cosmological parameters: H$_0$=71 km s$^{-1}$
Mpc$^{-1}$,          $\Omega_M$=0.27,          $\Omega_{\Lambda}$=0.73
\citep{spergel03}.

\section{The sample and the observations}
 
The  sample of  3CR radio  galaxies at  redshift $z\;<  0.3$  has been
entirely observed  by Chandra. In  particular we consider the  113 3CR
sources from \citet{spinrad1985} with $z\;< 0.3$ (having only excluded
3C~231,  aka M~82,  a star  burst galaxy  and 3C~71,  aka  NGC~1068, a
Seyfert galaxy). The sample  includes a significant number of powerful
classical edge-brightened  FR~II radio galaxies,  as well as  the more
common      (at       low      redshift)      edge-darkened      FR~Is
\citep{fanaroffRiley1974}. In  this paper we  focus on the  51 objects
classified as HEGs or BLOs in \citet{buttiglione10,buttiglione11}.  We
include these two  classes in our study because  their differences can
be ascribed only to selective  nuclear obscuration and they are likely
to differ only by their orientation  with respect to our line of sight
(e.g., \citealt{antonucci93}).  Indeed, although  the HEGs and BLOs differ in
the  presence of  the  broad  component of  the  permitted lines,  the
luminosities   and   ratios  of   the   narrow   emission  lines   are
indistinguishable.
 
We downloaded  from the Chandra  archive the observations for  all HEG
and BLO  sources. The Chandra  Obs. ID and  exposure times for  the 51
sources of our sample are listed in Tab. \ref{bigtable}. Most of these
data sets (31) were obtained as part of a two cycle snapshot proposal
(PI Dan Harris, exposure time 8 ks per target).

We  applied the standard  reduction data  procedure using  the Chandra
Interactive   Analysis  of   Observations  CIAO   4.3,   with  Chandra
Calibration  Database CALDB  version  4.4.1.  We  set the  observation
specific bad pixel files and when necessary we filtered data to remove
flares,  then  we  generated  a  new level=2  event  file,  using  the
chandra\_repro  reprocessing script. The  standard grade,  status, and
good time filters were applied.

\section{Imaging analysis} 

 As the  aim of  this paper is  the study  of the extended  soft X-ray
 emission,  we need  a quantitative  method to  establish when  such a
 structure  is  indeed  present  in  addition  to  the  nuclear  point
 source. In order  to define a source as  ``extended'', we compare its
 surface brightness  profile (SBP) with that of  the theoretical Point
 Spread Function (PSF).

  The Chandra Point Spread Function (PSF) depends on both the incident
  spectrum and on the source location. We produce the appropriate
  PSF  utilizing Chandra  PSF  simulator (ChaRT)  for each  individual
  source.  The input  parameters of this software are  the position of
  the observed source on the  detector (the off-axis and the azimuthal
  angle),  the  exposure  time,  and  the model  spectrum.  The  model
  spectrum is  obtained from a fit  to the point  source emission. The
  result of the simulation are a collection of rays that MARX projects
  onto the  detector plane obtaining  a pseudo-event file that  can be
  directly compared with the observations.

  We then obtain the SBP of each source by measuring the net counts in
  a series  of equally spaced annuli  centered on the  nucleus, with a
  step of 1 pixel, and dividing by the respective areas. We normalize
  the PSF to match the observed SBP  in the first 2 pixels and we look
  for an  excess in the bins at  larger radii. In the  case of sources
  affected  by pile-up the  normalization is  instead performed  in an
  annulus of  radius between 3 and  5 pixels. We consider  a source as
  ``extended'' when  we measure an excess  of the SBP  with respect to
  the  PSF with  a  significance greater  than  3$\sigma$. We  instead
  classify a source as ``possibly  extended'' when we derive an excess
  at a significance level between 2$\sigma$ and 3$\sigma$. The objects
  that do not show any  evidence of extended features, were classified
  as ``non  extended'' objects or  ``non detected'' if no  emission is
  present at all.

  However, the SBP  method is not always well  suited to capturing the
  presence  of the  faint, elongated  structures we  often see  in our
  images.  We therefore  considered a  second criterion.  In  order to
  emphasize  the  departure from  circular  symmetry  of any  extended
  emission we compare the counts in two rectangular boxes, one aligned
  with  the axis  of optical  Narrow Line  Region, NLR  (or eventually
  along the  radio axis)  and the second  box orthogonal to  the first
  one. The central square formed by the intersection of the two boxes,
  which contains  the nuclear source,  is excluded from  analysis.  We
  consider as  ``extended'' (or ``possibly extended'')  the sources in
  which  the difference between  the counts  in the  two boxes  have a
  significance  larger than  3$\sigma$ (or  between 2  and 3$\sigma$).
  The reduced  uncertainty, associated  with the lower  background and
  PSF contribution  due to the smaller  integration areas, corresponds
  to a higher number of  objects with a positive detection of extended
  emission with this latter method.

  This classification is reported  in Table \ref{bigtable}. Soft X-ray
  images of  the ``extended'' and  ``possibly extended ''  sources are
  shown in Fig.~\ref{images}.

  The CIAO tool  {\it srcextent} is commonly used  to measure the size
  and   the    position   angle   (PA)   of    X-ray   emission   (see
  Tab. \ref{bigtable}).  However, this measurement is dominated by the
  structure of the nuclear source  and it does not return satisfactory
  measurements for  our sources because  we are instead  interested in
  the properties of the extended emission alone. We therefore measured
  the  source   extension  and   the  PA  inspecting   the  iso-counts
  contours. In Table~\ref{bigtable} we also provide a short additional
  description  of  the  soft  X-ray  morphology for  a  few  cases  of
  interest.
\input{bigtable.tex}

\section{Flux measurements}
\label{fluxes}

We extract the spectrum  of the ``extended'' and ``possibly extended''
sources shown in Fig.  \ref{images}. The spectral extraction region is
formed by one or two rectangular boxes, whose position has been chosen
based  on the  X-ray morphology, or,  in a  few cases,

\begin{figure*}
\centering
\includegraphics[width=7cm]{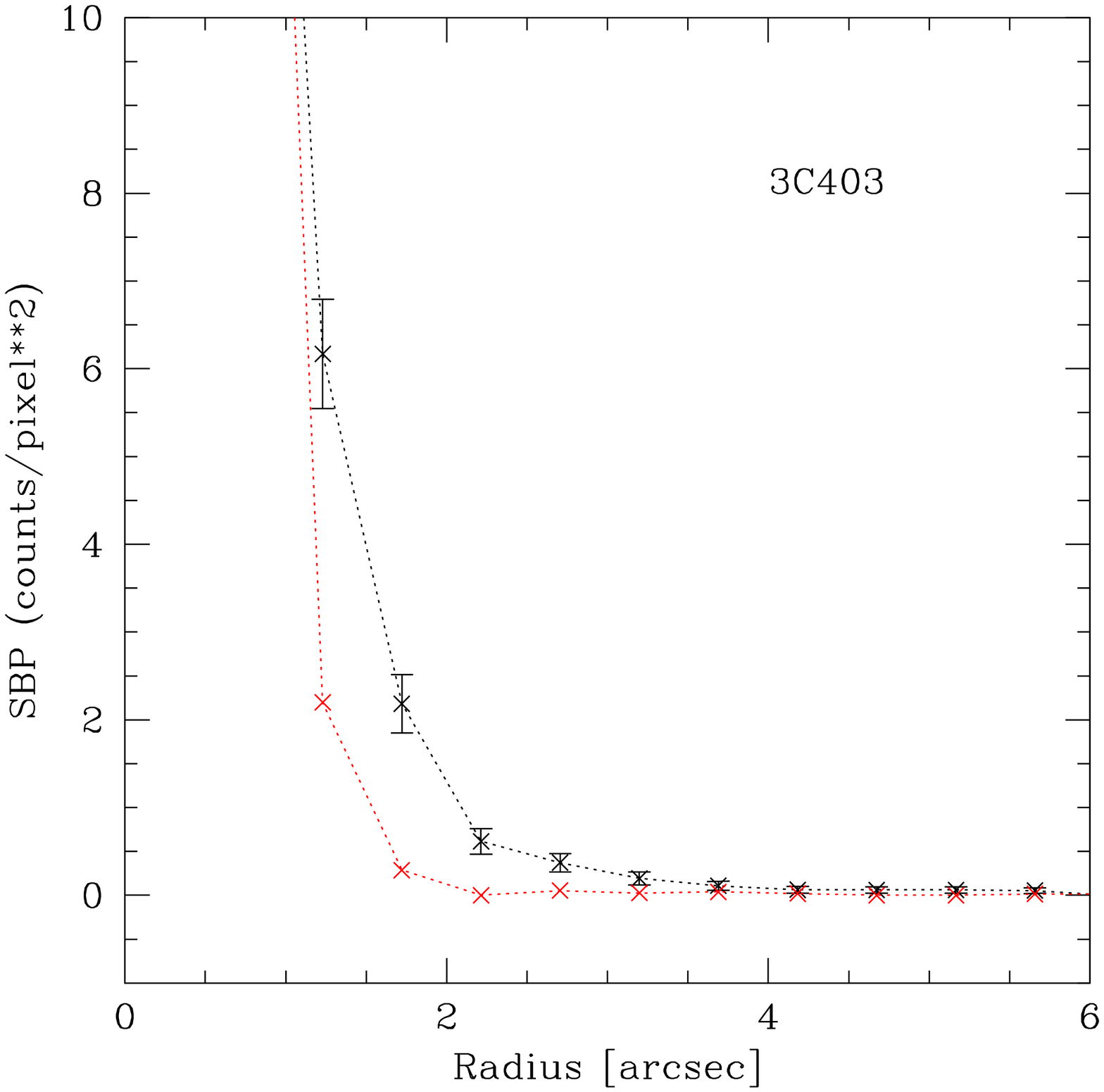}
\includegraphics[width=7cm]{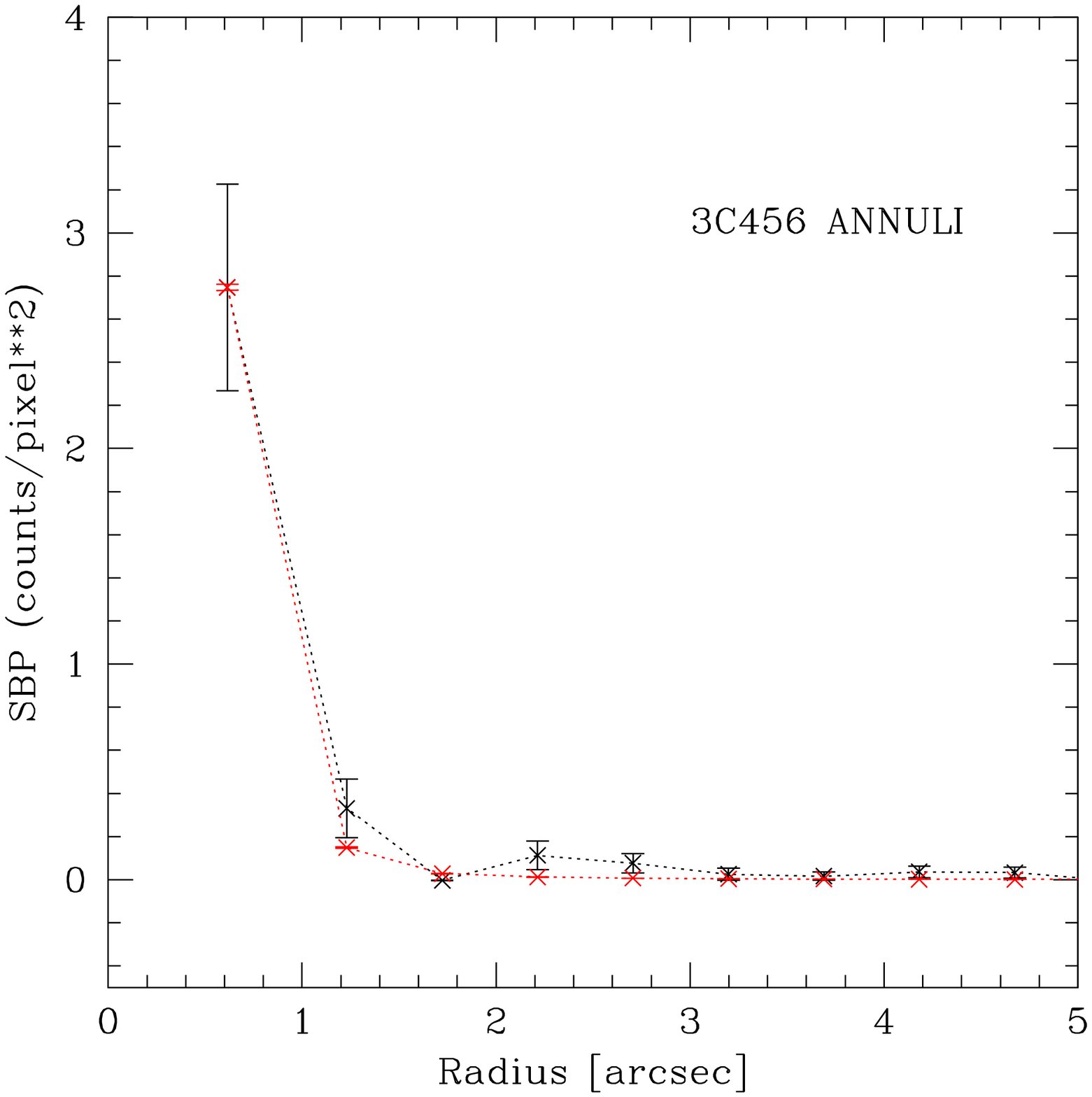}
\includegraphics[width=7cm]{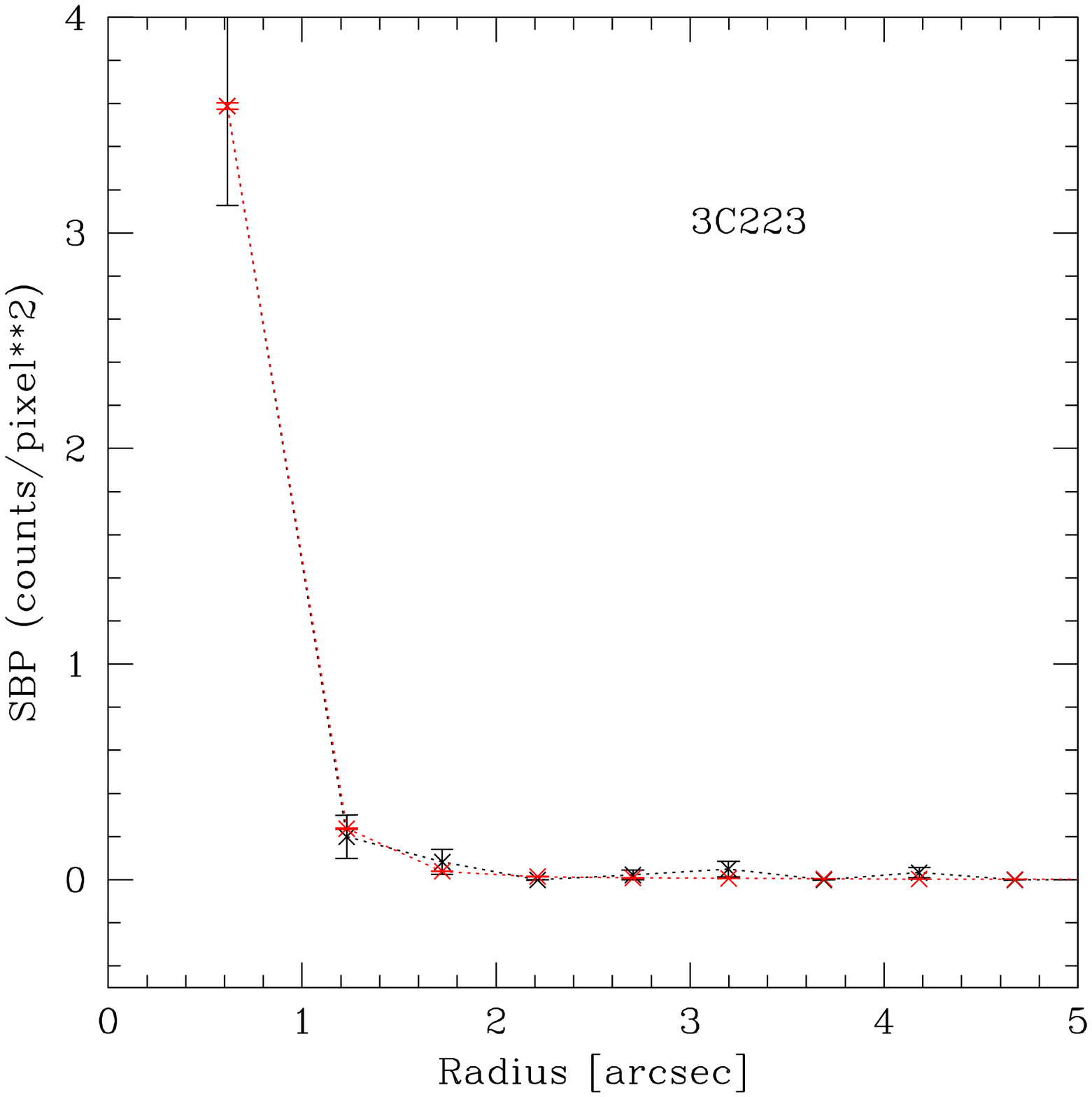}
\includegraphics[width=7cm]{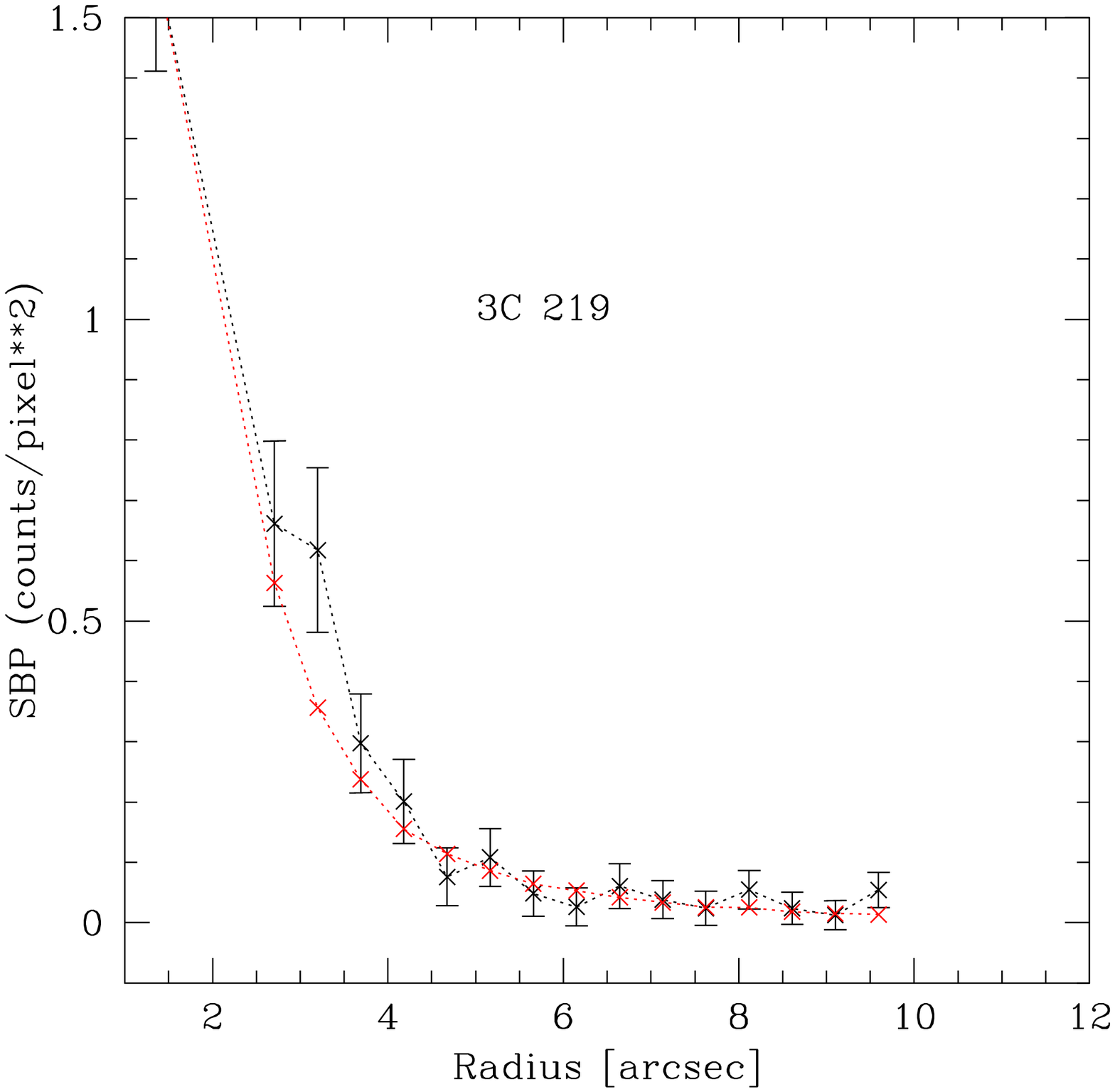}
\caption{Comparison of the observed SBP (black) with the PSF profile (red) for
  4 representative examples. Top left: 3C~403, an "extended" object. Rebinning
  from 3 to 5 pixels we obtain an excess significant at 9.4 $\sigma$. Top
  right: 3C456, an "asymmetric" object.  Rebinning from 5 to 9 pixels, we
  obtain an excess with a significance of 2.5 $\sigma$, but the difference of
  the counts considering two orthogonal boxes (one parallel to the radio axis,
  the second perpendicular to it) centered on the nucleus, is significant at
  3.5 $\sigma$. Bottom left: 3C223, an ``unresolved'' object. Bottom right:
  the BLO 3C~219. The emission seen at large radii is well reproduced by the
  wings of the nuclear point source, with only a marginal (1.4 $\sigma$)
  excess, and no significant asymmetry is found comparing the two orthogonal
  boxes. We consider this object to be ``unresolved''.}
\label{sbp}
\end{figure*}

\begin{figure}
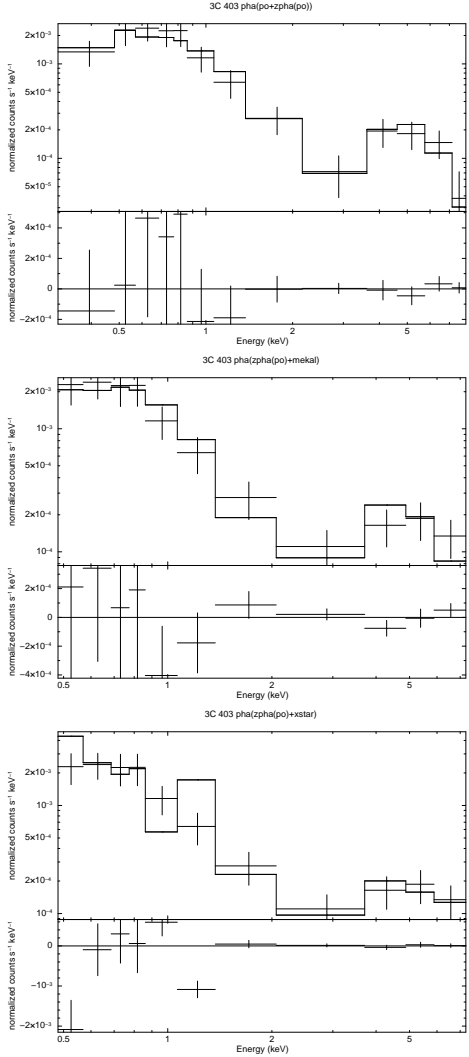

\centering
\includegraphics[width=4.66cm,angle=-90]{19561f17.ps}   
\includegraphics[width=4.66cm,angle=-90]{19561f18.ps}
\includegraphics[width=4.66cm,angle=-90]{19561f19.ps}
\caption{We report, as an example, the spectrum of 3C~403 in the extended
  regions fitted with three different models: power-law, \mekal, and \xstar,
  absorbed by a Galactic column density value.  We add to the model an
  intrinsically absorbed power-law to reproduce the emission seen at higher
  energies (due to the leaking of the nuclear emission).}
\label{3c403fit}
\end{figure}

\input fit.tex

    also the  NLR
morphology. To  avoid nuclear contamination we position  the inner boundary
of the boxes  at a distance of at least 1\farcs2  or 1\farcs5 from the
nucleus, depending on its brightness.
 
In 5 sources  (namely 3C~171, 3C~305, 3C~403, 3C~433,  and 3C~445) the number of
counts in the soft band (between 51 and 134) are sufficient to perform
a spectral analysis. We re-bin the spectra with at least 10 counts/bin
and apply the Poissonian Cash statistic (except for 3C~171 and 3C~445,
for which we re-bin the spectrum at 20 counts/bin and use the $\chi^2$
statistic). We consider  three models to fit the  entire energy range:
power-law, \mekal\footnote{We  also considered an apec  model, but the
  results are  indistinguishable from those obtained  with \mekal\ and
  are not reported.}, and  \xstar\ \citep{bautista01}. As discussed in
more  detail  below,  these   models  correspond  to  three  different
radiation mechanisms in which the  soft X-ray emission can be ascribed
to i) scattered  nuclear X-ray light from free  electrons, ii) hot gas
that is collisionally ionized, and iii) photoionized gas clouds.

In addition to the dominant soft X-ray emission, we note the presence of a 
high energy bump (above $\gtrsim$  2-3 keV) in all sources for which spectral 
analysis was possible (except for 3C 305). As  discussed below, this can be ascribed to the
leaking of the nuclear emission due to the larger PSF extent at higher
energies.

As an example,  we show in Fig.~\ref{3c403fit} the  spectrum of 3C~403
fitted  with the  3 different  models.   In addition  to the  extended
component, we  added an intrinsically absorbed power  law to reproduce
the emission above  2 keV.  Using simulated PSFs  from ChaRT and MARX,
we confirmed that although the  absorbed nucleus in 3C 403 contributes
only $\sim$ 0.3\% of the extended flux in the 0.5-2 keV range, it can produce
substantial leakage at higher energy. For  this source,  the predicted flux  value in  the extended
region  of the  galaxy is  consistent with  the value  of  $\sim$ 2\%,
obtained from the  spectral fit.  The fit for the  other 4 sources are
given in  the Appendix. The fit  results are given  in Table \ref{fit}
and will be discussed in Section \ref{discussion}.

Apart from  these five  sources, the observed  counts in  the extended
regions  are   not  sufficient  for  a   detailed  spectral  analysis.
Nonetheless, as  discussed below,  a robust conversion  between counts
and fluxes can be still obtained.

We convert  the measured  counts into X-ray  fluxes adopting  the same
models used above. For each model the normalization is set in order to
match  the model  count  rate to  the  observed one.   We correct  the
observed  fluxes of  each source  for the  Galactic column  density as
given  by  \xspec\ and  report  them  in  Table \ref{flux}.   For  the
power-law  model we  used  as  reference value  for  the photon  index
$\Gamma=1.7$.   For   \mekal\  model  we  fixed   the  temperature  at
$kT=1$\,keV, the  abundance at  0.5 of the  solar value,  the hydrogen
density at  1 cm$^{-3}$.  For the \xstar\  model, we generate  a table
with  XSTAR2XSPEC  that can  be  used in  the  model  fitting data  by
\xspec. The  electron gas  density of the  clouds is held  constant to
$N_{\rm  e}=10^3 \,  {\rm cm}^{-3}$  and the  temperature is  fixed to
$T=10^5$ K. We choose  as input configuration a simple photoionization
model comprising a  power-law continuum, $f_{\nu}\propto \nu^{\alpha}$
($\alpha=-0.7$ fixed),  that is photo-ionizing  optically thick, solar
abundance emitting clouds.

\begin{figure*}
\centering
\includegraphics[width=6cm]{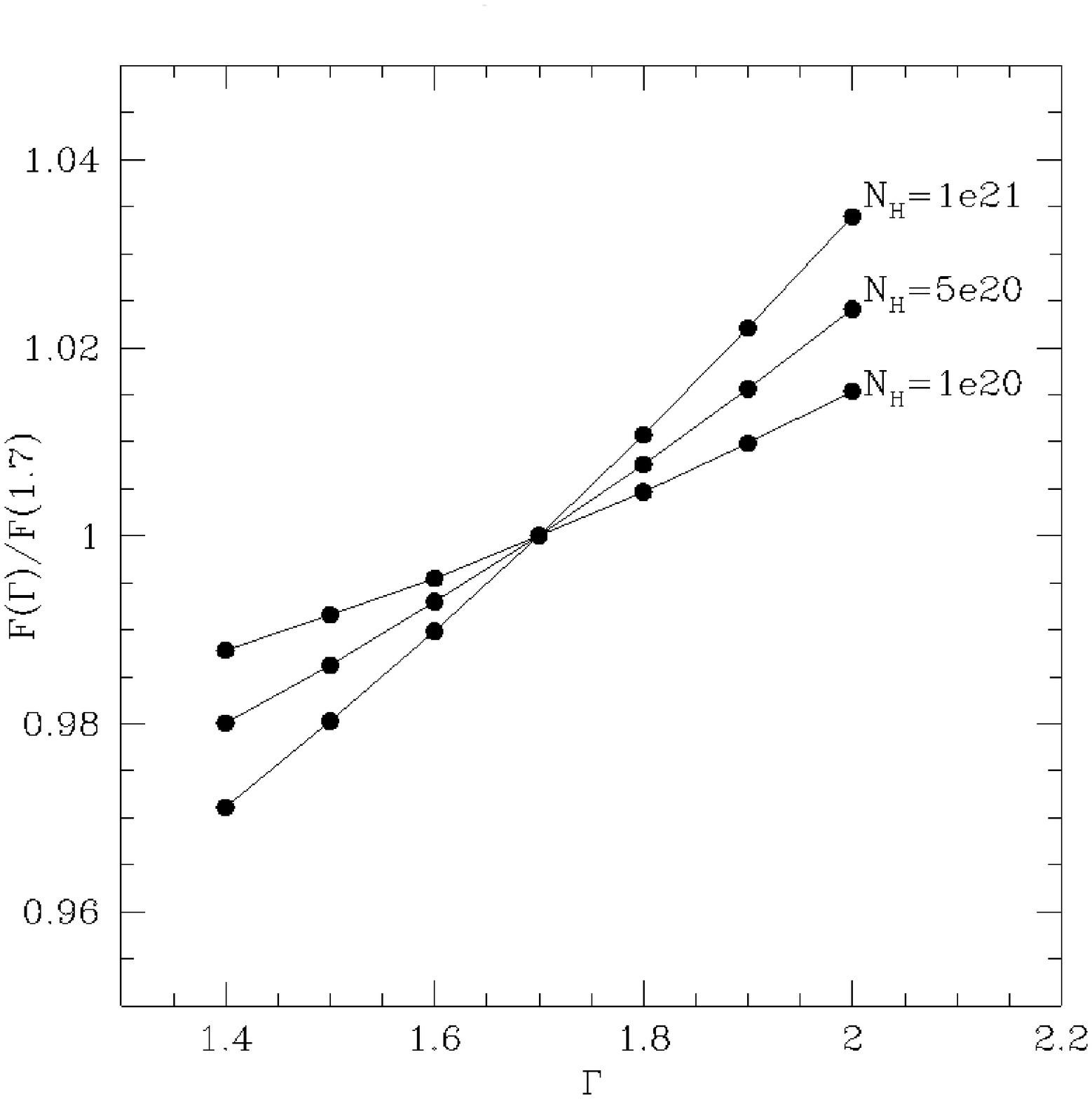}
\includegraphics[width=6cm]{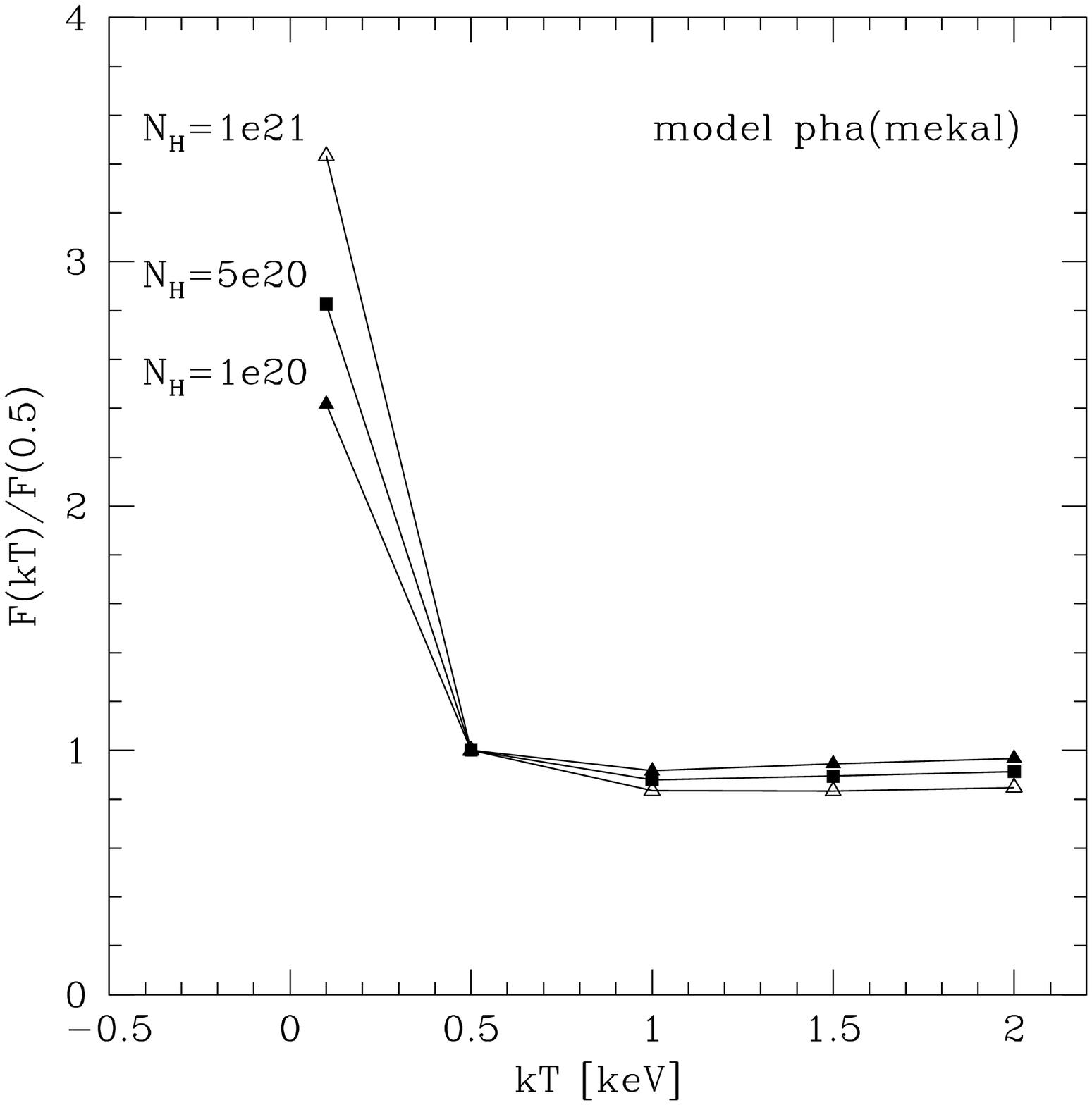}
\includegraphics[width=6cm]{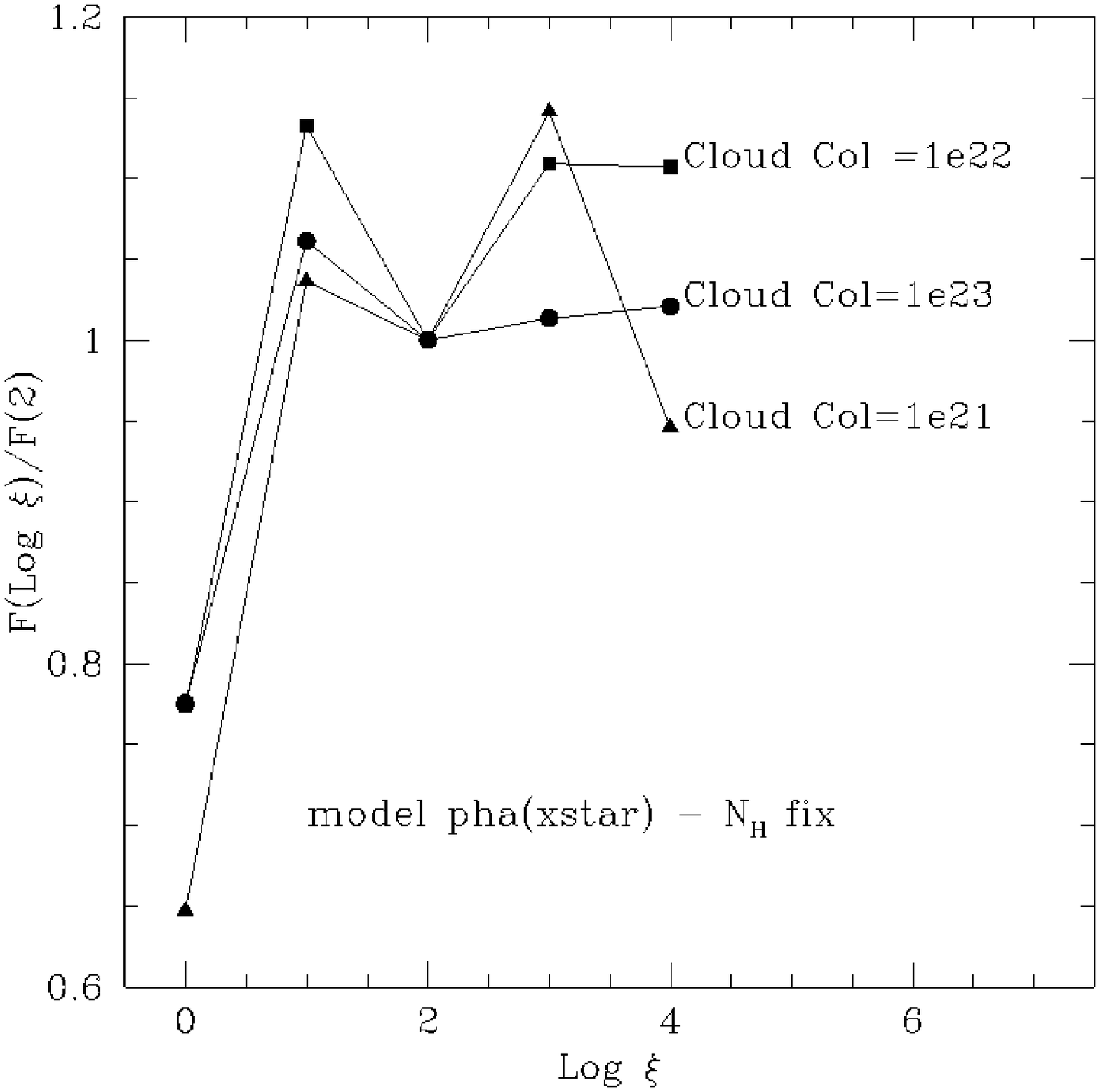}
\caption{For the three different models considered, phabs (pow), phabs(\mekal),
  and phabs(\xstar), we evaluate the flux variations in the 0.5-2 keV range
due to changes in the most
  important parameters of the models (see Section \ref{fluxes}). These values have been
  calculated for a source at the aim-point in ACIS/S.}
\label{compareMod}
\end{figure*}

Different  models  do  not   produce  relevant  discrepancies  on  the
resulting flux values: for a power-law, a collisional hot diffuse gas,
or a  photo-ionized gas, we  obtain marginal flux  differences usually
within a $\pm$ 10\% level (see Table \ref{flux}).

We also  explore how  our choice of  the model parameters  affects the
flux measurements.   We choose an  arbitrary source test  (we verified
that it is  observed near the aim-point in ACIS/S;  the result that we
obtain can be extended to  all the sources observed near this position
on  the  detector).   We  calculated  the  model  flux  for  different
combinations of  the varying parameters with respect  to the reference
value  and we  plot these  ratios in  Fig.  \ref{compareMod}.  For the
power-law  model we  used  as  reference value  for  the photon  index
$\Gamma=1.7$,   and   explored   the   range   $\Gamma=1.4-2.0$.   For
\mekal\ model we varied the temperature in the range $kT=0.1 - 2$ keV.
For  the  \xstar\ model,  we  varied  two  parameters: the  ionization
parameter\footnote{Defined   as  $\xi   =   \frac{L}{nR^2}$  [erg   cm
    s$^{-1}]$, where $L$ is the luminosity of the ionizing source [erg
    s$^{-1}$]  integrated between  1  and 1000  Rydberg,  $n$ is the  gas
  density [cm$^{-3}$], and r is the  distance of the ionizing source from
  the absorbing  gas [cm].}  Log  $\xi = 0  - 4$ and the  cloud column
density  from  10$^{21}$  to  10$^{23}$ cm$^{-2}$.   The  results  are
graphically  presented  in   Fig.   \ref{compareMod}.  With  the  only
exception of  a thermal model of  very low temperature (kT  = 0.1 keV)
the various models differ from each other by less than $\sim$ 20 \%.

The main uncertainties in the flux measurements are not related to the
choice of the model type  or model parameters, but from the (generally
low) number  of counts.  Indeed, in 3  sources (3C~180,  3C~379.1, and
3C~452) the emission  from the circumnuclear regions does  not reach a
3$\sigma$ significance. At this stage, we discard 3C~321  from any further analysis due to its
complex structure,  strongly contaminated  by a nearby  Seyfert galaxy
\citep{evans08}. Summarizing, this analysis allows us  to measure the soft X-ray fluxes
from the extra-nuclear regions in 14 3CR sources.

\input flux.tex

\section{In which sources do we observe extended soft X-ray emission?}

We  find extended  soft X-ray  emission in  14 out  of the  51 sources
considered  and for  5 of  them we  have performed  both  spectral and
photometric analysis.  By separating them in  spectroscopic classes we
have 9  extended (and 3 possibly  extended) images among  the 33 HEGs,
with a detection fraction of $\sim$36\%.

In the 18 BLOs, extended emission is seen only in 2 objects; the lower
detection  rate in  BLOs  with respect  to  HEG might  be  due to  the
presence of their bright X-ray nuclei, possibly outshining any genuine
diffuse emission.  We explore this interpretation,  by testing whether
the typical extended emission in HEGs would have been detected also in
BLOs.

  The count rates  seen in the extended soft X-ray  regions of HEGs is
  in  the range  $\sim0.2\times10^{-3}-2\times10^{-3}$  counts/s, with
  the  only  exception of  3C~305  where  it reaches  $2\times10^{-2}$
  counts/s. The  regions extend  from $\sim2\arcsec$ from  the nucleus
  out to a  radius of $\sim5-10\arcsec$.  We then  estimated the count
  rate  produced by  the PSF  wings  of the  BLO in  the same  spatial
  regions          which         is          typically         between
  $\sim4\times10^{-3}-6\times10^{-3}$ counts/s. Therefore, an extended
  X-ray emission with a surface brightness similar to that seen in the
  HEG  would  be a  factor  between  2 and  30  with  respect to  that
  associated with  the nuclear source.  This implies that even  in the
  presence of genuine soft X-ray extended emission in BLOs, similar to
  that observed in the HEGs, this would not be generally detected from
  the analysis of  the surface brightness profiles nor  would induce a
  significant asymmetry.

\begin{figure}
\centering
\includegraphics[scale=0.4]{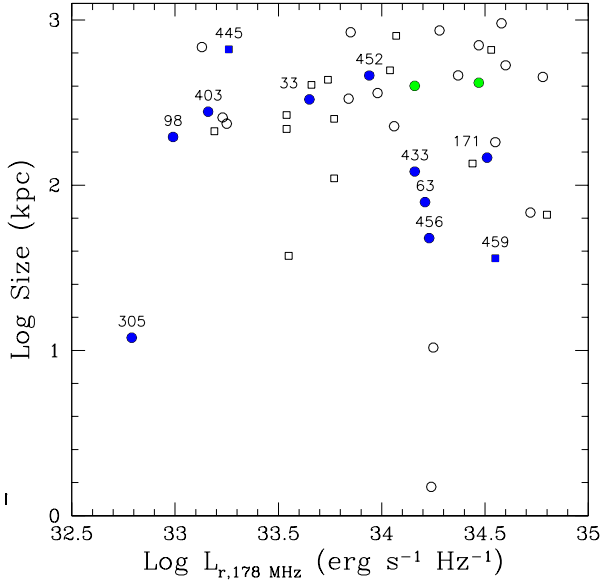}
\caption{Radio luminosity vs. radio size for the 3CR HEG and BLO
  sources. Circles are HEGs, squares are BLOs. We marked with a solid blue
  sign the sources with extended soft X-ray emission (for which we also show
  the galaxy's name) and with green symbols the partially extended sources.}
\label{radioprop}
\end{figure}

We then look  for a relation between the  soft X-ray extended emission
and the  radio properties,  i.e., the size  and the luminosity  of the
radio  source. In  Fig. \ref{radioprop}  we  plot the  radio size  and
luminosity and  we highlight  the sources that  show an  extended soft
X-ray emission. We find that the fraction of extended objects does not
depend on  either their size or radio-power.  Similarly, no dependence
on redshift is seen.

The stronger link  is found, rather obviously, with  the length of the
Chandra  observations.  Indeed, considering  only  the exposures  with
${\rm t_{exp}} >$  10 ks, we find that  75\% (6 out of 8)  of the HEGs
show  extended regions, against  13\% (3/23)  of the  sources observed
with  ${\rm  t_{exp}}  \lesssim$  10  ks. By  including  the  possibly
extended and the complex sources, the percentages increase to 80\% and
26\% for the two classes, respectively.

Summarizing, we find that when a 3CR source fulfills two requirements,
i.e.,  the  absence  of  a   bright  nucleus  and  when  long  Chandra
observations are available, a soft X-ray extended emission is normally
observed  implying that  this  is a  general  characteristic of  radio
galaxies. As shown in the next Section, however, the best predictor of
the X-ray properties is represented by the morphology and the strength
of the optical emission line region.

\section{Comparison between X-ray and optical emission}

\begin{figure*}
 
\includegraphics[width=6cm]{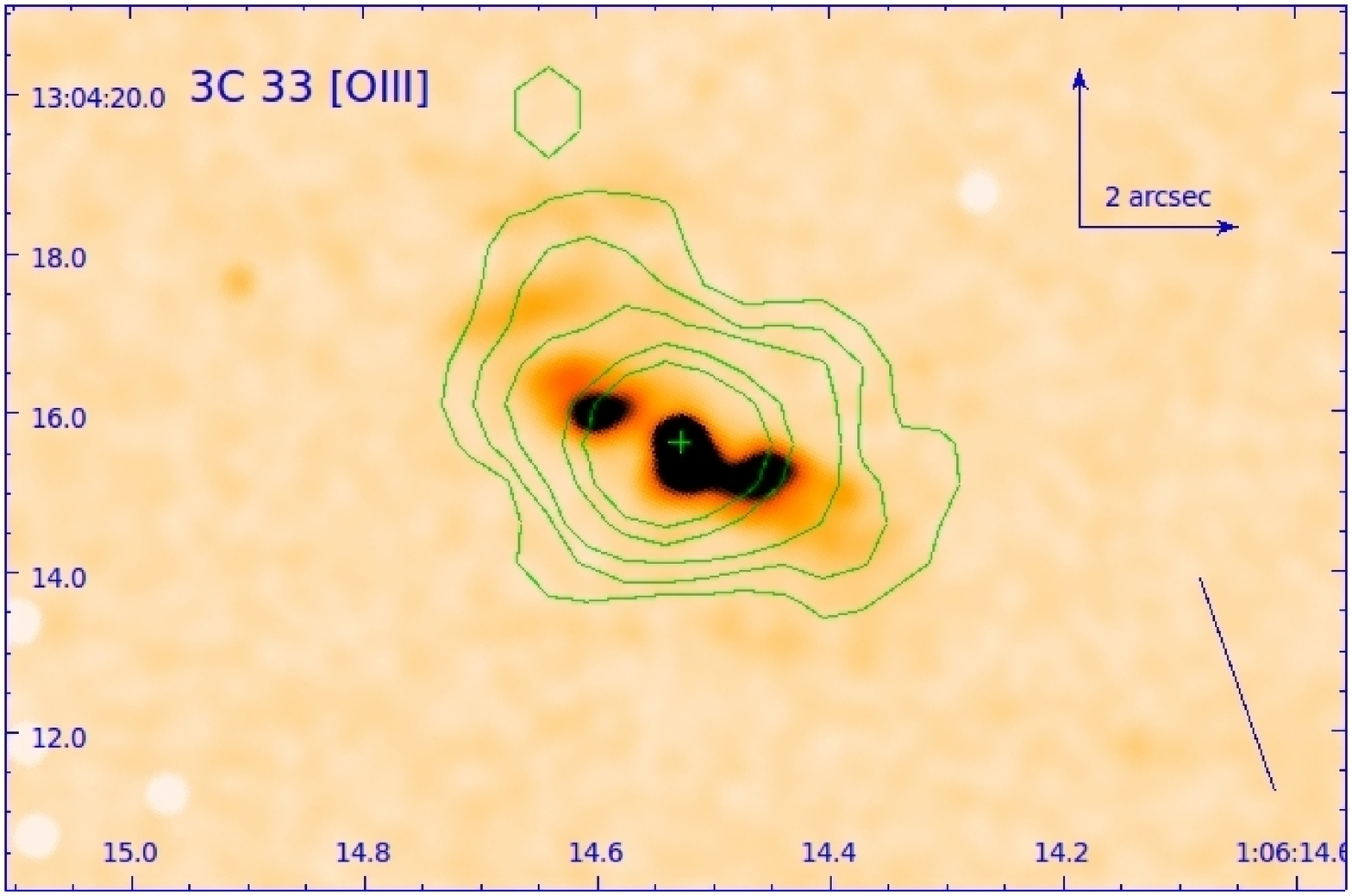}
\includegraphics[width=6cm]{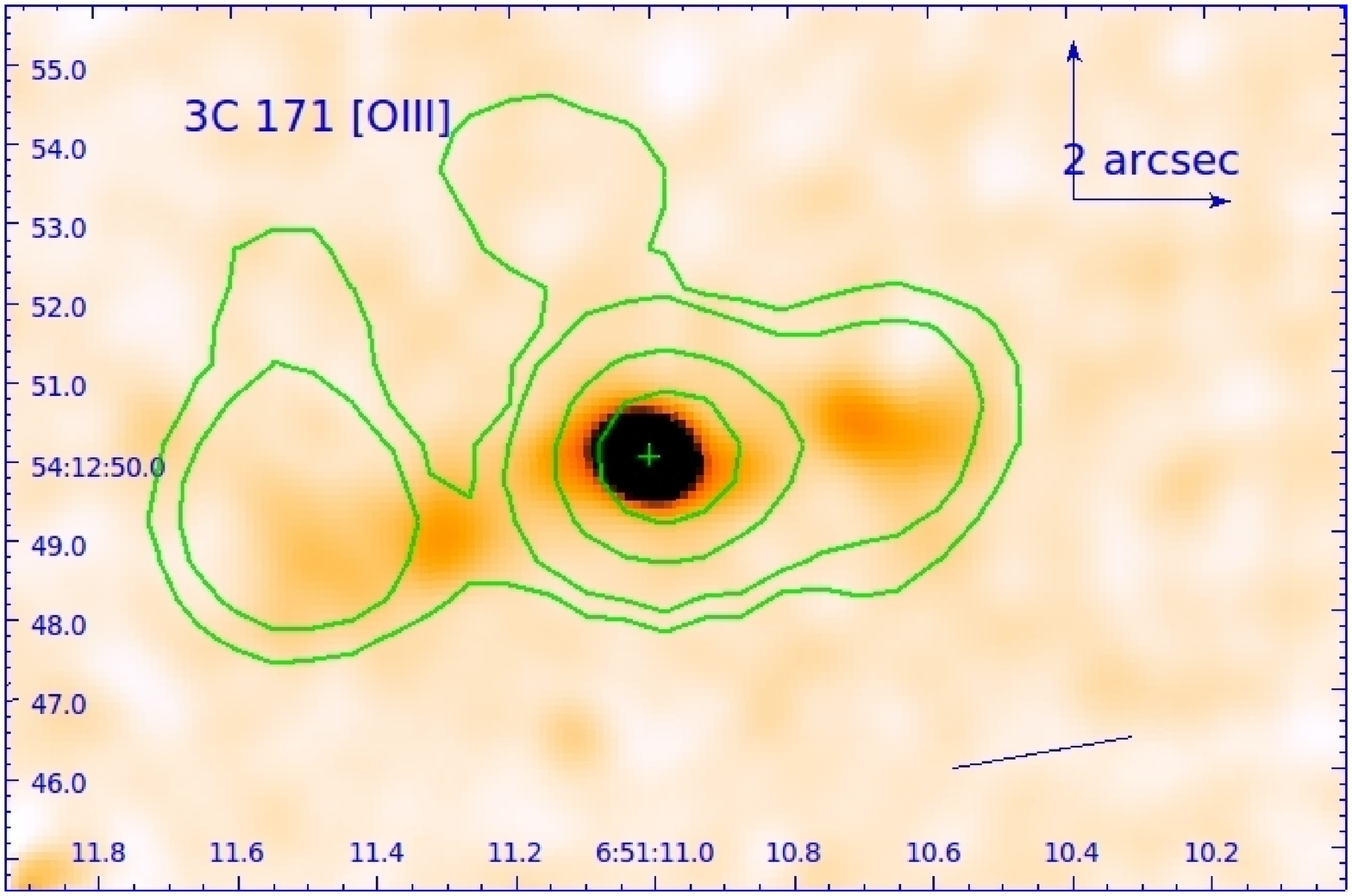}
\includegraphics[width=6cm]{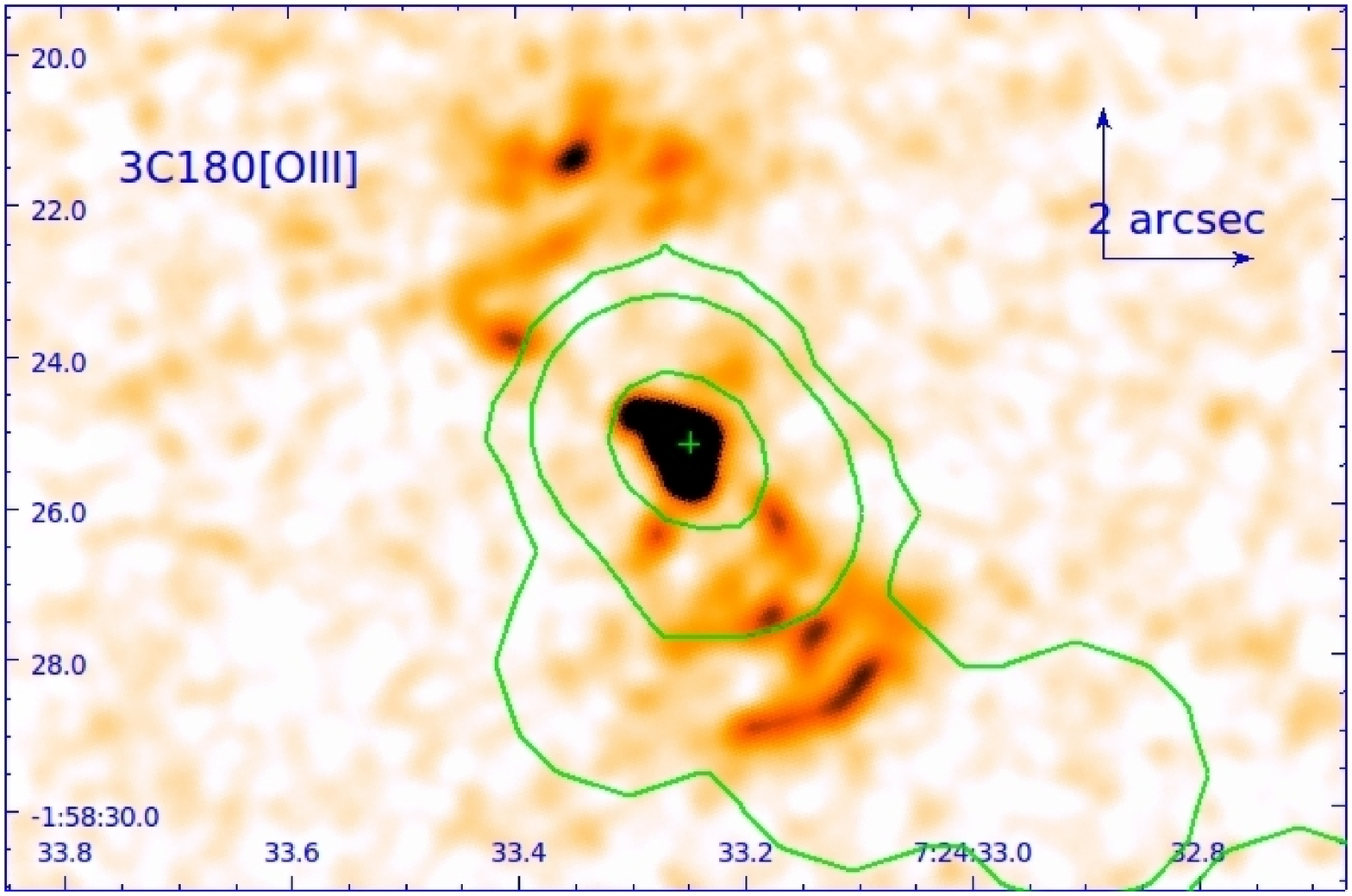}
                                        
\includegraphics[width=6cm]{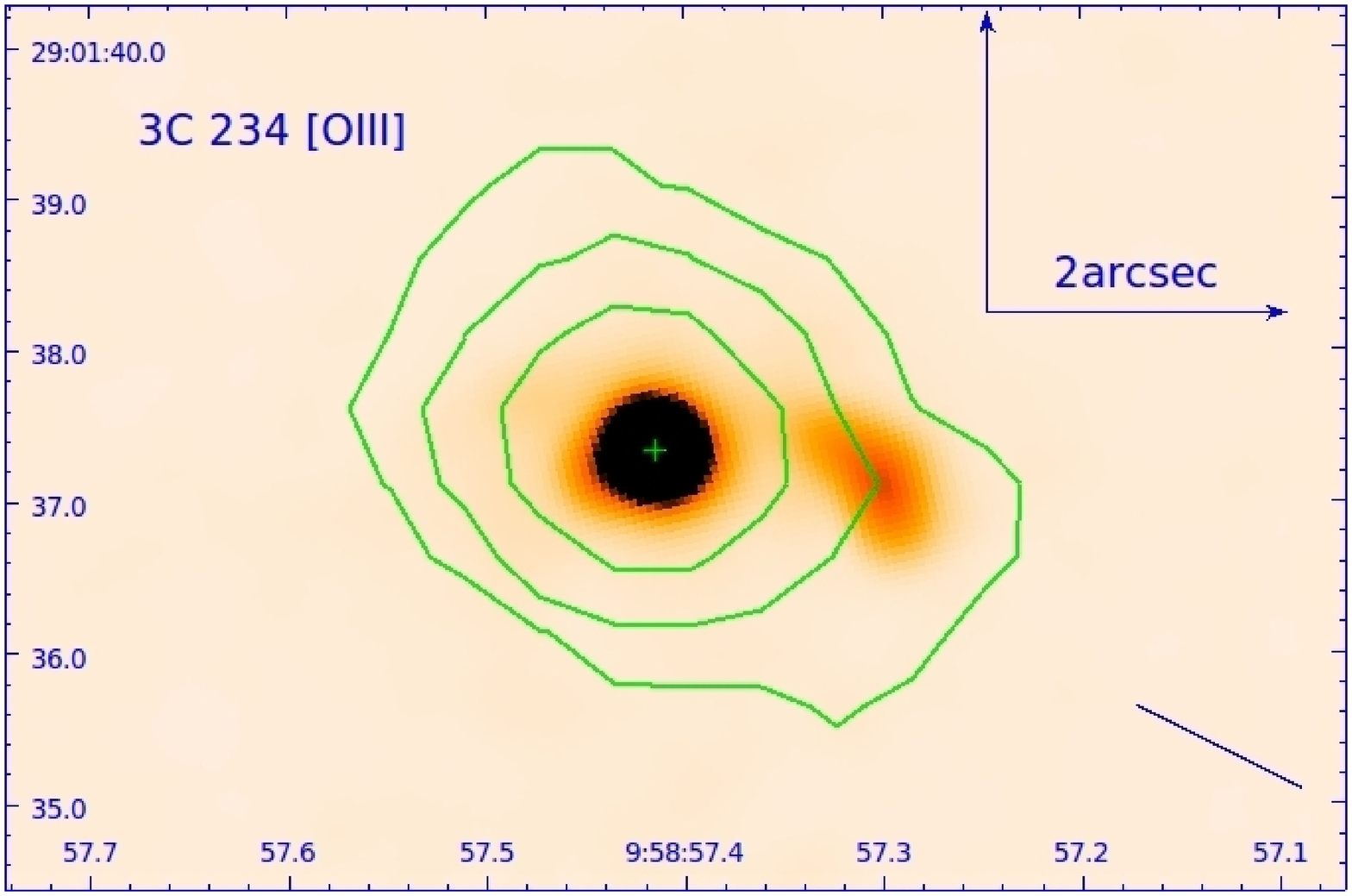}
\includegraphics[width=6cm]{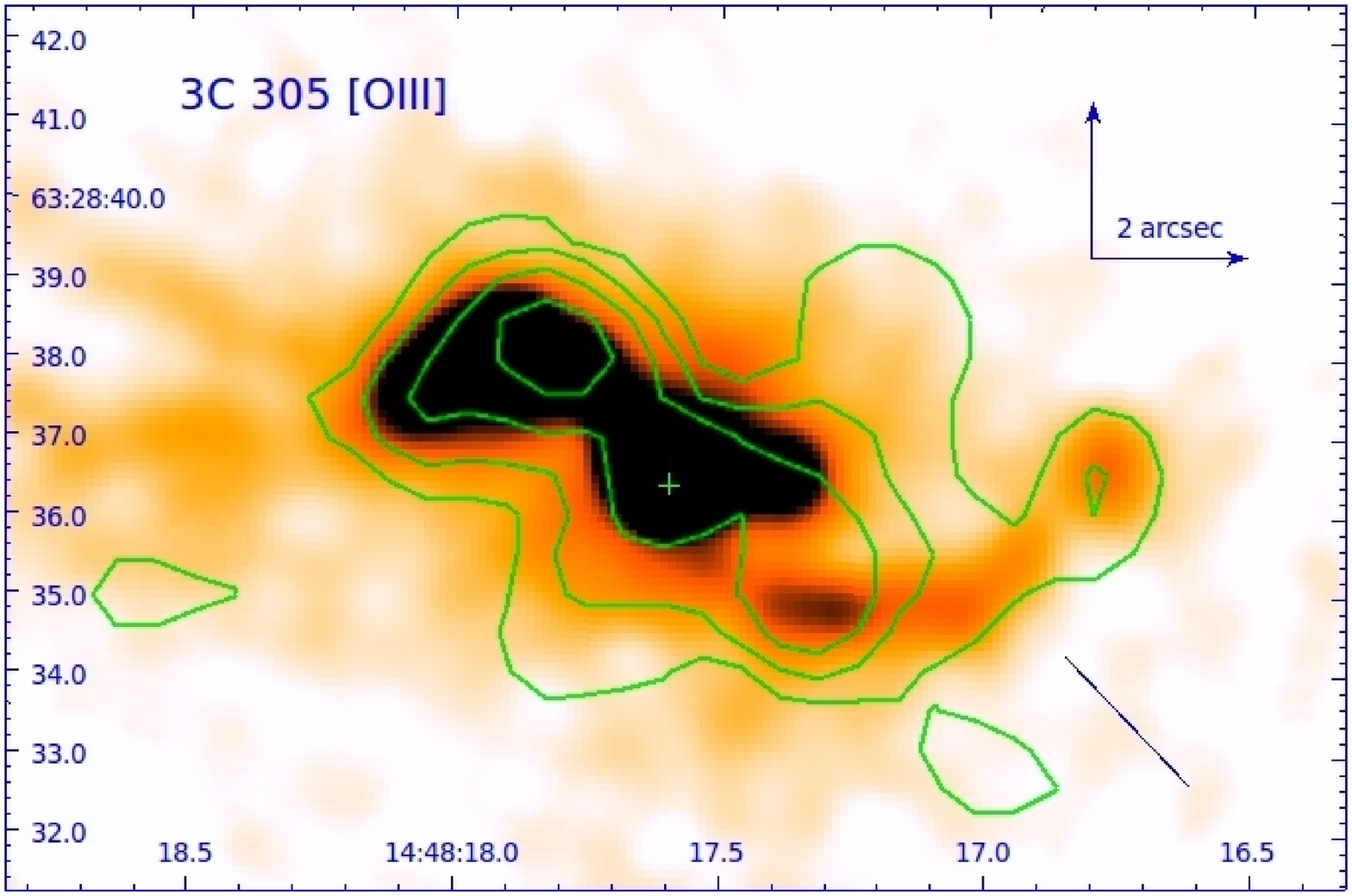}
\includegraphics[width=6cm]{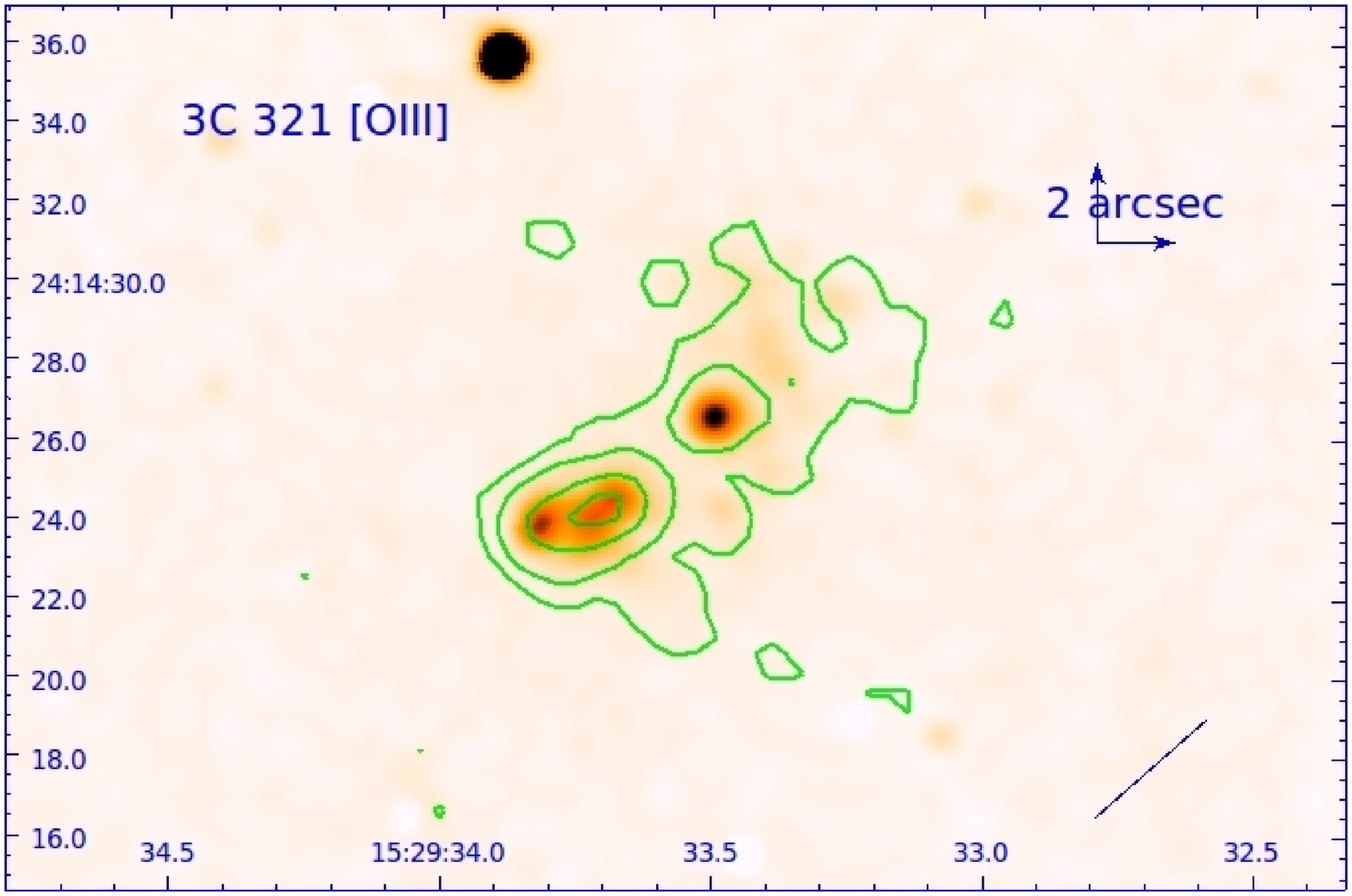}
                                        
\includegraphics[width=6cm]{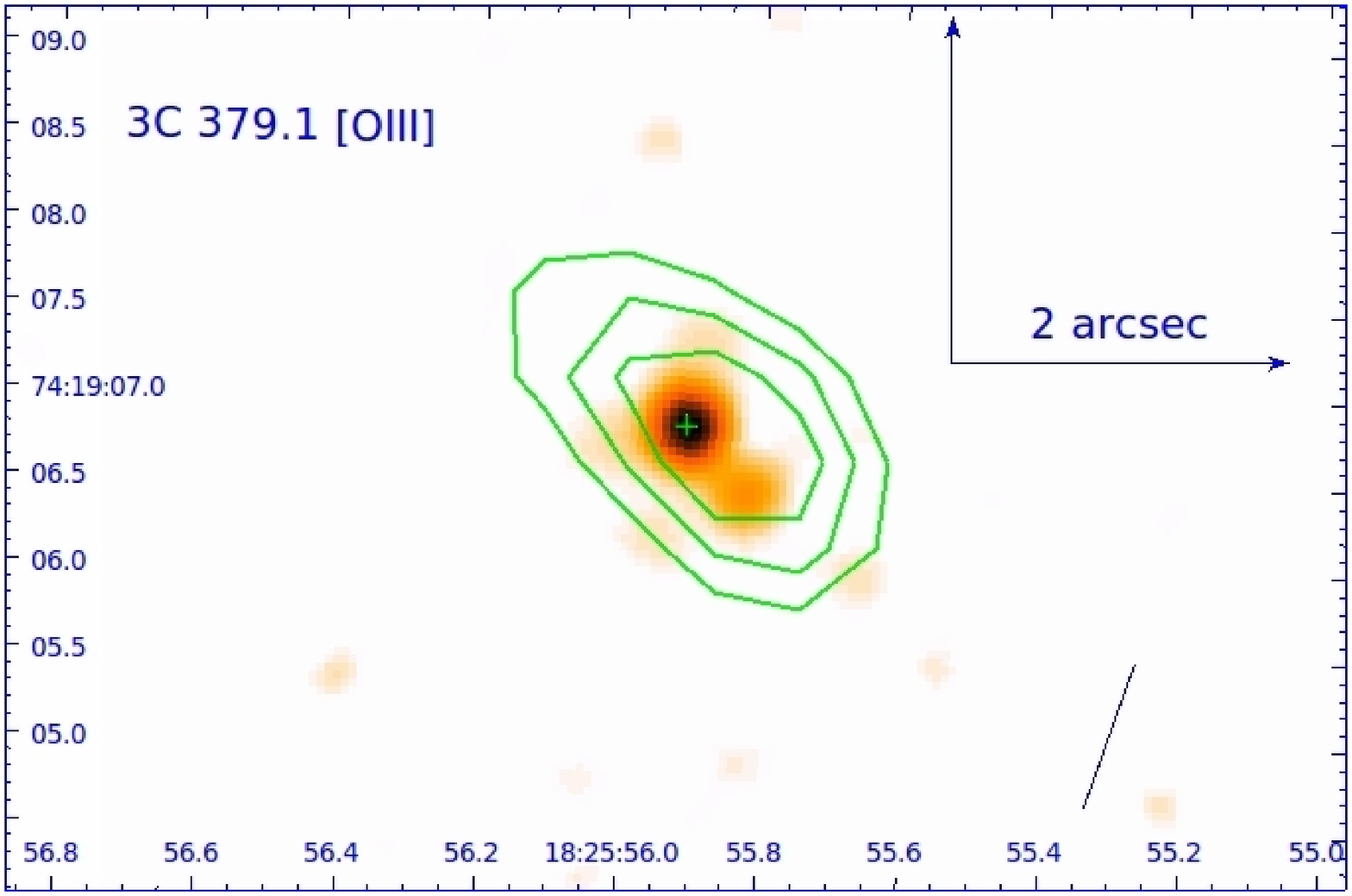}
\includegraphics[width=6cm]{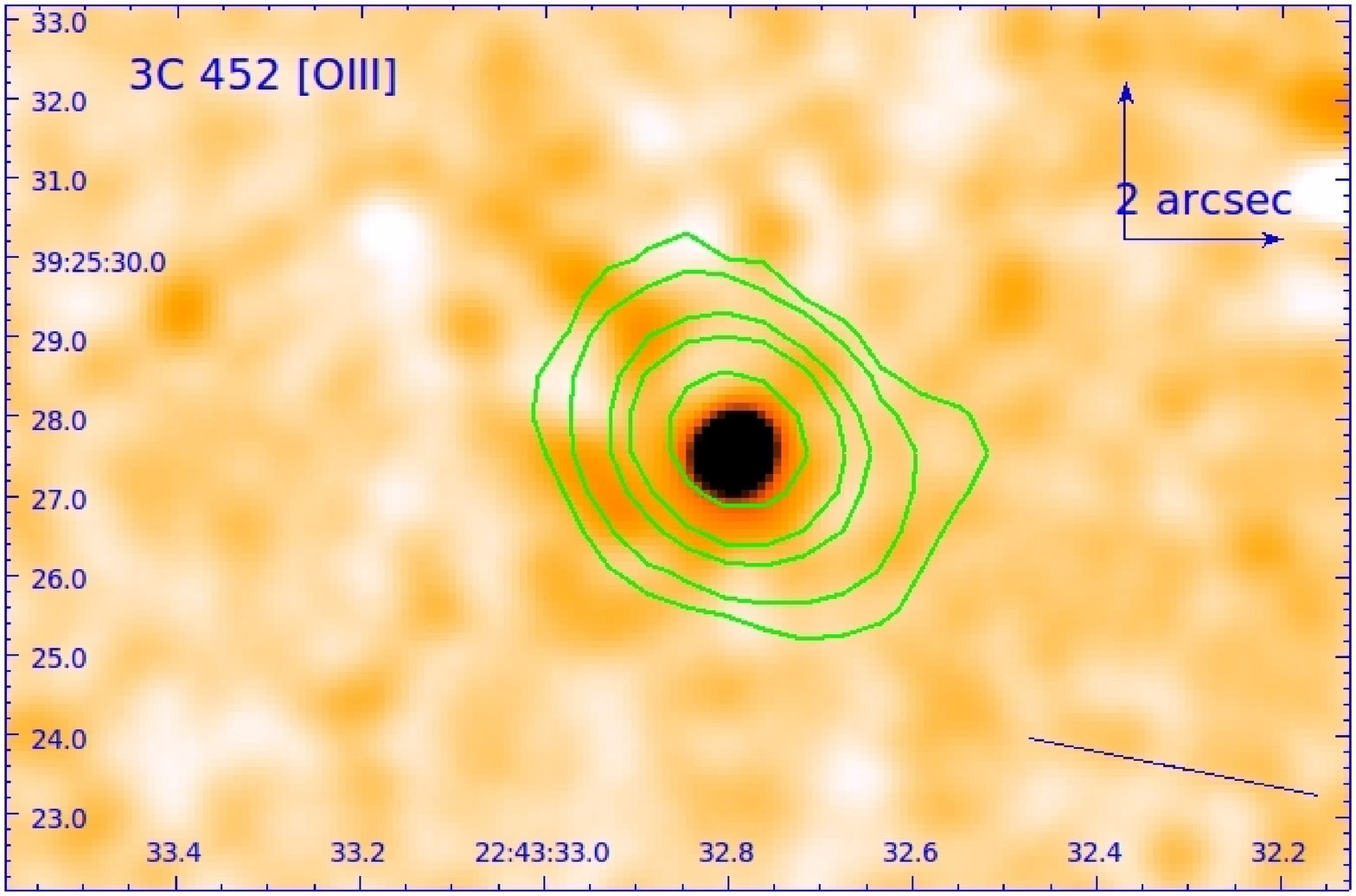}

\caption[]{\label{superposedHST} Green iso-counts contours of the soft X-ray emission 
  (0.5-2 keV), superposed onto the HST [O~III] images
  (smoothed with 0\farcs3 Gaussian kernel). Contours for the various sources
  are drawn at the following levels of counts pixel$^{-1}$: 3C~33 (0.2,0.5,1,3,5,20), 3C~171
  (0.05,0.1,0.3,0.5), 3C~180 (0.01,0.05,0.2), 3C~234 (0.2,1,4), 3C~305 (0.2,0.5,1,2), 3C~321
  (0.4,2,10,20), 3C~379.1 (0.1,0.2,0.3), 3C~452 (0.5,1,3,5,10).}
\end{figure*}

\begin{figure*}

\includegraphics[width=6cm]{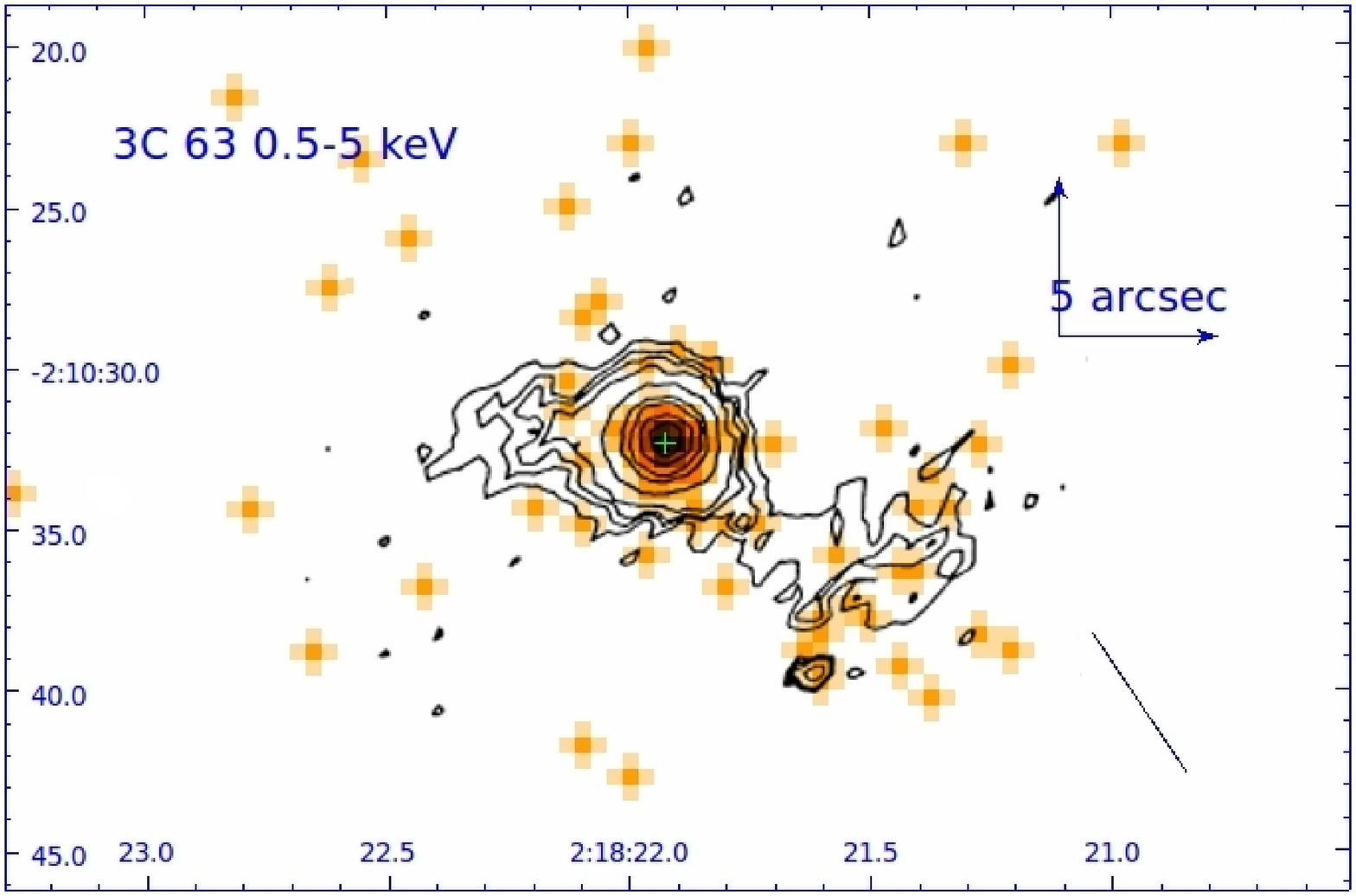}
\includegraphics[width=6cm]{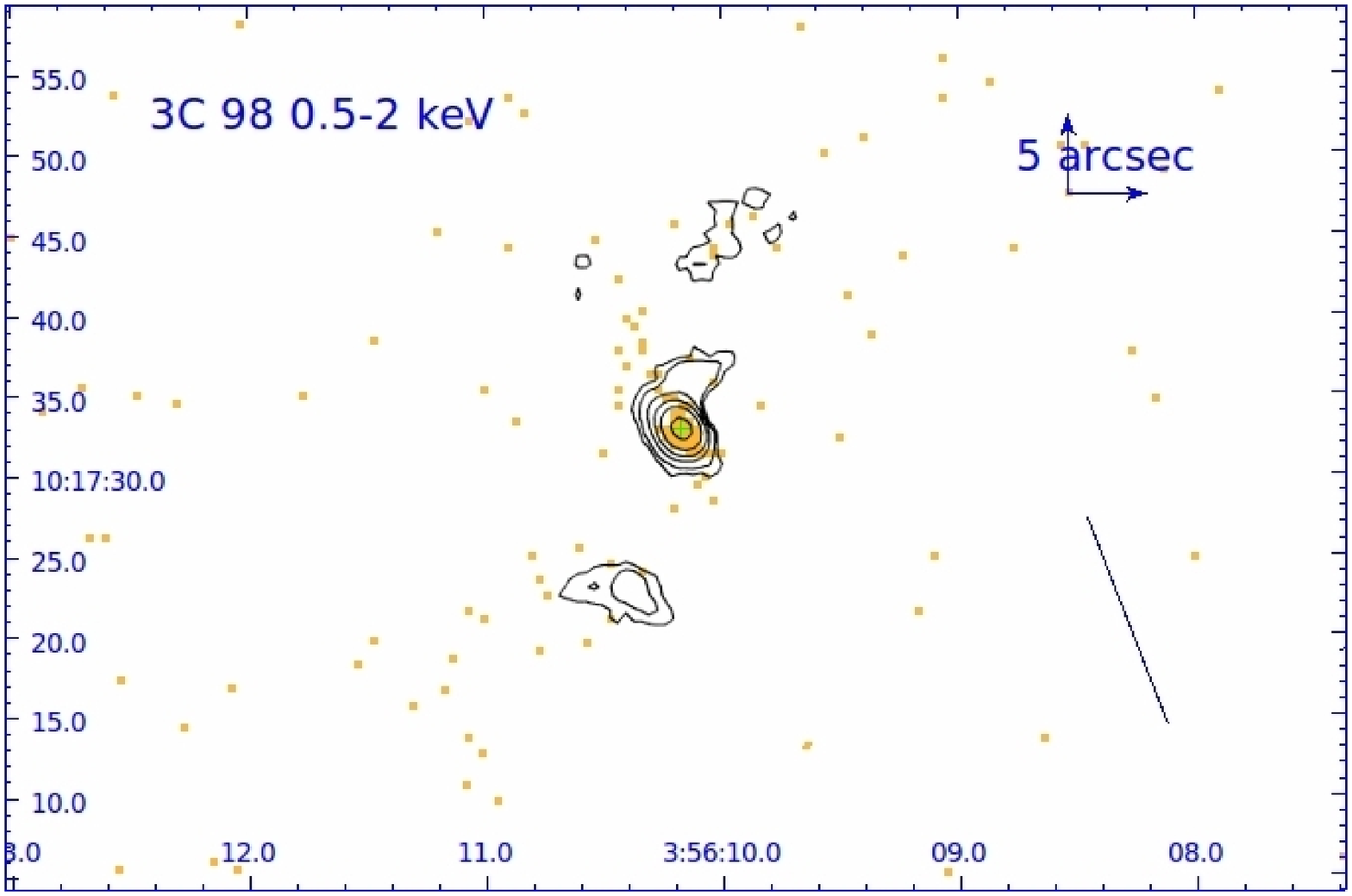}
\includegraphics[width=6cm]{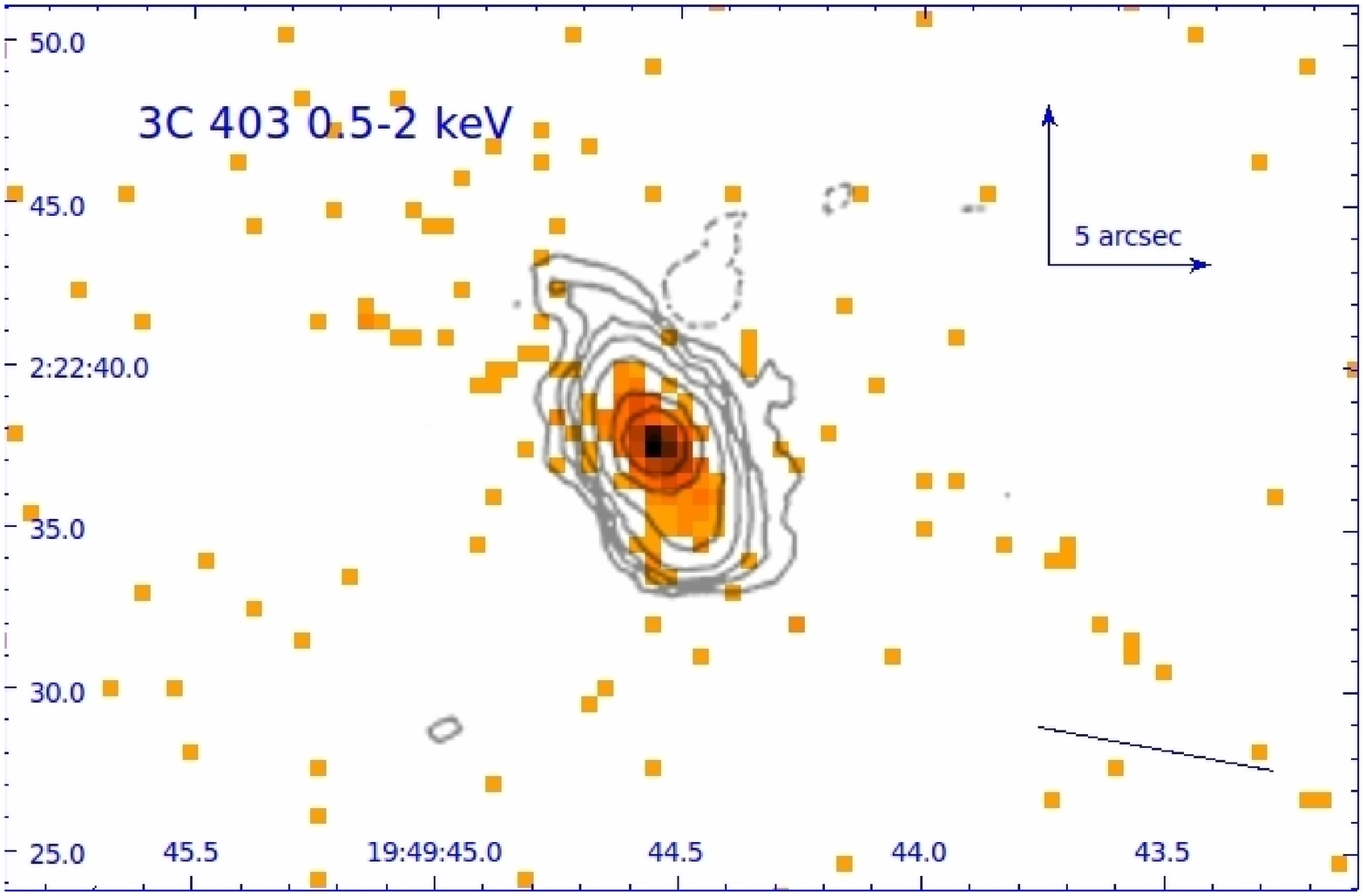}
                                        
\includegraphics[width=6cm]{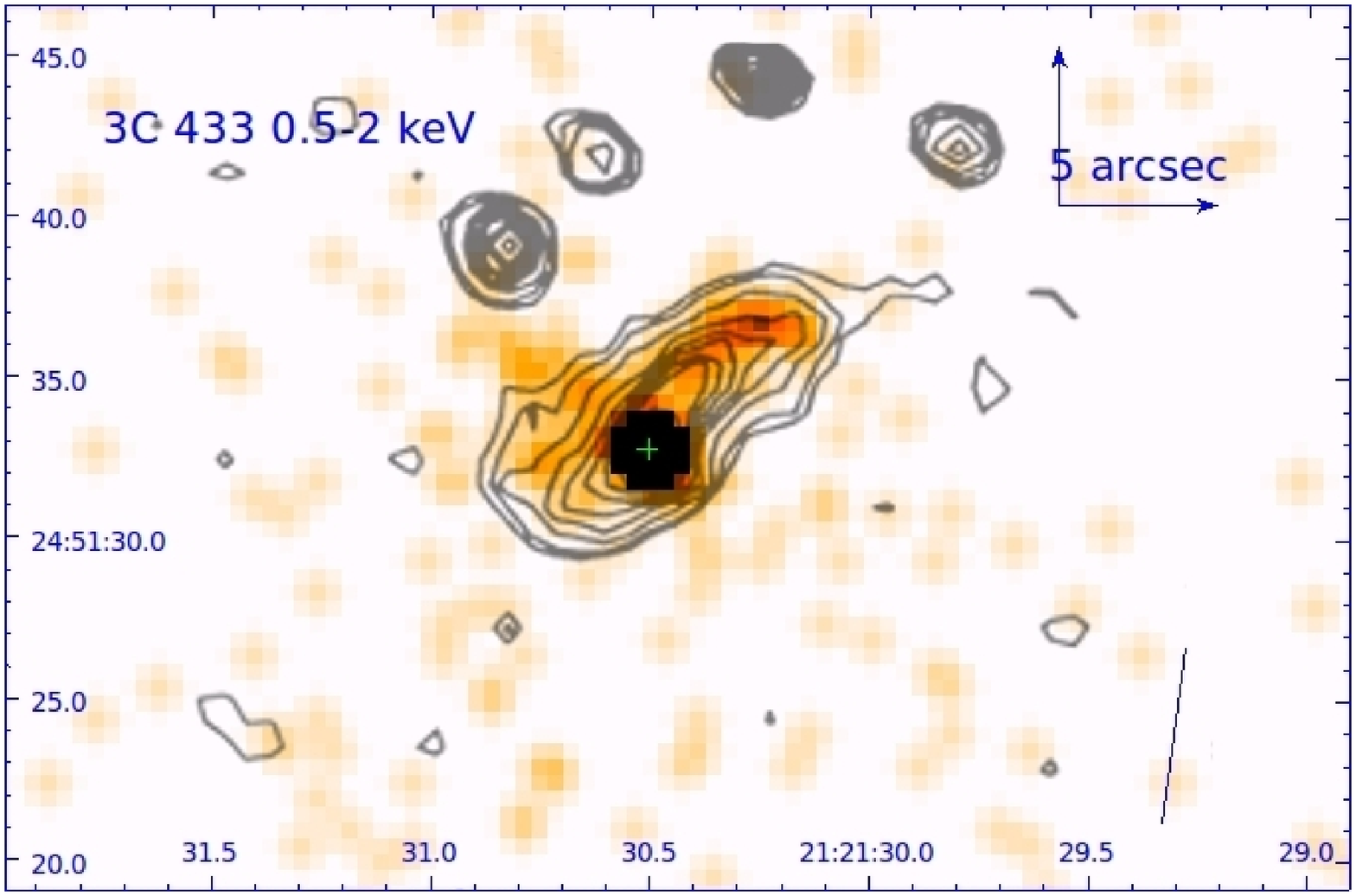}
\includegraphics[width=6cm]{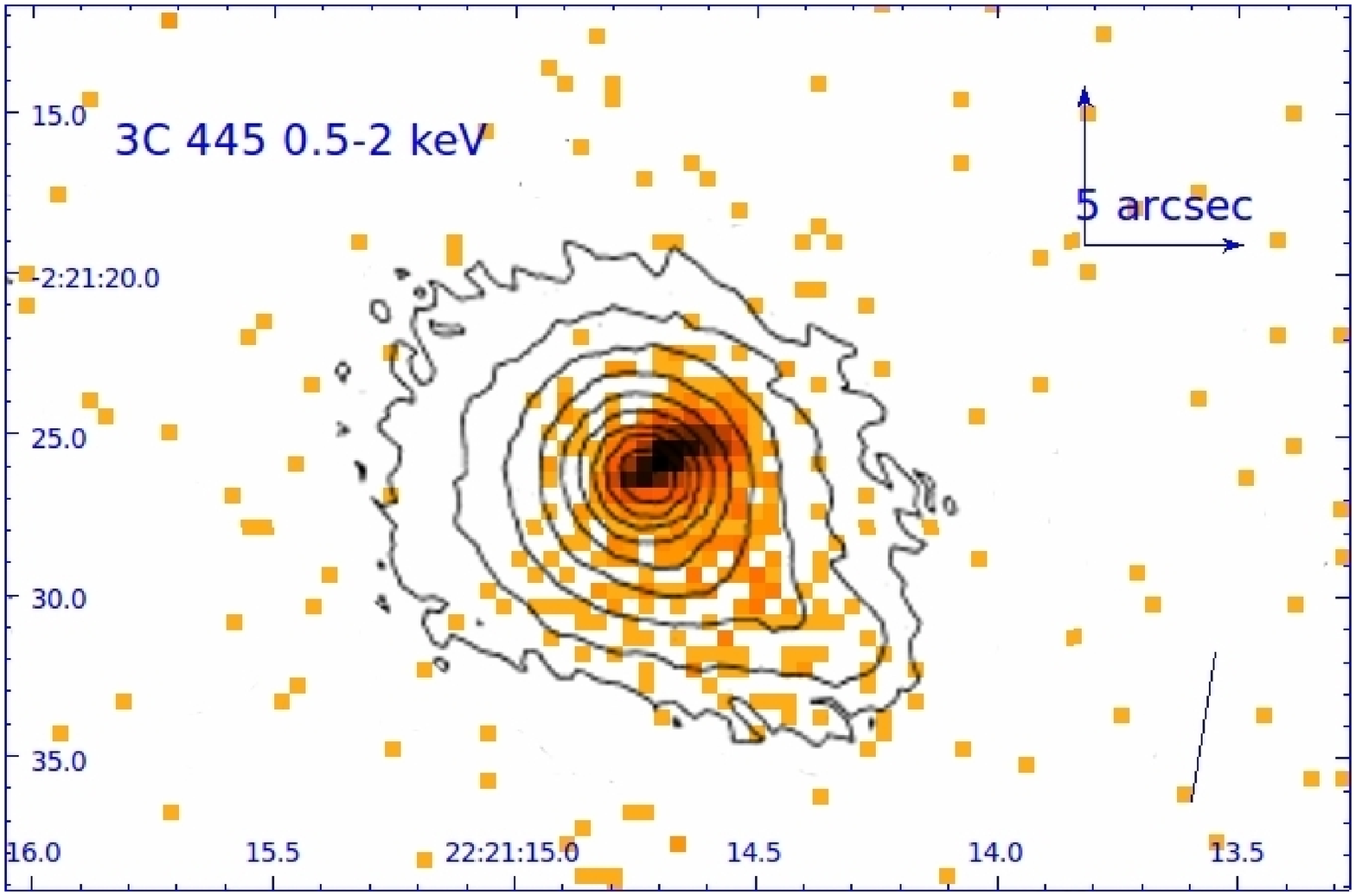}

\caption[]{\label{superposed0} Comparison  between soft X-ray emission
  (0.5-2 keV) and narrow line region (mapped usually by
  [O~III] or \Ha+[N~II]). We reproduced and superposed optical line contours from published
  images on the Chandra images.

3C~63, 3C~403, 3C~433: \Ha+[N~II] contours from \citet{baum88}.

3C~98:
Superposed [O~III] contours are from \citet{hansen87}.

3C~445: the Chandra image is off-axis causing a PSF elongation along the
north-west axis. Instead the extension of about 10 arcsec along P.A. $\sim$
135$^{\circ}$ is real and also visible in the V band ground based observation
published in \citet{heckman86} (contour superposed).}

\end{figure*}
As discussed in the Introduction, a connection between X-ray, radio, and
optical emission lines has been found in several active nuclei, both
radio-quiet and radio-loud. We now explore the possibility that this applies
in general to a large sample of radio galaxies.

\subsection{Optical emission-line images}
We search in  the literature for optical emission  line images for the
sources classified in  the soft X-rays images as  extended or possibly
extended,  according to Tab.  \ref{bigtable}. About  3/4 of  them have
been imaged  with narrow band  filter (usually centered on  [O~III] or
\Ha+[N~II]) with ground based  or HST observations. More specifically,
we   have   access   to   the   new   re-calibrated   HST/ACS   images
(\citealt{tremblay09}  and  Rodriguez Zaurin  et  al.  2012 in  prep.)
(3C~33, 3C~180, 3C~234, 3C~285)  and to the HST/WFPC2 images published
by  \citet{privon08} (3C~171,  3C~305, 3C~321,  3C~379.1,  3C~452). In
other cases  (3C~63, 3C~98, 3C~433, 3C~403, 3C~445)  we reproduced and
superposed    contours   from    published    images   (mostly    from
\citealt{mccarthy95} and  \citealt{baum88}). Only for  3C~456 we could
not find any optical line image.

In  order to register  the Chandra  and optical  images, we  align the
centroid  of the  hard X-ray  emission with  the peak  of  the optical
emission line structure.

In Fig. \ref{superposedHST} we  superposed the X-ray contours onto the
HST [O~III]  line images, while  in Fig. \ref{superposed0}  we present
soft  X-ray   images  with   superposed  ground  based   optical  line
contours.  In Tab.  \ref{bigtable}  we report  the main  morphological
parameters characterizing the sources in the different bands.

\subsection{Relationship between X-ray, lines, and radio structures}
\label{xor}

Overall, we  find a very strong correspondence  between the morphology
of  the  NLR and  the  soft  X-ray emission.  In  most  cases the  two
structures are  closely co-spatial, an effect  particularly clear when
we  focus  on the  objects  classified  as  ``extended''. Despite  the
difficulties  in defining the  size of  structures measured  in images
with widely different levels of resolution and sensitivity, we find an
excellent match between the NLR and soft X-ray extension.

\begin{figure*}
\centering
\includegraphics[width=6cm]{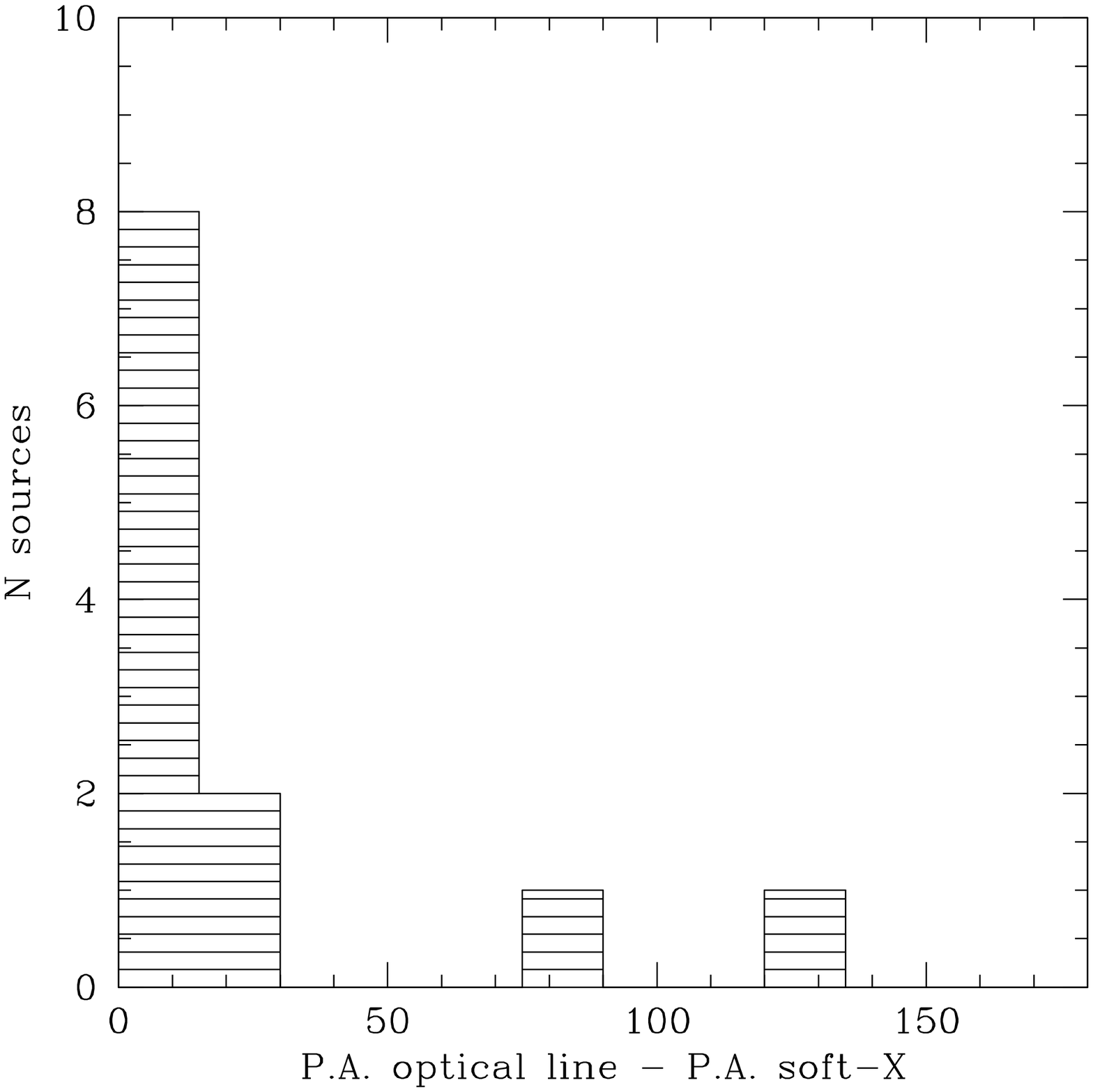}
\includegraphics[width=6cm]{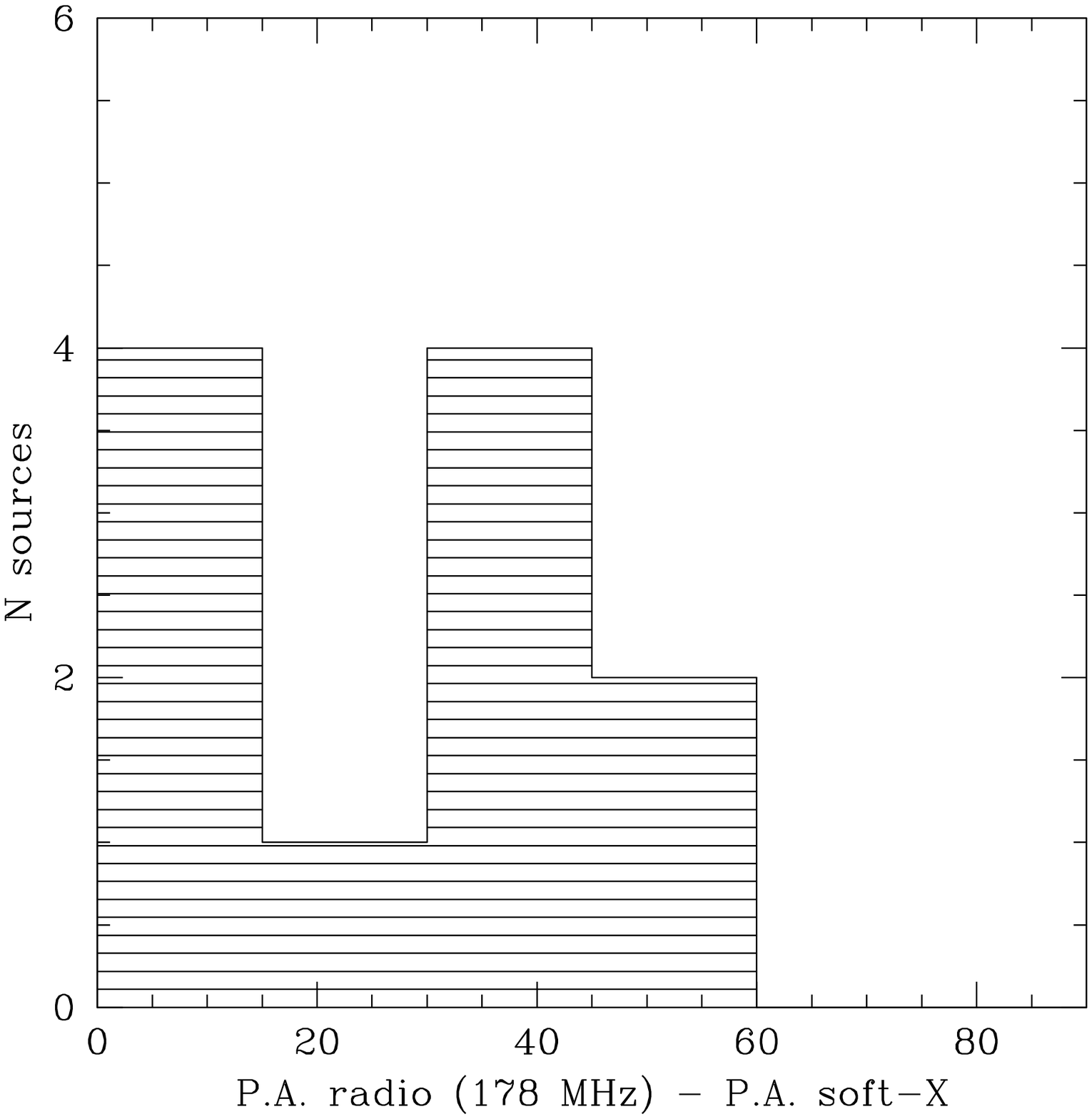}
\includegraphics[width=6cm]{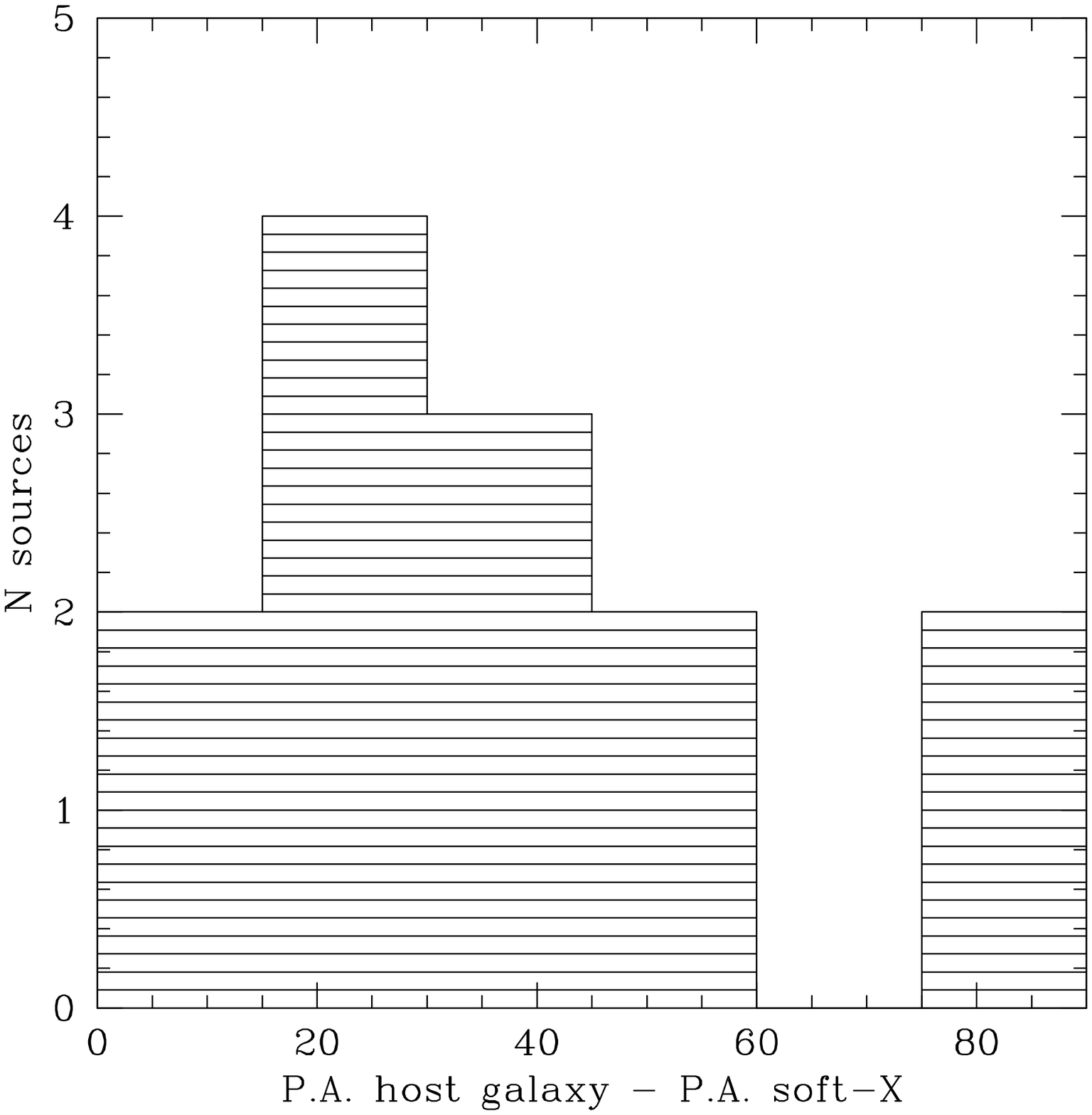}
\caption{Histograms of the offset between the position angle (P.A.)  of
  extended regions measured in soft X-ray, with respect to the narrow line
  region (left), the radio emission (middle), and the host galaxy's
  major axis (right).}
\label{ist}
\end{figure*}

In Fig. \ref{ist} (left panel) we show the histogram of the difference
between the  direction of the  extended emission in optical  lines and
soft X-ray;  this offset can be  measured reliably in  12 sources (see
Table \ref{bigtable}).  A very strong  alignment is found,  with 10/12
sources aligned within 30$^\circ$.

Only two sources  show a high apparent offset between  the NLR and the
X-ray  emission.   In  3C~452  the   NLR  is  very  compact   and  its
P.A.  measurement is  highly uncertain  and  in 3C~98  the X-rays  are
produced in  a narrow structure, $\sim$ 13\arcsec\  long, well aligned
with   the  Northern   radio  jet,   while  the   NLR  is   offset  by
$\sim$30$^\circ$.

In  Fig. \ref{ist}  (middle  panel) we  show  instead the  differences
between the position  angle of the radio emission  and of the extended
soft X-ray region. Although a clear preference toward small offsets is
seen,  the distribution  is  much  broader than  that  of line  versus
X-ray. In  this respect six objects are  particularly notable (namely,
3C~33, 3C~379.1, 3C~403, 3C~433, 3C~445 and 3C~452) all showing a very
well  defined  soft  X-ray  extended  region which  is  misaligned  by
30$^\circ$  to 60$^\circ$  from the  radio  axis, while  they are  all
aligned within a few degrees with the NLR.

Finally, we consider the possibility  of a connection between the host
galaxy major  axis and  the soft X-ray  emission (see  Fig. \ref{ist},
right panel);  such a  link can  be envisaged as  the result  of X-ray
emission  associated with  the host  hot corona  or with  star forming
regions. The  distribution of  their differences is  essentially flat,
arguing against a link between them.

\subsection{Notes on individual sources}
\label{note}
3C~33:  in  the soft  X-ray  band  the  nuclear structure  is  clearly
extended, with  the emission  elongated along NE-SW  axis. There  is a
remarkable spatial coincidence with  the [O~III] emission line region,
as already noticed by \citet{kraft07} and \citet{torresi09}.

3C~98:  the  soft  X-ray  emission  forms  a  $\sim$13\arcsec\  narrow
structure well aligned with the Northern radio jet \citep{leahy97}.

3C~171: the soft X-ray emission  is clearly extended and aligned along
the  radio   axis.  This  source   has  been  studied  in   detail  by
\citet{hardcastle2010}.

3C~180: the biconical structure seen  in optical line imaging does not
have  a clear  X-ray counterpart  in our  short exposure  time Chandra
image.  It just  reveals  a few  counts  aligned in  the same  general
direction.

3C~234:  the emission  line  region  and the  soft  emission are  both
aligned with the radio axis.

3C~277.3:  two blobs  of X-ray  emission  are seen  to the  SW of  the
nucleus.   They are  cospatial with  the two  radio knots  K1  and K2,
associated with regions of line emission \citep{vanbreugel85} that are
also    visible     in    optical    and     infrared    HST    images
\citep{capetti00,madrid06}.   The   same  association  between  radio,
X-ray, and emission line is seen  North of the nucleus at the location
of the Northern radio hot-spot.

3C~305: the soft X-ray emission extends beyond the radio one and it is
cospatial with the optical emission line region \citep{massaro09}.

3C~321: this source  has a companion galaxy, that  also harbors an AGN
\citep{evans08}. The  X-ray emission extends over a  large region that
includes both  galaxies. Due to the  complexity of this  source, it is
not  trivial to separate  the extended  emission associated  with each
individual active nucleus.

3C~379.1: the soft X-ray and  the [O~III] emissions are rather compact
and cospatial, but not aligned with the radio axis.

3C~445:  the Chandra  image  is  off-axis (the  aim-point  was on  the
Northern  hot-spot) and  the  PSF is  elongated  along the  North-West
axis.  Instead, the  extension of  about  10 $\arcsec$  along P.A.  of
$\sim$ 135$^{\circ}$ is real. A similar elongation is visible in the V
band ground based image  published in \citet{heckman86}. These authors
claim that this  region is a source of emission  lines. This result is
confirmed by  our TNG spectroscopic observations  (see Fig. \ref{tng})
that were coincidentally obtained with the long-slit aligned with this
feature and that shows emission  line extending $\sim$ 8 \arcsec\ from
the nucleus.

\subsection{A quantitative comparison of emission lines and soft X-ray
  structures.}
\label{comparison1}

\begin{figure}
\includegraphics[width=4.4cm]{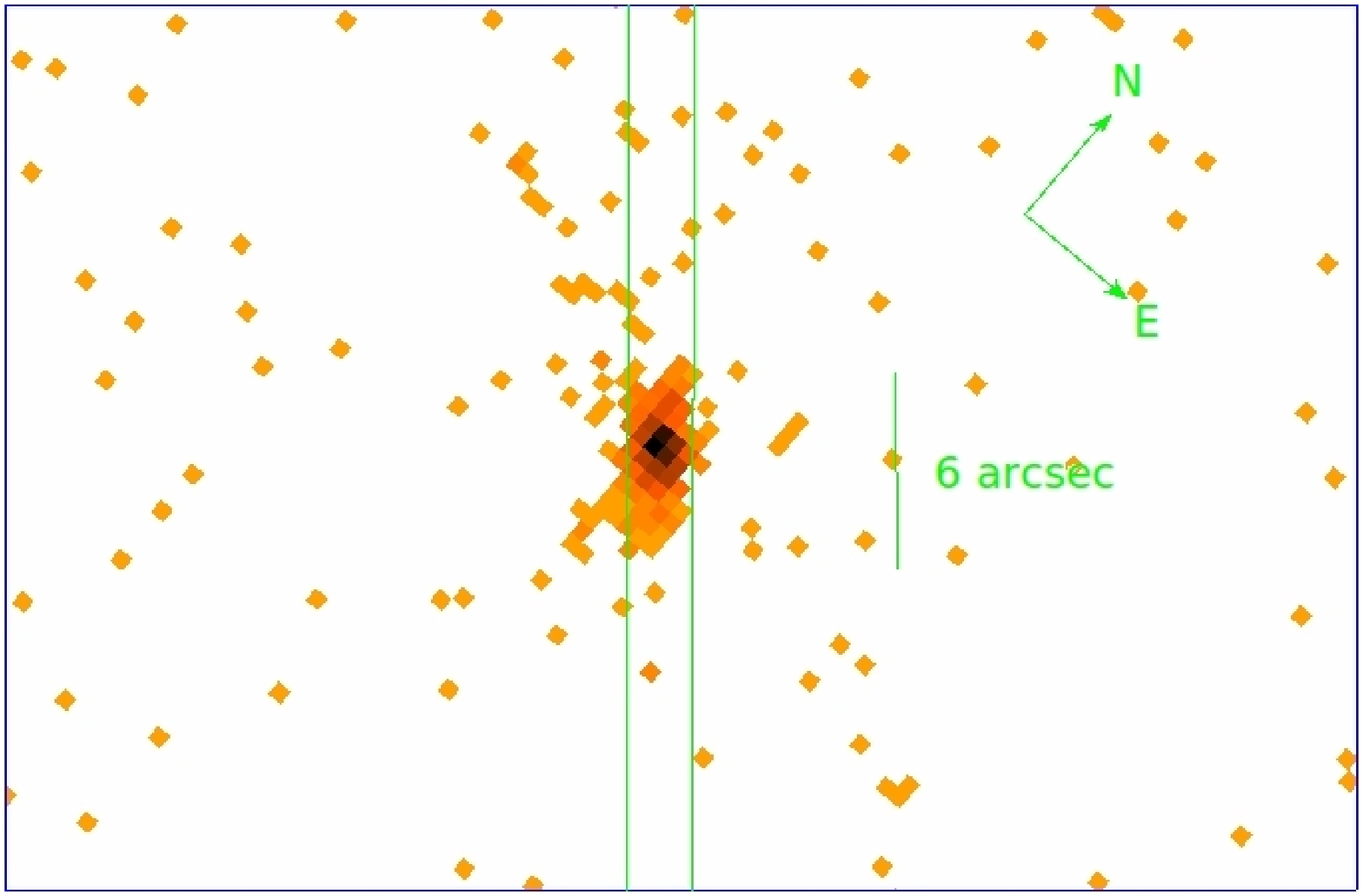}
\includegraphics[width=4.5cm]{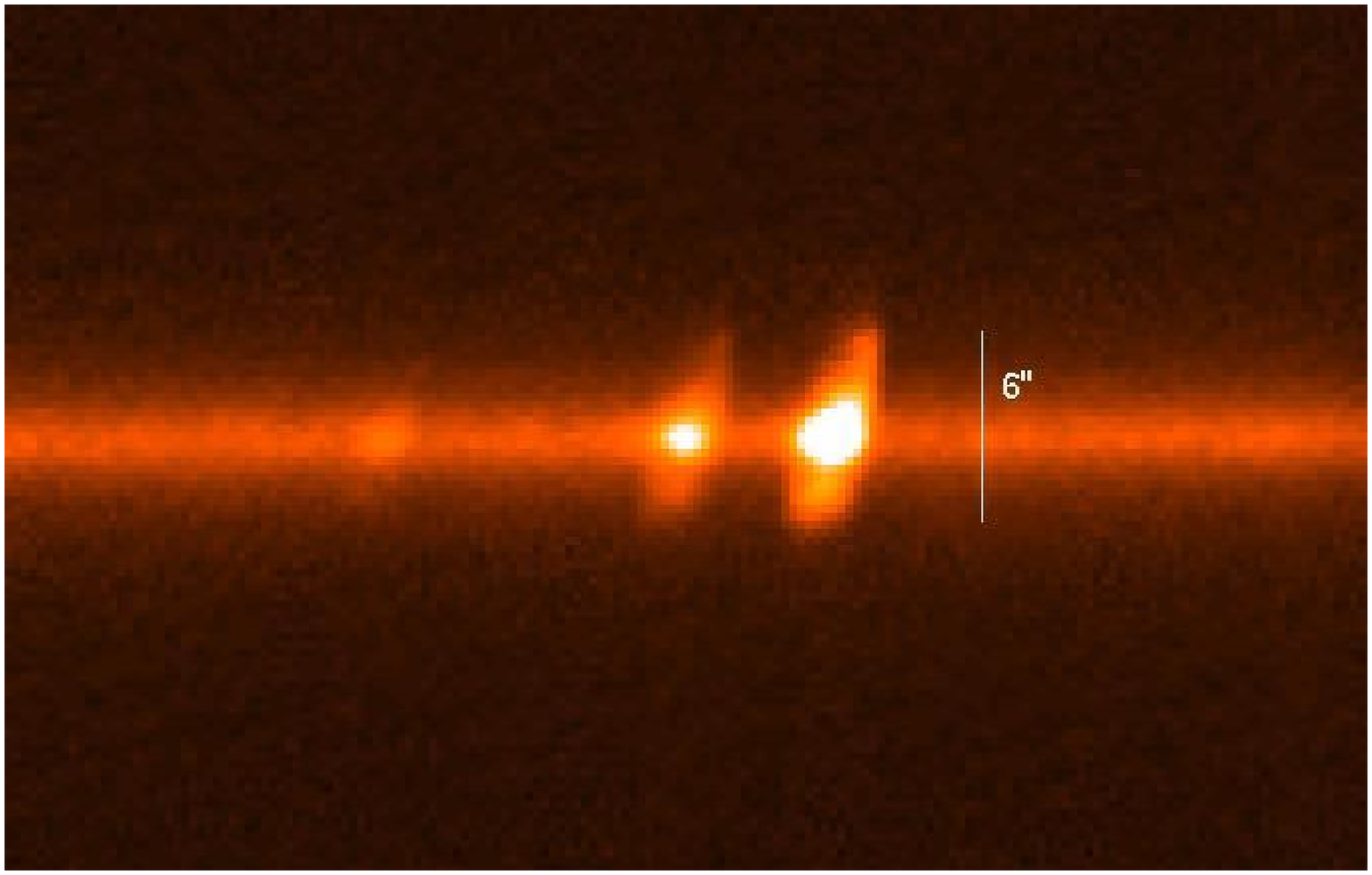}
\includegraphics[width=4.4cm]{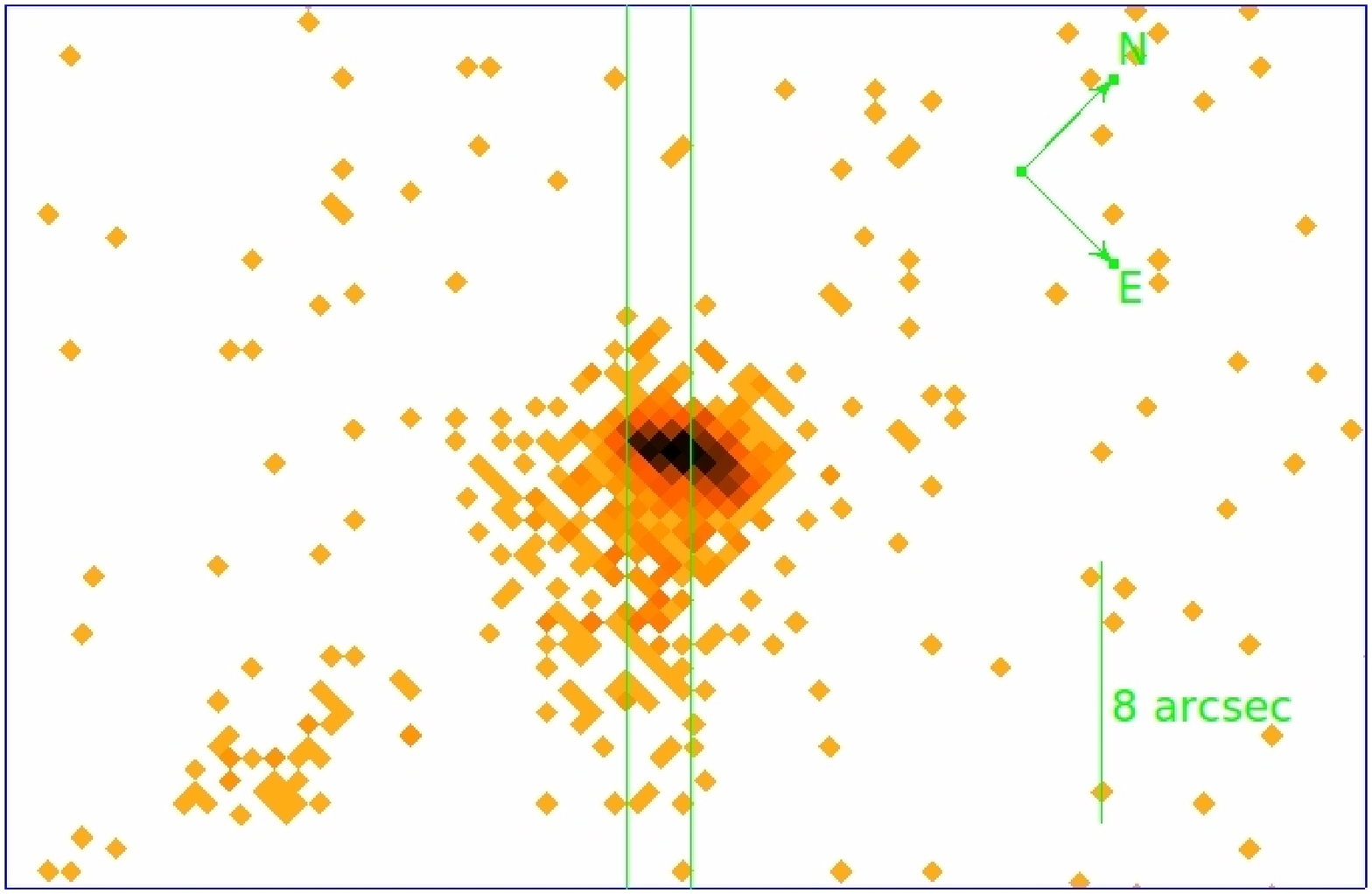}
\includegraphics[width=4.5cm]{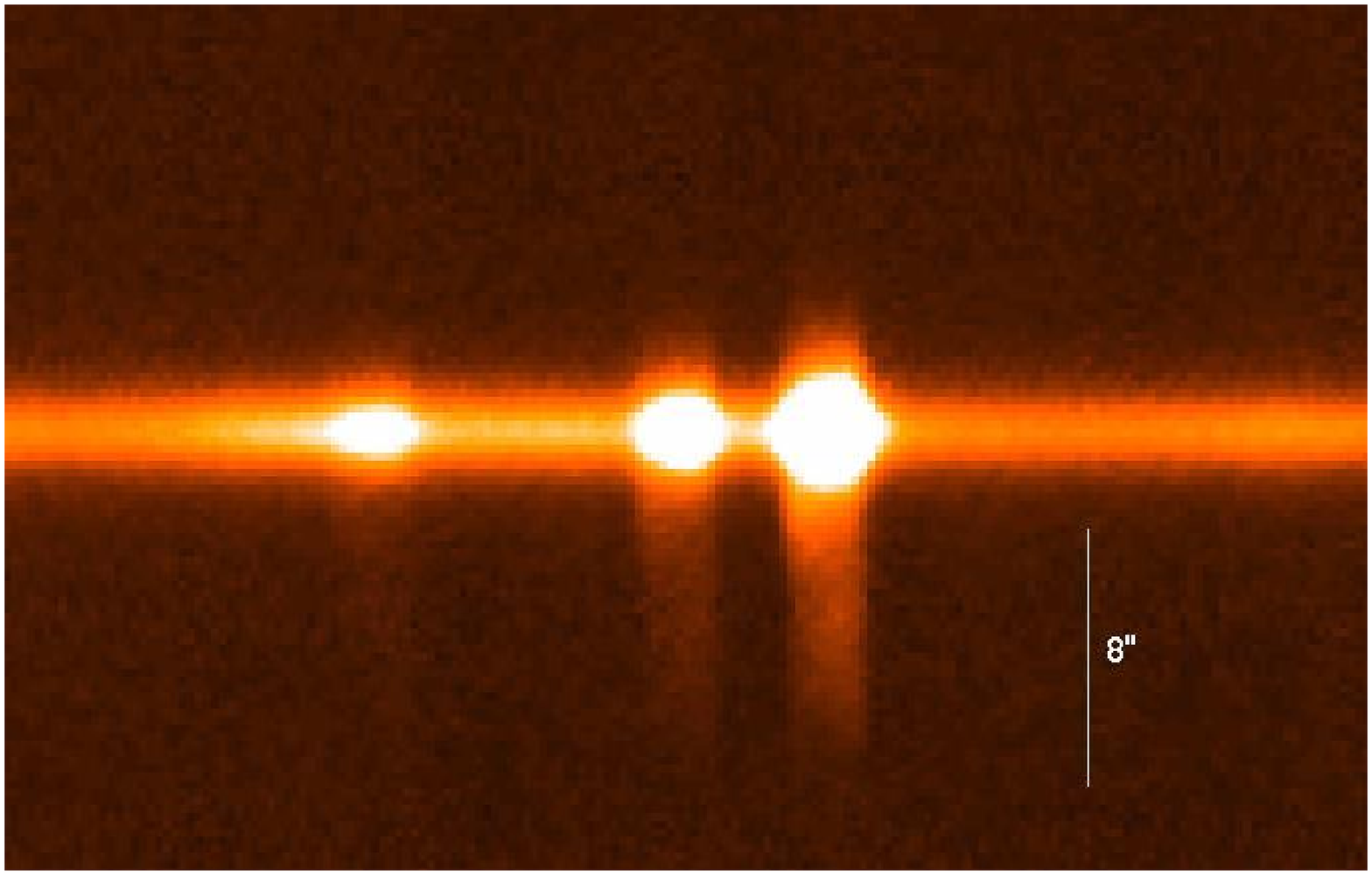}
\caption{Left panels: rotated Chandra soft X-ray images for 3C~403 (top) and
  3C~445 (bottom) with the superposed 2$\arcsec$\ wide slit used to obtain
  the spectra at the TNG shown in the right panels. The three emission lines
  are \Hb\ and the [0~III]$\lambda\lambda$ 4959,5007 doublet. Substantial line
  emission is produced along the direction of the soft X-ray emission. The
  sizes of the two emitting regions (in line and X-rays) are similar. }
\label{tng}
\end{figure}

In order to quantitatively compare the optical line and the soft X-ray
emission,  we measure  the [O~III]  flux in  the available  HST images
using the same regions chosen for the X-ray spectral analysis drawn in
Fig. \ref{images}.
 
Due to the  difficulty to obtain an accurate  flux calibration for the
HST/WFPC2 ramp filter images  (available for 3C~171, 3C~305, 3C~379.1,
and  3C~452),   we  decided  to  use  the   information  published  by
\citet{buttiglione09} obtained  from the ground based  TNG spectra. We
derive a conversion factor between  counts and fluxes by measuring the
counts in the HST images in the same 2\arcsec$\times$2\arcsec\ nuclear
extraction region  used by  \citet{buttiglione09} to measure  the line
fluxes (accurate to $\sim$ 10\%). The resolution of the HST images has
been  previously degraded,  with a  Gaussian smoothing,  to  match the
resolution  of each  TNG observation.   This scaling  factor  has been
applied  to  the  counts  measured  in the  extra-nuclear  regions  to
estimate line fluxes.

Two of the sources of our  sample (namely, 3C~445 and 3C~403) were not
observed with HST  with the ramp filters. However,  these objects were
observed by \citet{buttiglione09}. In those observations the long slit
was, by chance, oriented in the  same direction as the PA of the X-ray
extended emission.  For these  sources we extracted the X-ray spectrum
in  the  same   region  covered  by  the  2\arcsec\   wide  slit  (see
Fig. \ref{tng}).

We corrected all line fluxes for reddening due to the Galaxy using the
extinction  law of  \citet{cardelli89} and  the galactic  E(B  $-−$ V)
value  for each  object as  tabulated by  \citet{buttiglione09}, taken
from the NASA Extragalactic Database (NED).

\begin{figure}
\centering
\includegraphics[width=7cm]{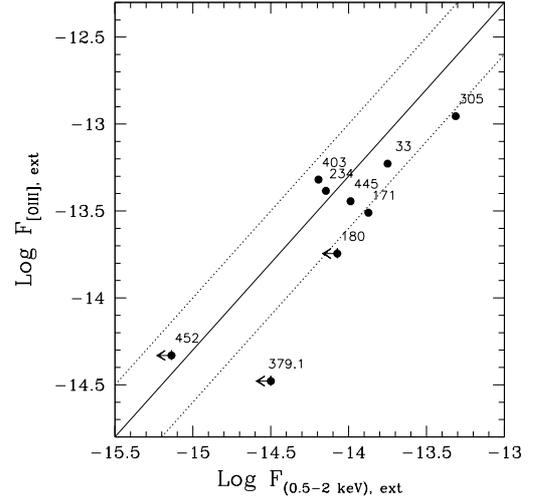}
\caption{Comparison of the [O~III] and soft X-ray fluxes measured in the
  extended regions drawn in Fig. \ref{images}. The solid line corresponds to a
  constant ratio F([O~III])/F(soft X-ray) = 5; the two dashed lines mark a ratio equal to
  2.5 and 10.}
\label{oiiivsx}
\end{figure}

In  Tab. \ref{flux}  we report  the resulting  [O~III] and  soft X-ray
fluxes,   the    corresponding   luminosities,   and    their   ratio,
$\cal{R}$([O~III]/sX).  The median  ratio of  $\cal{R}$([O~III]/sX) is
5.6 and  the scatter  is low,  with all sources  confined to  within a
factor   $\sim$  2   from   the  median   value.   No  dependence   of
$\cal{R}$([O~III]/sX)  on  luminosity is  found.  The  line and  X-ray
fluxes are graphically compared in Fig. \ref{oiiivsx}.

For this same group of 9 sources, we estimated $\cal{R}$([O~III]/sX) also
on the nucleus. We find values even more closely clustered, all falling 
within a factor of $\sim$1.3 from the median value of 2.4, a factor of $\sim$
2 lower than in the extended regions.

\section{Discussion}
\label{discussion}
\subsection{Origin of the soft X-ray emission}

The  presence of  extended soft  X-ray emission  can be  attributed to
various mechanisms:

-- thermal  emission  from   hot  gas,  collisionally  ionized  and/or
compressed by a jet or an outflow;

-- nuclear X-ray emission scattered by an extended medium;

-- non  thermal  emission from  i)  synchrotron  (or inverse  Compton)
radiation  from the radio-jets  and hot-spots  or ii)  inverse Compton
scattering of low energy photons  by the relativistic electrons in the
radio lobes;

-- emission from gas photoionized by the nuclear source.

In order  to identify the  emission process responsible for  the X-ray
emission, the most powerful tool is clearly a direct spectral fit. The
model that better reproduces the observed spectrum not only identifies
the radiation  mechanism but also constraints  the physical parameters
of  the emitting  medium. With  respect to  previous studies  based on
individual sources we can now  look for a common process that accounts
for the X-ray  spectra observed for this whole  group of galaxies. For
the 3CR sources  considered here, in 5 cases it  is possible to follow
this approach (see Sect. \ref{fluxes} and Table \ref{fit}).

Nonetheless,  the  results presented  in  Sect.  \ref{fluxes} are  not
sufficient  to securely  identify the  emission mechanism  at  work in
these sources, mainly due to the  low number of counts. There is not a
unique model that  proves to be superior in  fitting every source, or,
on the contrary, that does not provide a reasonable agreement with the
data in all of them. However, some general trends can be seen.

We  find that  power-law  models  require rather  high  values of  the
spectral index,  $2.5 < \Gamma <  3.9$, while extremely  low values of
the abundance are  highly favored for the thermal  models and the best
fit  are often  obtained  with an  effectively  null metallicity.  Our
results  are   fully  consistent  with  the  analysis   of  3C~171  by
\citet{hardcastle2010} and  of \citet{bianchi06} in  their analysis of
the extended X-ray emission from Seyfert 2 galaxies.

These  findings cast  doubts  on the  physical  sensibleness of  these
models.   Indeed  in  case  of  a power-law  emission  resulting  from
scattering of nuclear light from  gas located in the extended regions,
the spectral index is expected to be preserved and to be equal to that
observed  in the  nucleus.  The  measured values  of  $\Gamma$ in  our
sources are  instead much larger  than those usually observed  in both
radio-quiet  and  radio-loud  AGN  (e.g.  \citealt{bianchi09}).   High
energy  emission from  the  radio jets  or  hot spots  can already  be
generally excluded on a morphological basis (since it requires a close
cospatiality  of X-ray  and radio  emission, not  observed in  these 5
sources) but also on the  typical spectral index measured from objects
in   which   this   is   clearly   the   dominant   emission   process
\citep{harris06}. Similarly,  the location of the  soft X-ray emission
in our sources  in concentrated within a few kpc  from the nucleus and
it does not originate from the  large scale radio lobes as expected in
the  case of  IC emission  described  above. Concerning  the value  of
metallicity derived from the  spectral fits, note that the metallicity
of  the interstellar medium  in early-type  galaxies is  only slightly
sub-solar  \citep{fabbiano89,humphrey06}.   Furthermore,  the  optical
narrow line regions that, as  shown in Sect. \ref{xor}, are co-spatial
with the  soft X-ray emission, usually have  super solar metallicities
\citep{storchi96}.

In order to  assess whether these models can still  provide a good fit
to the data with more realistic  parameters, we forced them to be in a
physically  acceptable range, fixing  (for the  thermal model)  $\mu =
0.5$ solar and (for the  power-law models) $\Gamma =1.7$.  For the two
sources where we have a sufficient  number of counts to use a $\chi^2$
statistics we find that quality of the fits worsens significantly (see
Tab. \ref{fit}).

When considering the \xstar\ model, it generally provides a reasonably
good   fit   to   the   data   of  all   five   sources.   In   3C~171
\xstar\  corresponds to  the smaller  $\chi^2$ value,  and  the better
quality with respect to those  obtained with \mekal\ and power-laws is
strongly  increased when  we  limit  to models  with  values fixed  to
acceptable values,  as explained above.  In 3C~445 the  preference for
\xstar\ emerges  only against the latter set  of models. Unfortunately
the parameter values  obtained with the \xstar\ have  large errors and
thus we obtain only very limited information on the physical condition
of the  emitting gas.  This is the  same general result  obtained from
similar   studies    that   can    be   found   in    the   literature
(e.g. \citealt{ogle03}).

Nonetheless,  in the  light of  the  results of  the somewhat  limited
spectral  modeling  possible  with  the available  data,  the  favored
mechanism to account for the extended soft X-ray regions appears to be
emission from a photoionized plasma.

\medskip Although no spectral  analysis was possible for those sources
with low count rates, the  soft X-ray fluxes in their extended regions
can still generally be estimated (as discussed in Section \ref{xor}).
 The comparison of  the soft X-ray and
[O~III]  line  fluxes,  possible  for  11  galaxies,  shows  a  strong
connection between  the emission  in the two  bands, with  all sources
confined  to  within  a factor  $\sim$  2  from  the median  value  of
5.6.  Interestingly,  this  same  result is  found  considering  their
nuclear regions, see Table  \ref{fit}, with an only slightly decreased
median ratio of 2.4.  Furthermore, the $\cal{R}$([O~III]/sX) ratios in
the 3CR sources  are in remarkable coincidence with  those measured in
the  Seyfert 2  galaxies  by \citet{bianchi06}  that  cover the  range
$\cal{R}$([O~III]/sX)  =  2.8  -   4.8  (with  only  one  object  with
$\cal{R}$([O~III]/sX)  = 11). The  small range  in the  ratios between
optical lines  and soft X-ray  emission and the similarity  with those
measured in their nuclear regions points to a common mechanism for the
emission in  these two  bands over both  the nuclear and  the extended
regions.

As already mentioned in  the Introduction, the few X-ray spectroscopic
studies  for the  nuclear  regions  of radio  galaxies  show that  the
dominant  radiation  mechanism  is  associated with  the  presence  of
photoionized gas.  This same  result is obtained  from the  studies of
both  the  nuclear and  extended  regions  in  Seyfert galaxies.  This
indicates that, in presence of photoionized gas, a strong link between
soft X-ray  and optical emission lines fluxes  arises naturally.  This
argues in favor of an interpretation in which the extended X-rays seen
in the 3CR sources are also due to a collection of emission lines.
 
A  further element  in  favor of  a  line origin  for  the soft  X-ray
emission is  the close spatial  connection with the optical  lines. We
found  only one  clear exception  (3C~98)  and a  second complex  case
(3C~277.3)  that we  discussed  in Section  \ref{note}.  In all  other
sources the  available optical line  images show a  remarkable spatial
coincidence with the morphology of the extended X-ray emission. Again,
in Seyfert galaxies there is similar close spatial association between
[O~III] and X-ray morphology.

\section{Summary and conclusions}

 We analyzed  the diffuse soft X-ray  emission (0.5-2 keV)  of the 3CR
 radio galaxies  at z$<$  0.3. Extended soft  X-ray emission  has been
 observed  in Seyfert  galaxies  and  also in  a  few radio  galaxies,
 usually  matching very well  the morphology  of the  optical emission
 line  region. In  order to  characterize and  constrain  the physical
 mechanism  that produce  the observed  soft X-ray  emission,  we take
 advantage of this large and complete sample, now entirely observed by
 Chandra.  We here  focused  on the  properties  of the  spectroscopic
 sub-classes of high excitation galaxies and broad line objects.

 We performed  an imaging analysis,  describing the morphology  of the
 extended features  and measuring their sizes and  position angles. In
 the 33 HEGs  we detected extended (or possibly  extended) emission in
 about  40\%  of the  sources;  the  fraction  is even  higher  (8/10)
 restricting the  analysis to the  objects with exposure  times larger
 than 10 ks.

In the  18 BLOs,  extended emission  is seen only  in 2  objects; this
lower detection rate  can be ascribed to the  presence of their bright
X-ray nuclei that easily outshine any genuine diffuse emission.

 The count rates seen in the extended soft X-ray regions of HEGs is in
 the  range $\sim0.2\times10^{-3}-2\times10^{-3}$  counts/s,  with the
 only   exception  of   3C~305  where   it   reaches  $2\times10^{-2}$
 counts/s. The regions extend from $\sim2\arcsec$ from the nucleus out
 to  a radius  of usually  $\sim5-10\arcsec$.  We  estimated  that the
 count rate produced  by the PSF wings of the BLO  in the same spatial
 regions  is   typically  between  $\sim4\times10^{-3}-6\times10^{-3}$
 counts/s.  Therefore,  an  extended  X-ray emission  with  a  surface
 brightness similar to that seen in  the HEG would be a factor between
 2  and  30   with  respect  to  that  associated   with  the  nuclear
 source. This implies that even  in the presence of genuine soft X-ray
 extended emission in BLOs, similar to that observed in the HEGs, this
 would  not be  generally detected  from the  analysis of  the surface
 brightness profiles nor would induce a significant asymmetry.

Thus, a  soft X-ray extended  emission is a general  characteristic of
radio galaxies.  With respect  to previous studies based on individual
sources, we  can here look for  a common mechanism  that can reproduce
the  properties of  these galaxies,  revealing a  common  process that
accounts for the X-ray spectra observed.

 We  consider different scenarios:  scattered nuclear  light, emission
 from  collisionally   ionized  plasma,  non-thermal   radiation,  and
 emission from photoionized gas.  Among these various possibilities we
 argue that the most plausible dominant process is AGN photoionization
 based on the following results:

 -- the first clue is  the impressive morphological similarity between
 the  optical NLR  and the  soft X-ray  emission regions.  The offsets
 between soft X-ray  and optical lines axis are  generally very small,
 $\lesssim 20^{\circ}$. Instead  the offsets distribution between soft
 X-ray, radio and optical galaxy axis are much broader.

 -- the  fluxes of  the [O~III]  optical line  and of  the  soft X-ray
 emission  in the  extended regions  are strongly  correlated  (with a
 scatter  of  less than  a  factor of  2),  suggesting  that they  are
 produced  by   the  same   mechanism.   Furthermore,  the   ratio  of
 [O~III]/soft X-ray  fluxes is  very similar to  that observed  in the
 nuclear  regions of  radio  galaxies  and in  both  the extended  and
 nuclear  regions of Seyfert  galaxies. For  these latter  sources the
 origin  of the  soft X-ray  emission  from photoionized  gas is  well
 established.

 -- in 5 sources the counts in the soft band are sufficient to perform
 a spectral  analysis. Although we  are not able to  securely identify
 the emission mechanism  at work in these sources,  a photoionized gas
 model (\xstar\  in \xspec) always  provides an acceptable fit  to the
 data. Unfortunately the resulting parameters values have large errors
 and provides  us with only  very limited information on  the physical
 condition of the emitting gas.  Nonetheless, the resulting values for
 the ionization parameter obtained in  this framework are in the range
 log$\xi\sim2.5-5$, consistent  with the  median value of  $\sim$\ 4.8
 erg   s$^{-1}$  cm  found   by  \citealt{tombesi10},   exploring  the
 properties of  ultra fast  outflows in a  sample of  broad-line radio
 galaxies.   Given the large  change in  radius going  from pc  to kpc
 scale, this requires a strong decrease in the gas density to maintain
 an approximately constant ionization  parameter, that can be obtained
 with   a   gas   density   decreasing   roughly   proportionally   to
 r$^{-2}$. This is  in line with the limited  information of spatially
 resolved changes of $\xi$  with radius (e.g. \citealt{bianchi06}). In
 this scenario,  the soft X-rays radiation arises  from emission lines
 originating from tenuous, hot gas cospatial with the clouds producing
 the optical lines.

 The weaknesses  of the  alternative scenarios are  mainly due  to the
 results  of the  limited spectral  fitting  we can  perform with  the
 available data.  Indeed, in case of scattered  nuclear radiation from
 free electrons we expect the spectral index of the incident power-law
 to be  preserved; instead we  obtained values much larger  than those
 usually observed in AGN.  Similarly, emission from hot gas reproduces
 the   5  available   spectra  only   adopting  an   effectively  null
 metallicity.  However,  the main  strength  of  the photoionized  gas
 scenario  is that  it naturally  reproduces the  close correspondence
 between  optical lines  and soft  X-ray  emission, both  in terms  of
 morphology  and  of  relative   intensity,  being  due  to  a  common
 mechanism.  In case  of scattering  or of  emission from  hot  gas, a
 certain  level  of  co-spatiality  can  be  expected  (see  e.g.  the
 correspondence between polarized/scattered light and optical lines in
 Seyfert  galaxies, \citealt{ngc1068pol,ngc5728,kishimoto02})  but the
 constant  flux ratio  with the  optical  lines appears  to be  rather
 contrived.

 Finally, we remind that in one source (3C~98) the soft X-ray emission
 forms  a narrow  structure well  aligned with  the jet,  in  a region
 devoid of any  line emission, a clear evidence  that we are observing
 the high energy counter-part of its radio emission; in another galaxy
 (3C~277.3) there are  two blobs of X-ray emission  cospatial with two
 radio knots, which  are however also associated with  regions of line
 emission. These  two sources warn us  that our approach  based on the
 assumption of a common emission mechanism in all radio galaxies might
 be oversimplified. For example, we note that the sources with a radio
 size smaller than  $\sim$ 100 kpc have a  slightly lower [O~III]/soft
 X-ray ratio  ($\cal{R}$([O~III]/sX)$<$ 2.3), suggesting  that the gas
 is  in a  higher  state of  ionization,  with respect  to the  bigger
 sources ($\cal{R}$([O~III]/sX)$>$3). This suggests a possible role of
 the  shocks produced by  the interaction  of the  gas with  the jets,
 particularly important in the smaller radio sources, in which the jet
 is  still expanding through  the denser  regions of  the interstellar
 medium \citep{best00}.  This strong  interaction is witnessed also by
 the close  alignment of  radio, optical, and  X-ray emission,  and on
 their disturbed kinematics \citep{baum92}.

\begin{acknowledgements} 

We wish to honor the memory of our great friend and colleague David 
Axon, who has been the steadfast inspiration and participant in this and 
many other key papers that through many years of dedicated efforts have 
led to significant breakthroughs and greater understanding of the 
physics of active galaxies. He will be greatly missed by all of us.

We are grateful to the anonymous 
referee for usuful comments that significantly improved the paper.
We thank also S. Bianchi and G. Matt for 
helpful comments on the manuscript.
This work was mainly supported by 
the Italian Space Agency through contract ASI-INAF I/009/10/0 and
ASI/GLAST I/017/07/0.

\end{acknowledgements}


\appendix
\section{Spectral fits}
\label{appendix}

\begin{figure*}
\includegraphics[width=4.66cm,angle=-90]{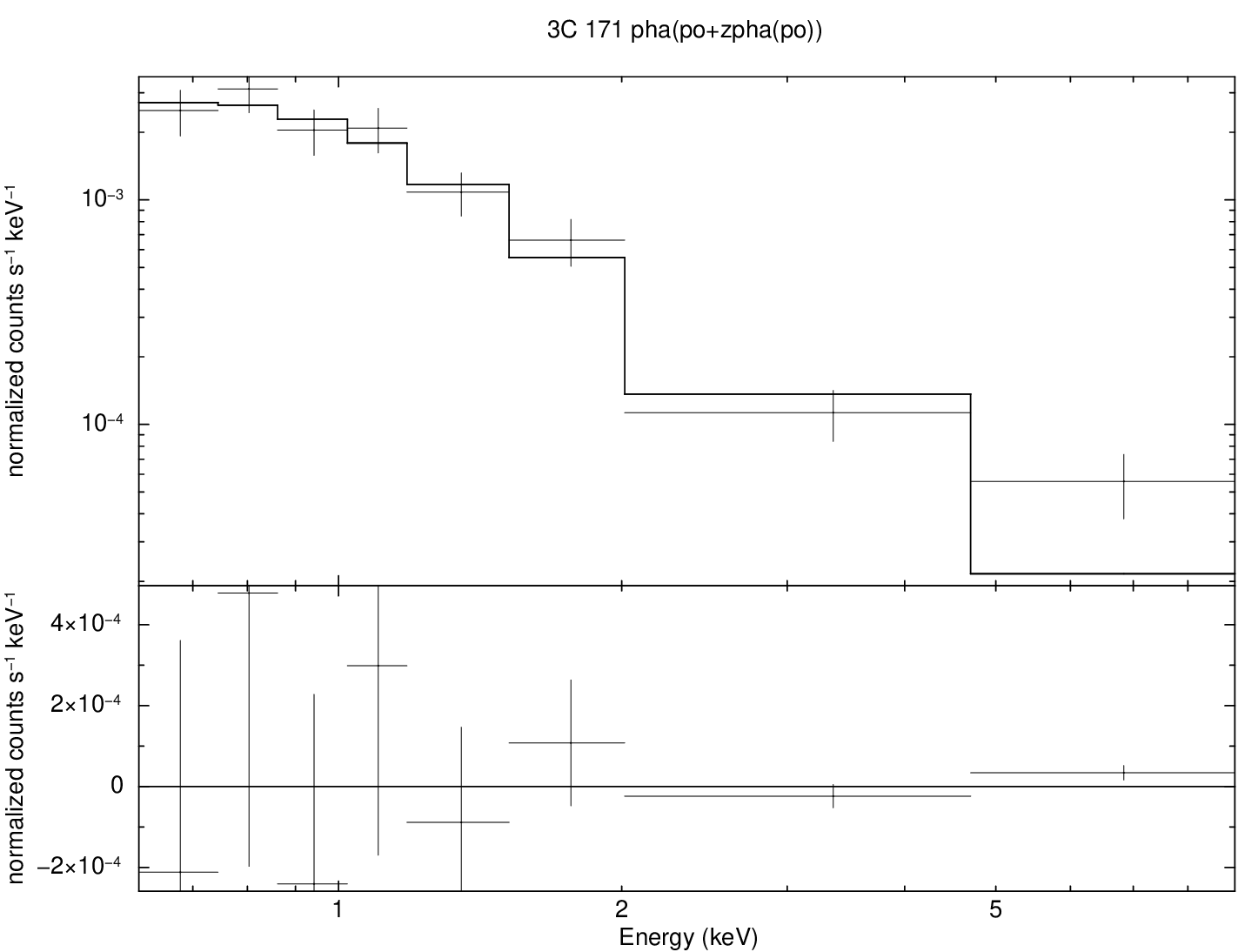}   
\includegraphics[width=4.66cm,angle=-90]{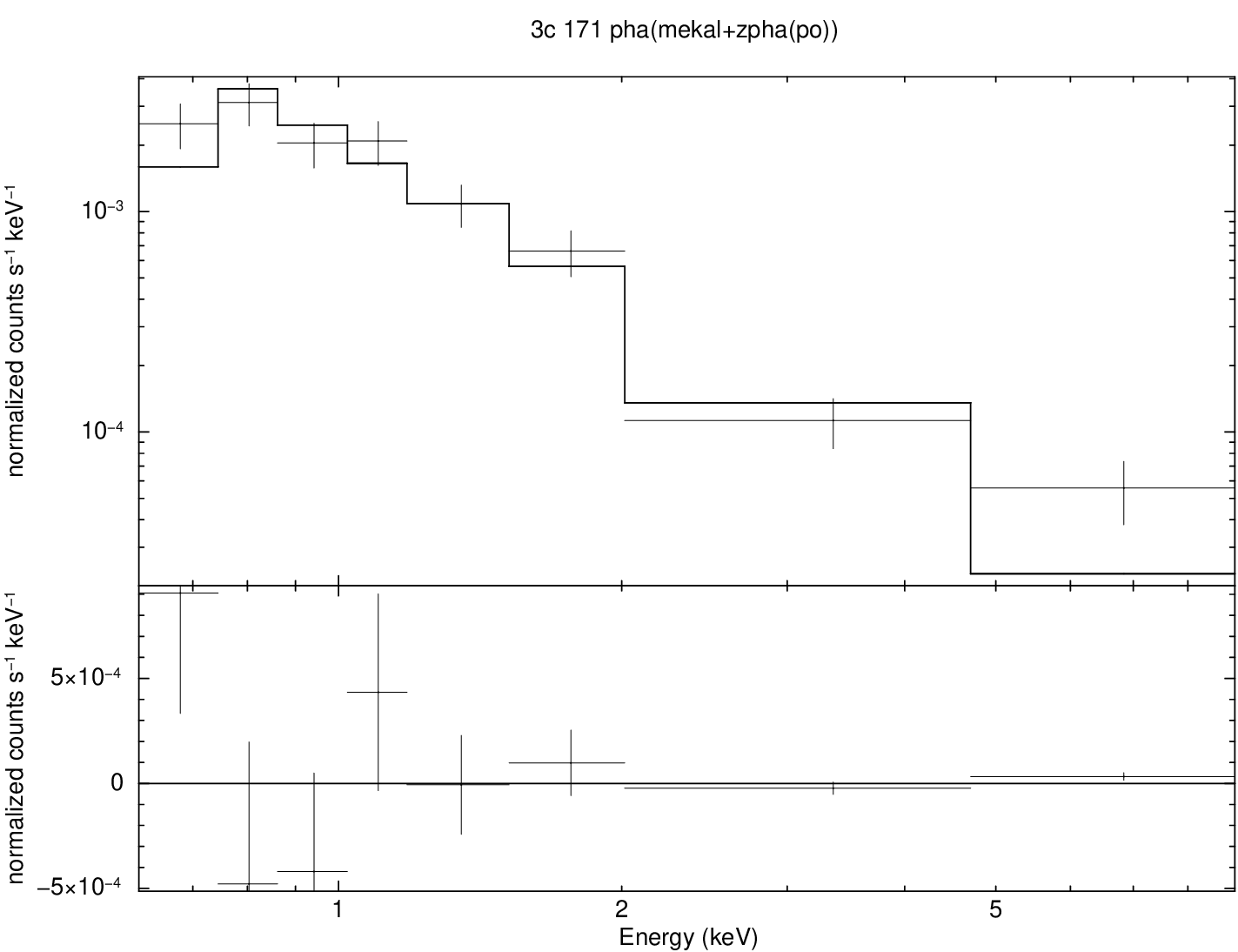}
\includegraphics[width=4.66cm,angle=-90]{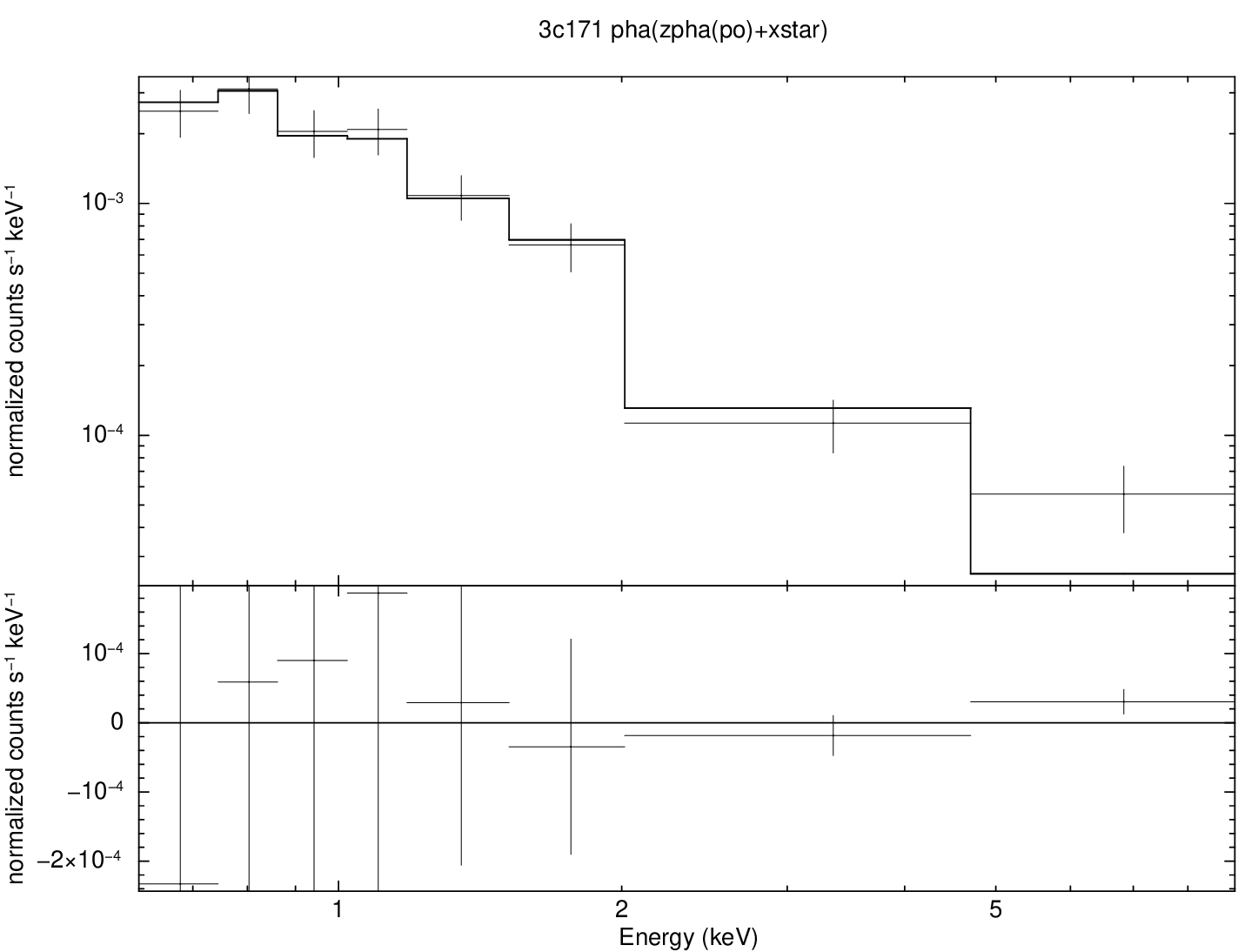}
\includegraphics[width=4.66cm,angle=-90]{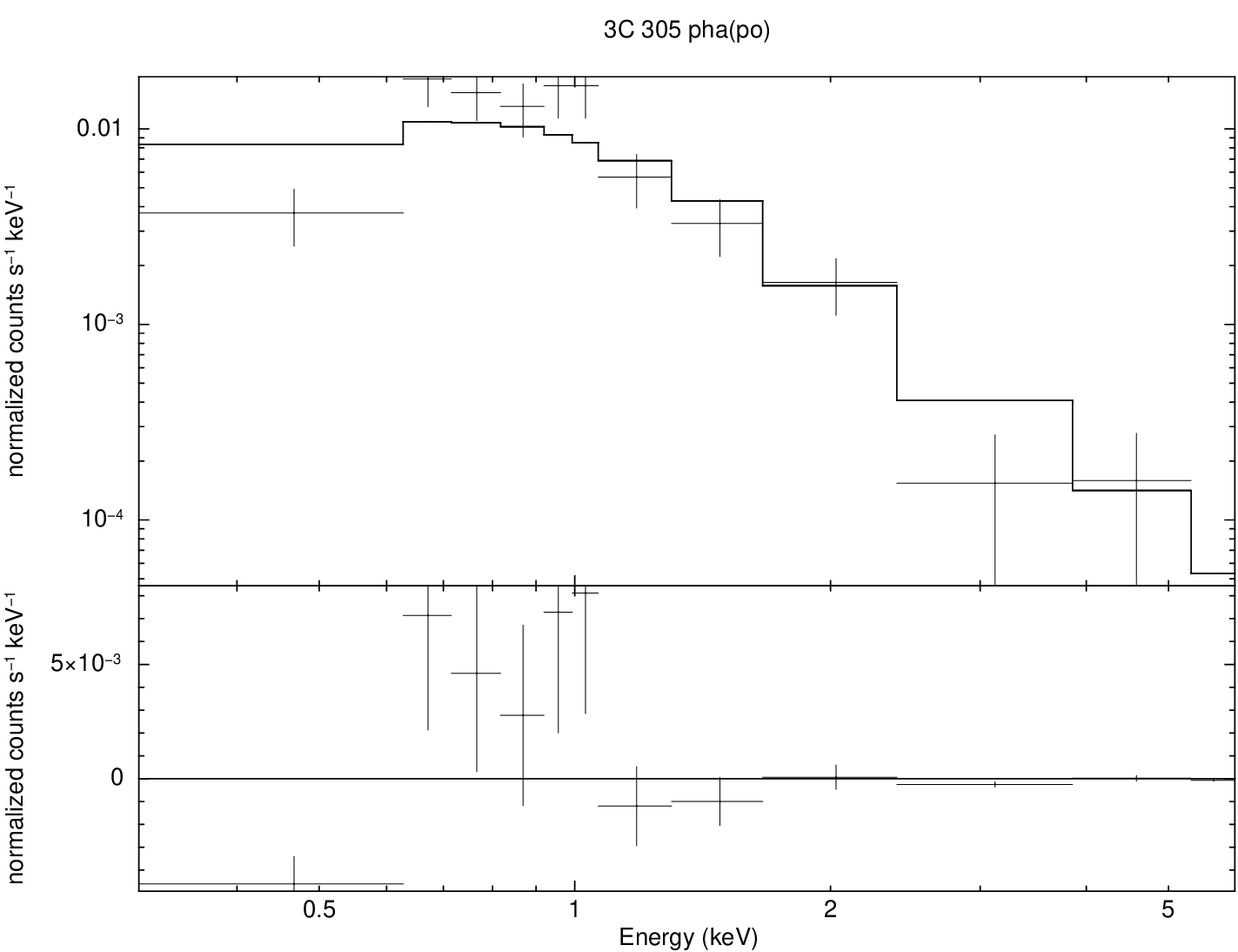}   
\includegraphics[width=4.66cm,angle=-90]{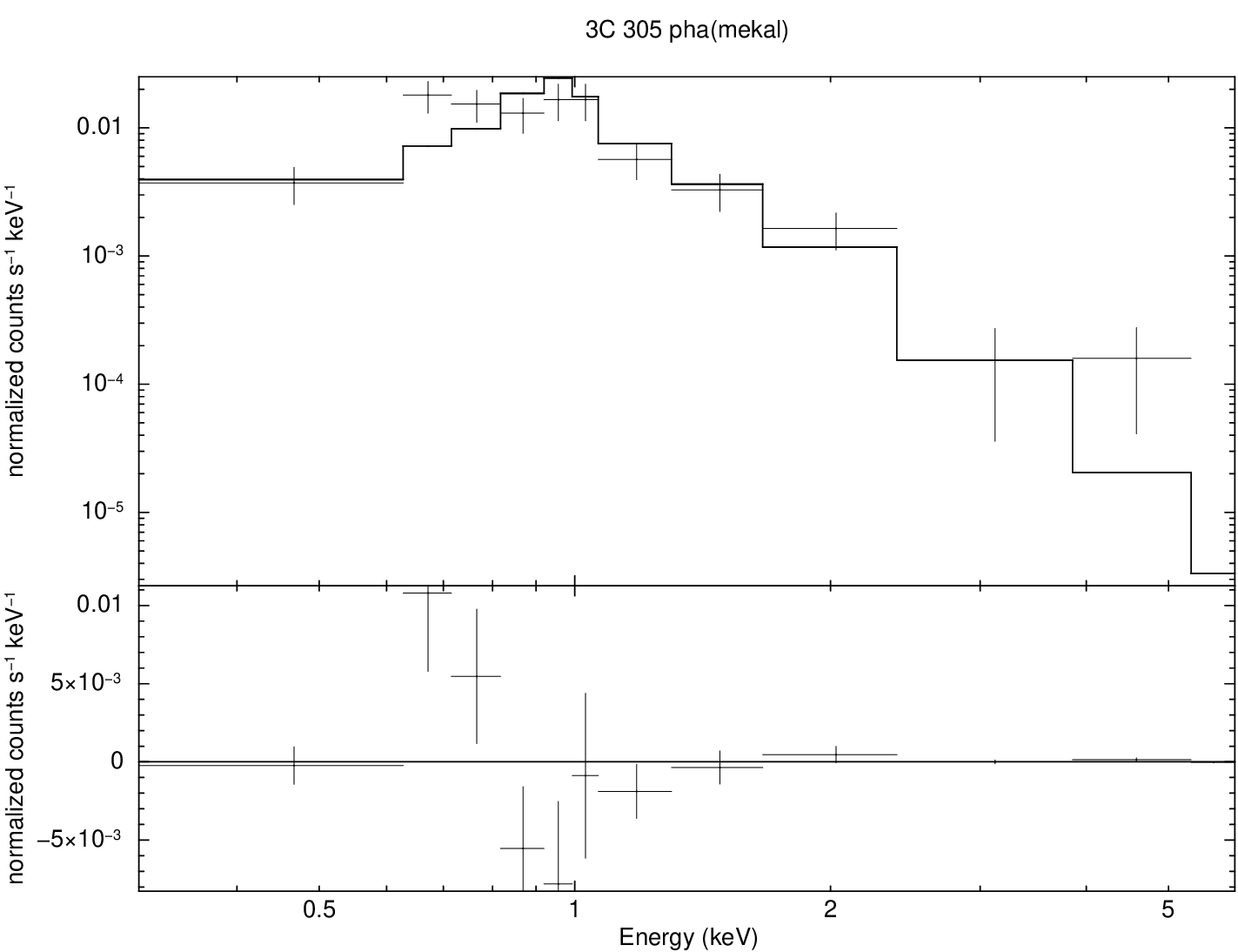}
\includegraphics[width=4.66cm,angle=-90]{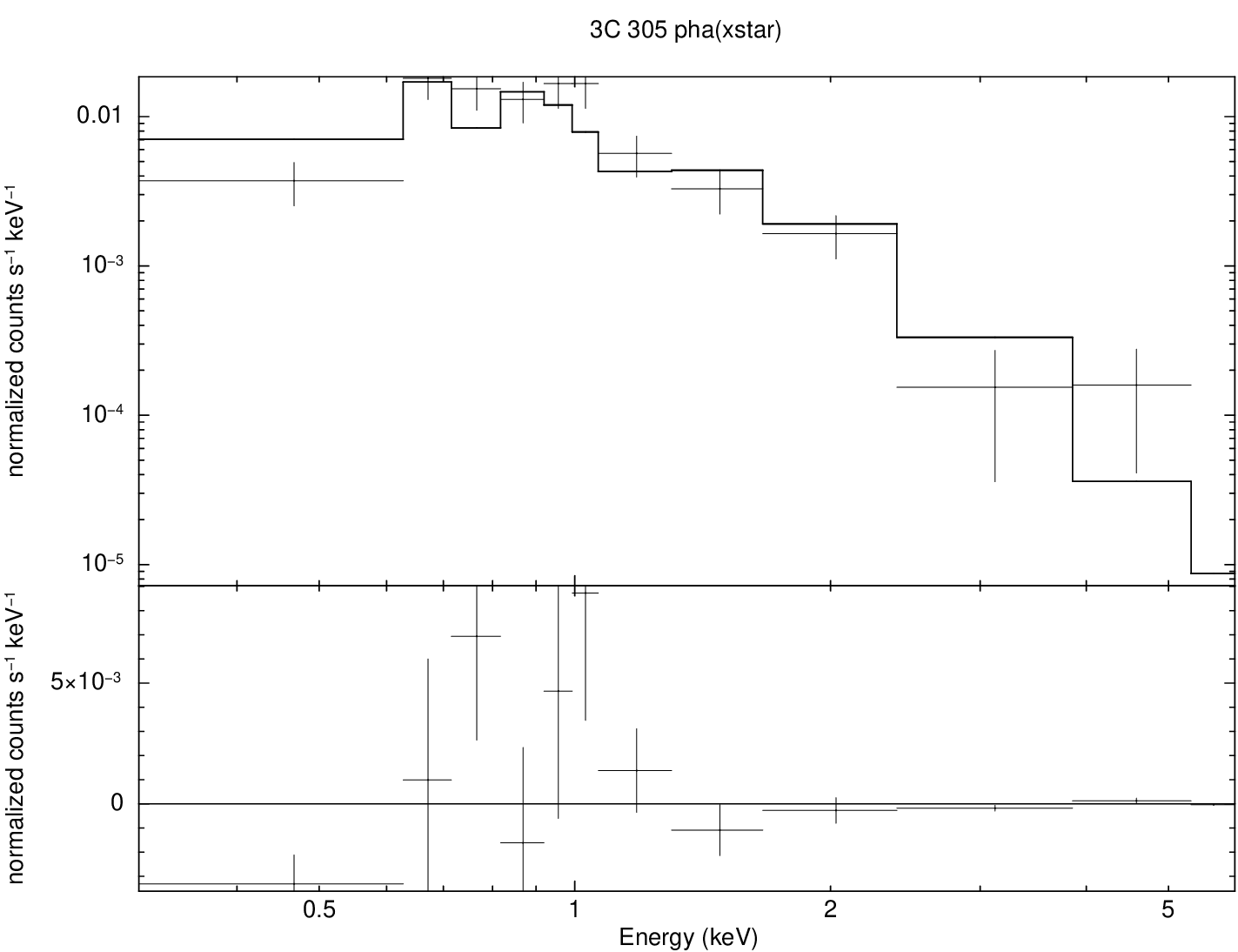}
\includegraphics[width=4.66cm,angle=-90]{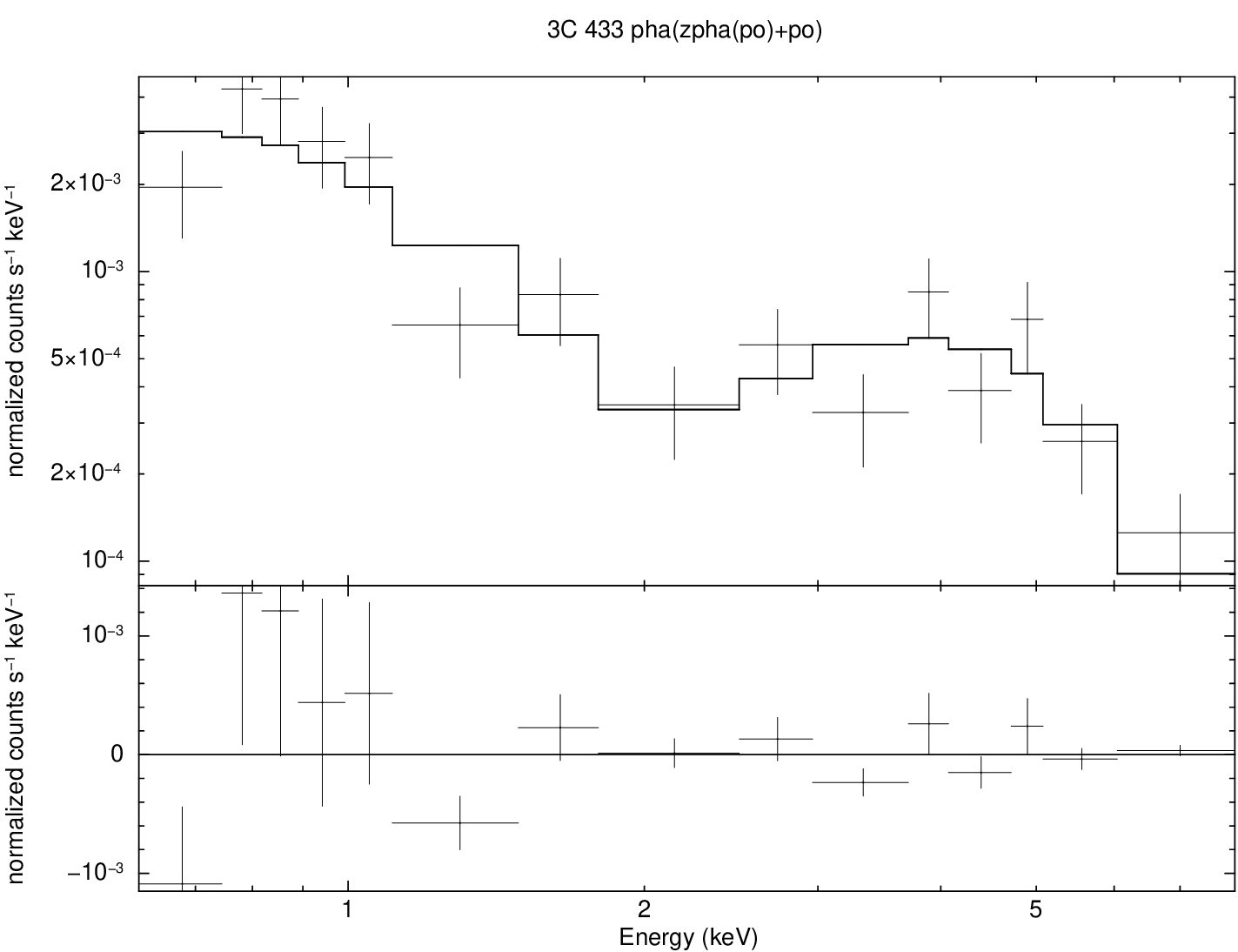}   
\includegraphics[width=4.66cm,angle=-90]{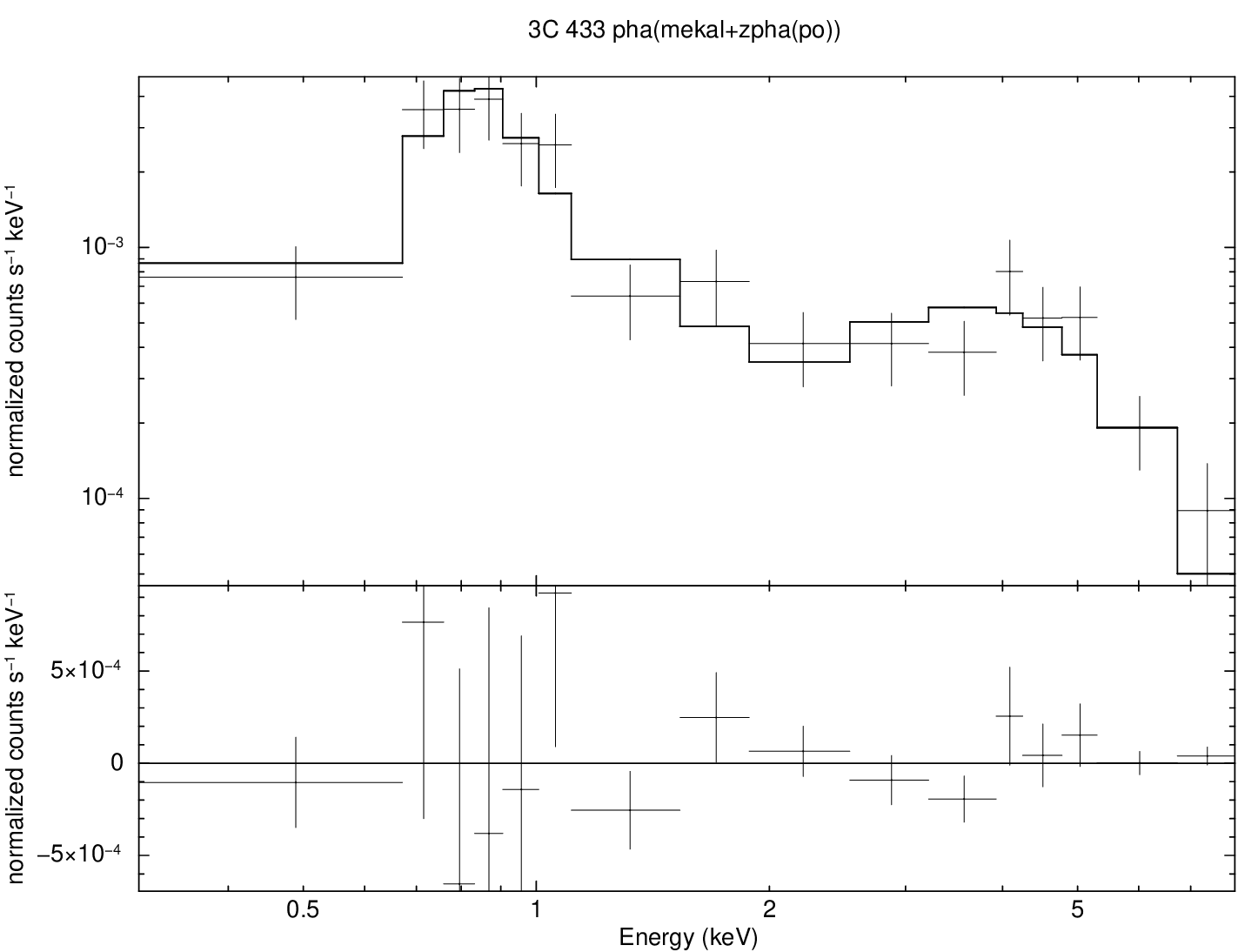}
\includegraphics[width=4.66cm,angle=-90]{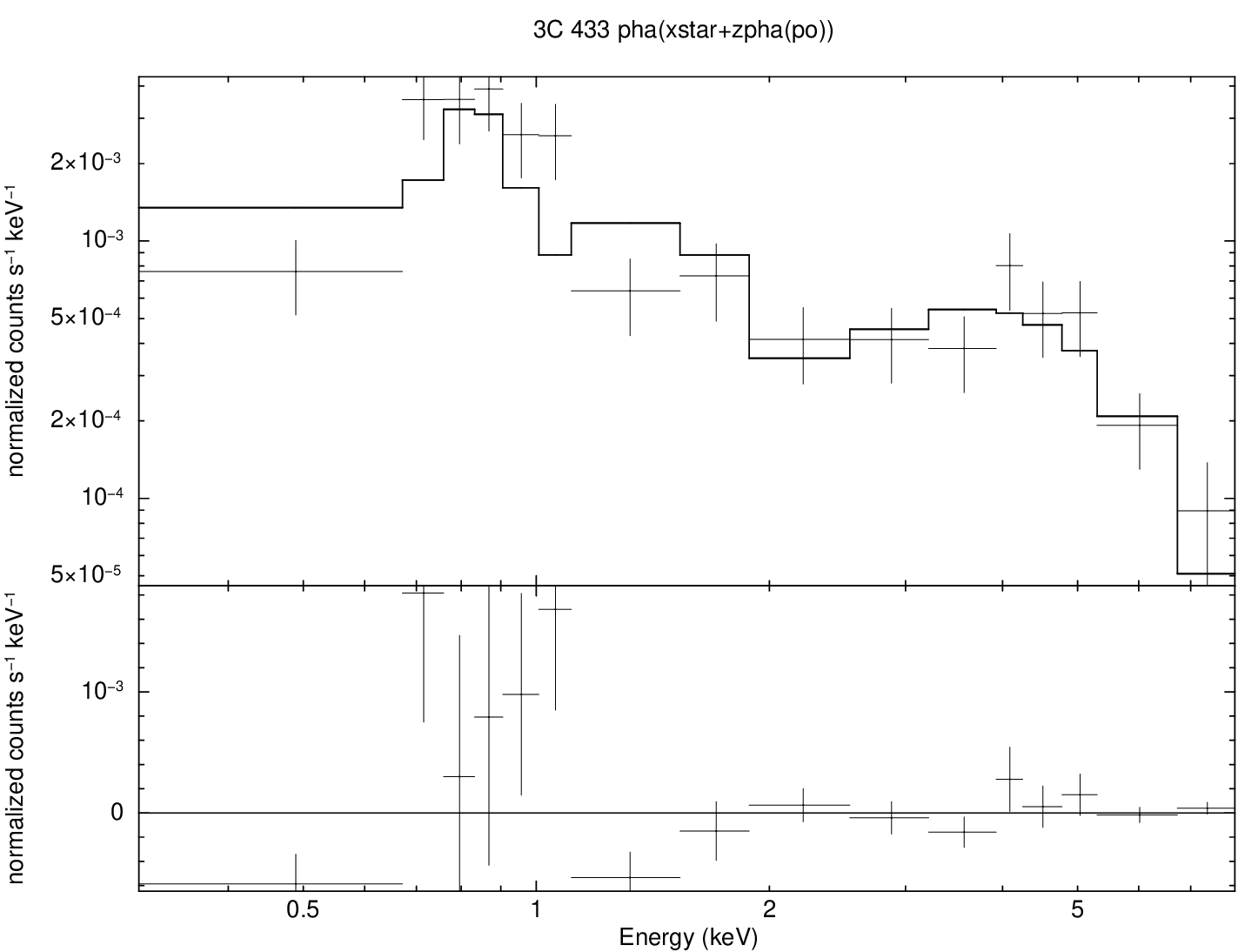}
\includegraphics[width=4.66cm,angle=-90]{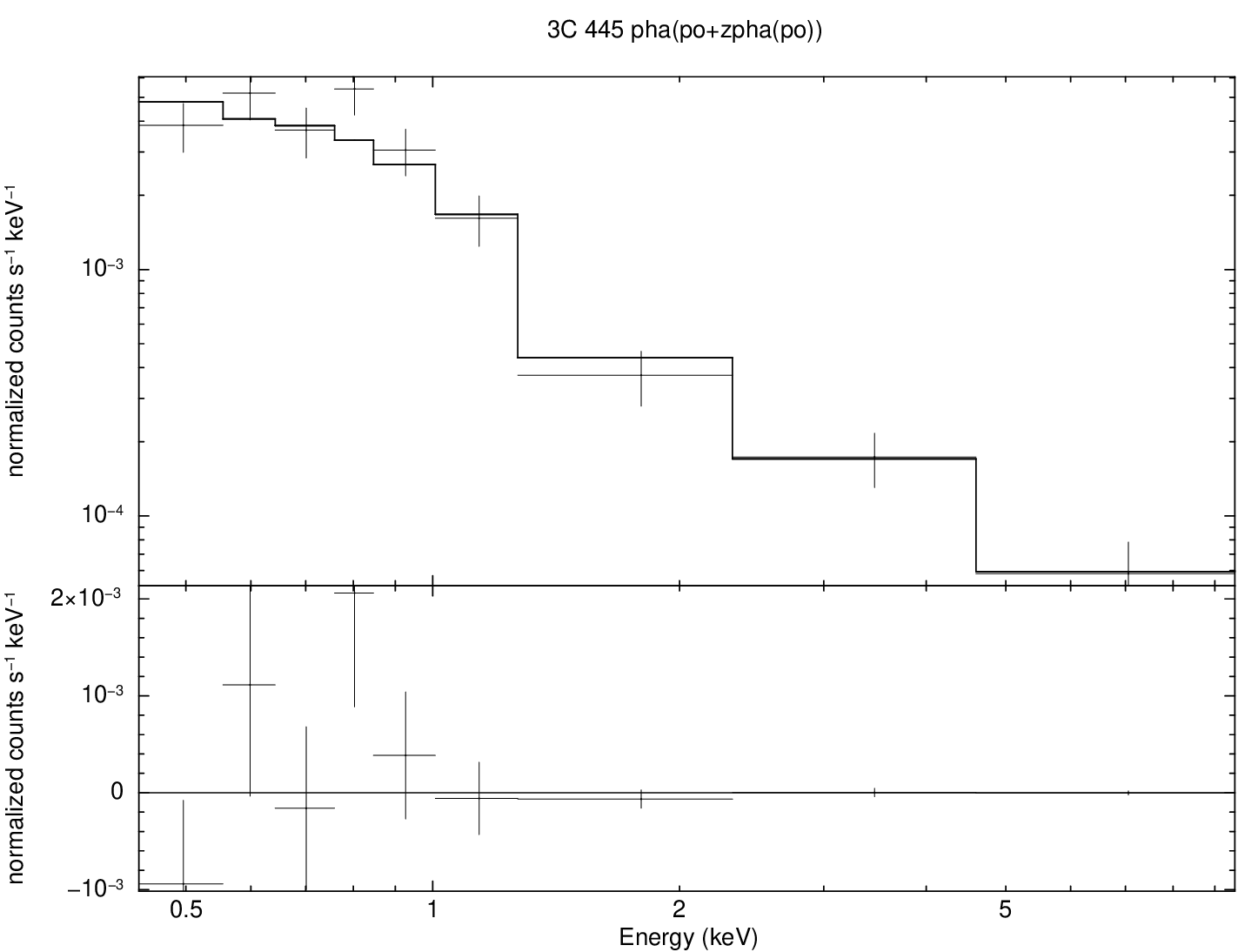}   
\includegraphics[width=4.66cm,angle=-90]{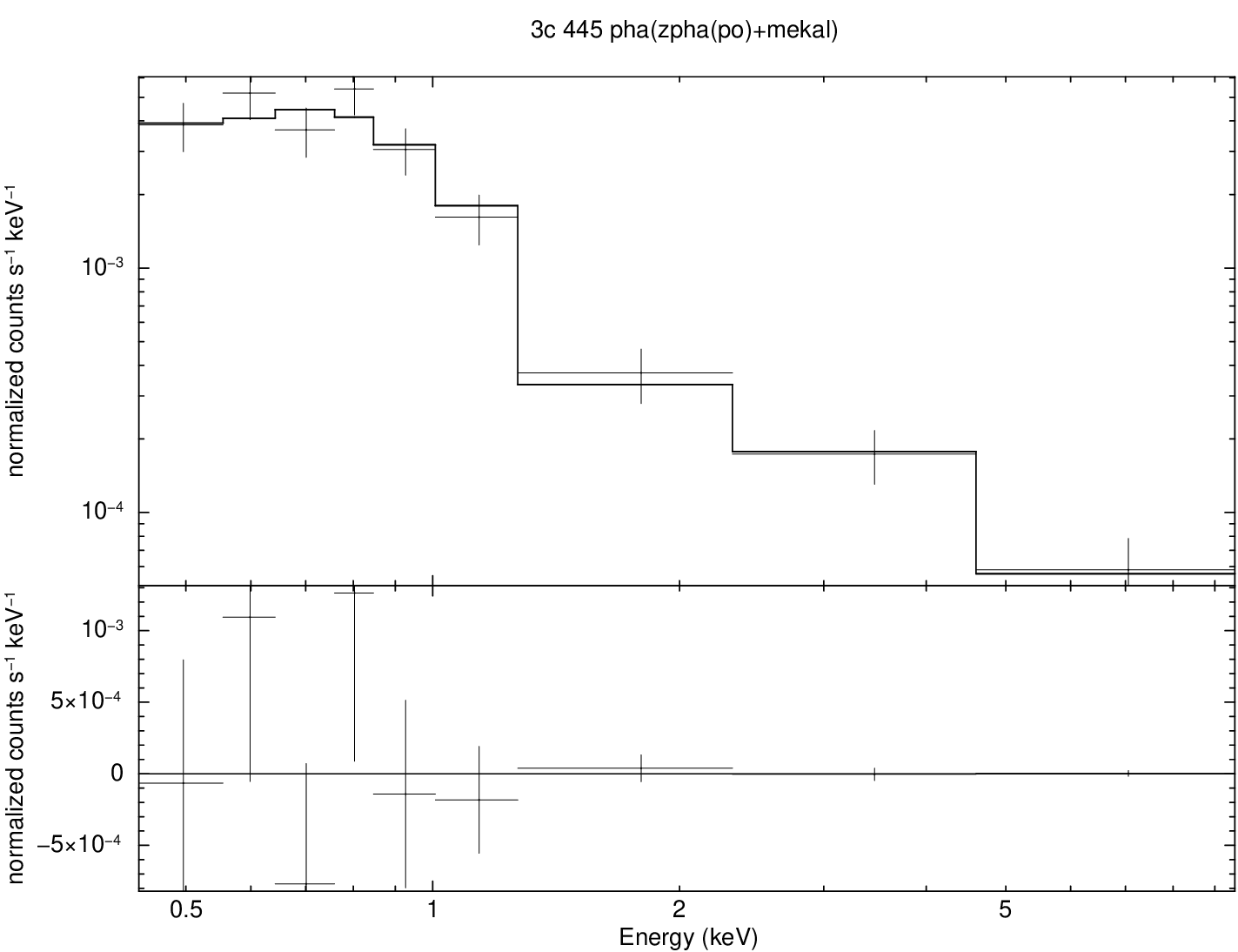}
\includegraphics[width=4.66cm,angle=-90]{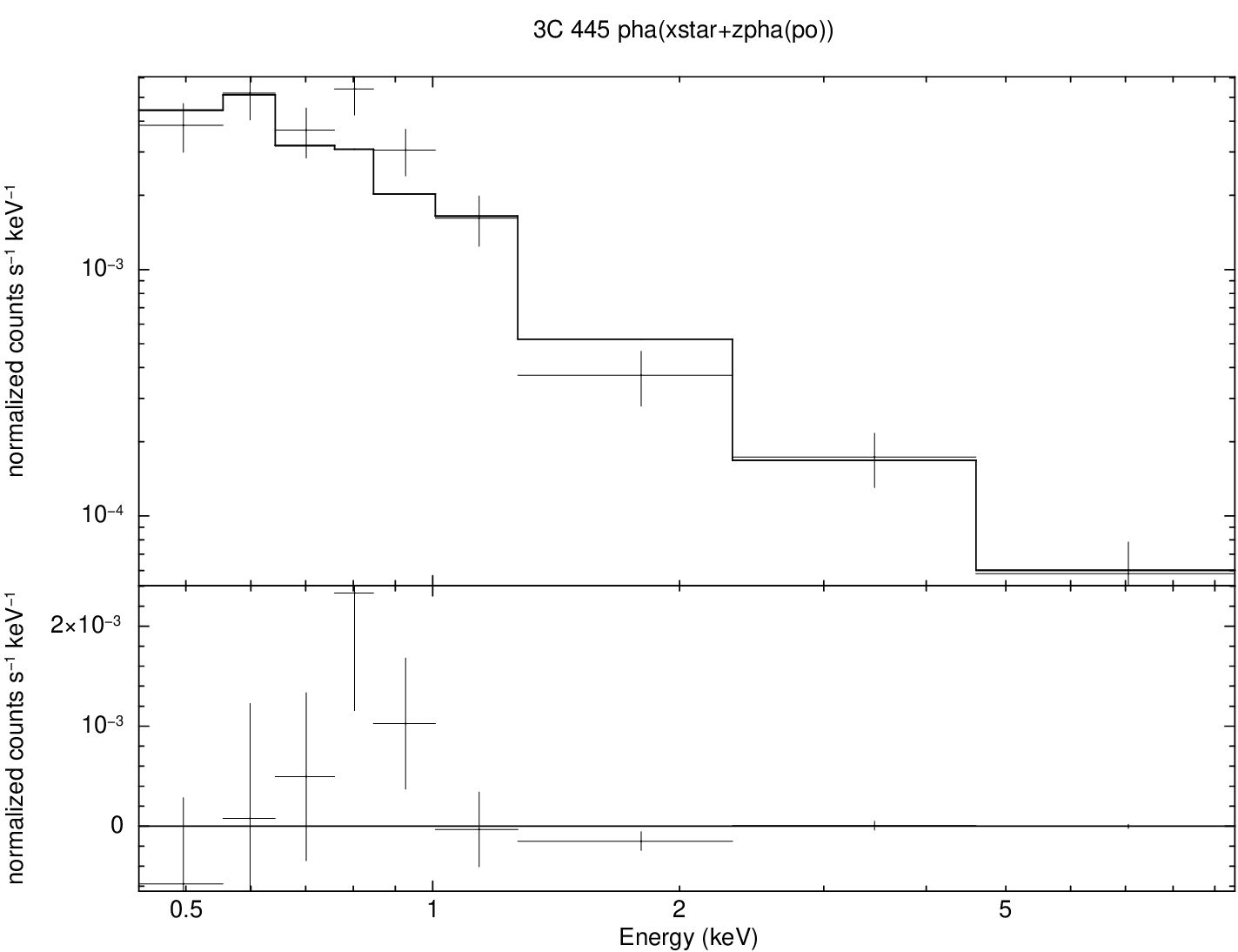}
\caption{We report, as an example, the spectrum of 3C~171, 3C~305, 3C~433 and 3C~445 in the extended
  regions fitted with three different models: powerlaw, \mekal, and \xstar,
  absorbed by a Galactic column density value.  We add to the model an
  intrinsically absorbed powerlaw, to reproduce the emission seen at higher
  energies (due to the leaking of the nuclear emission).}
\label{allfit}
\end{figure*}

\end{document}

%% file: bigtable.tex
\onecolumn
\begin{landscape}
\begin{longtable}{l|l|r|c c c |l c c l c c|l c c c| c}
\caption[Radio, soft-X and emission line region properties for HEG objects.]{Radio, soft X-ray and optical emission line properties for HEG and BLO objects of the sample.} 
\label{bigtable} \\

\hline \hline 
Name& redshift &  Classif. &\multicolumn {3}{|c}{Radio properties}&
 \multicolumn{6}{|c}{SoftX properties} &  \multicolumn{4}{|c}{Opt.lines properties} &  Host PA\\
 & &FR/spec & Size & PA & Ref.  &ObsId. & Exp. & Off angle &morph. & Size &PA 
    &morph.& Size  &PA& Ref &\\  
   & & &  ($\arcsec$)     & ($^\circ$) &  & &(ks) & ($\arcmin$) & &($\arcsec$) & ($^\circ$)& & ($\arcsec$) & ($^\circ$)& & ($^\circ$)\\
(1)& (2) & (3) & (4) & (5)      &(6)       &(7)&(8)&  (9)     &(10)     &(11)&(12)  &(13)&(14) & (15) & (16)&(17)   \\
\hline	
\endfirsthead

\multicolumn{4}{c}{{\tablename} \thetable{} -- Continued} \\[0.5ex]
\hline \hline 
Name& redshift&  Classif. &\multicolumn {3}{|c}{Radio properties}&
 \multicolumn{6}{|c}{SoftX properties} &  \multicolumn{4}{|c}{Opt.lines properties} &  Host PA\\
 & &FR/spec & Size & PA & Ref.  &ObsId. & Exp. &  Off angle &morph.& Size &PA 
    &morph.& Size  &PA& Ref &\\  
  & & &  ($\arcsec$)     & ($^\circ$) &  & &(ks) & ($\arcmin$) & &($\arcsec$)& ($^\circ$)& & ($\arcsec$) & ($^\circ$)& & ($^\circ$)\\
(1)& (2) & (3) & (4) & (5)      &(6)       &(7)&(8)&  (9)     &(10)     &(11)&(12)  &(13)&(14) & (15) & (16) &(17)   \\
\hline
\endhead
\hline
  \multicolumn{12}{c}{{Continued on Next Page}} \\
\endfoot

  \\ 
\endlastfoot

3C~020    &  0.174    & 2/HEG   & 51  & 101  & 3 & 9294  & 8 &0.3 &  non det  &      &       &                &    &        &       &130\\                                                          
3C~033	  &  0.0596   & 2/HEG   & 270 & 19.2 & 4 & 6910  &20 &1.8 &  ext      & 6.9  & 51    & ext            &    &65      & 3     &148$^a$\\                                   
3C~061.1  &  0.184    & 2/HEG   & 186 & 2    & 3 & 9297  & 8 &0.5 &  unres    &      &       &                &    &        &       &165\\                                               
3C~063    &  0.175    &   HEG   & 22  & 34   & 1 & 12722 & 8 &0.3 &  ext      & 11.8 & 55    & ext(unres)     &    &74      & 3     & 79 \\                                 
3C~079	  &  0.2559   & 2/HEG   & 86  & 105  & 3 & 12723 & 8 &0.3 &  unres    &      &       & ext            &    &146     & 3     &71\\                                         
3C~093.1  &  0.2430   &   HEG   & 0.3 & 165  & 1 & 12725 & 8 &0.3 &  unres    &      &       & part.res.      &    &65      & 1     &132  \\                  
3C~098    &  0.0304   & 2/HEG   & 314 & 21   & 2 & 10234 & 30&0.5 &  ext      & 13   & 21    & ext            &    &148     & 3     &55$^b$ \\                          
3C~105	  &  0.089    & 2/HEG   &     &      &   & 9299  & 8 &3.0 &  unres    &      &       &                &    &        &       &\\                                                          
3C~133	  &  0.2775   & 2/HEG   & 12  & 107  & 3 & 9300  & 8 &0.3 &  unres    &      &       &                &    &        &       &87\\                                               
3C~135	  &  0.1253   & 2/HEG   & 130 & 70   & 1 & 9301  & 8 &0.3 &  unres    &      &       & ext(part.res.) &3.71&54      & 1     & 141\\             
3C~136.1  &  0.064    & 2/HEG   & 522 & 108  & 1 & 9326  & 10&0.3 &  unres    &      &       & part.res.      &0.53&175     & 1     &117$^b$\\                            
3C~171	  &  0.2384   & 2/HEG   & 30  & 99   & 1 & 10303 & 60&0.2 &  ext      & 17   & 97    & ext            &3.44&85      & 1     &66  \\                       
3C~180    &  0.22     & 2/HEG   &     &      & 3 & 12728 & 8 &0.3 &  poss.ext & 3.0  & 50    & ext            &    &26      & 3     & 28\\                                      
3C~192	  &  0.0598   & 2/HEG   & 192 & 124  & 1 & 9270  & 10&0.3 &  unres    &      &       & non det(ext)   &    &161     & 3     & \\                                 
3C~223	  &  0.1368   & 2/HEG   & 300 & 164  & 1 & 12731 & 8 &0.2 &  unres    &      &       & ext            &2.1 &128     & 1     &93 \\                     
3C~223.1  &  0.107    & 2/HEG   & 117 & 15   & 3 & 9308  & 8 &0.3 &  unres    &      &       &                &    &        &       &40\\                                                     
3C~234    &  0.1848   & 2/HEG   & 110 & 64   & 1 & 12732 & 8 &0.3 &  poss.ext & 5.0  & 64    & ext            &2.37&79      & 1     &80 \\                 
3C~277.3  &  0.0857   & 2/HEG   & 29  & 158  & 1 & 11391 & 25&0.3 &  blobby   &      &       & part.res.      &    &147     & 3     &170$^b$\\                        
3C~284    &  0.2394   & 2/HEG   & 176 & 101  & 1 & 12735 & 8 &0.3 &  unres    &      &       & ext(unr)       &2.9 &74      & 1     &151\\                    
3C~285	  &  0.0794   & 2/HEG   &     & 82.8 & 4 & 6911  & 40&1.0 &  unres    &      &       & ext            &    &81      & 3     &129$^a$ \\                              
3C~300    &  0.27     & 2/HEG   & 96  & 130  & 3 & 9311  & 8 &0.3 &  unres    &      &       & ext            &    &126,90  & 3     &89\\                                          
3C~303.1  &  0.267    & 2/HEG   & 1.9 & 130  & 1 & 9312  & 8 &0.3 &  unres    &      &       & ext            &1.56&140     & 1     &169 \\                  
3C~305	  &  0.0416   & 2/HEG   & 14  & 44   & 1 & 9330  & 8 &0.3 &  ext      & 7.3  & 54    & ext            &6.37&46      & 1     &76$^b$ \\                   
3C~321    &  0.096    & 2/HEG   & 309 & 131  & 1 & 3138  & 50&0.8 &  complex  &      &       & ext            &6.21&108     & 1     &\\                                
3C~327    &  0.1041   & 2/HEG   & 169 & 100  & 3 & 6841  & 40&1.7 &  unres    &      &       & ext            &    &22      & 3     &135\\                                               
3C~379.1  &  0.256    & 2/HEG   & 76  & 161  & 1 & 12739 & 8 &0.3 &  poss.ext & 2.1  & 40    & ext            &0.72&43      & 1     &      \\            
3C~381    &  0.1605   & 2/HEG   & 69  & 4    & 1 & 9317  & 8 &0.3 &  unres    &      &       & ext            &5.69&150     & 1     &156\\                   
3C~403	  &  0.0590   & 2/HEG   & 230 & 79   & 2 & 2968  & 50&0.5 &  ext      & 6.6  & 34    &                &    &23      & 3     & 39$^b$   \\                                          
3C~433	  &  0.1016   &   HEG   & 58  & 172  & 1 & 7881  & 40&0.5 &  ext      &6.5,10.3&138,49& part.res.     &5.87&135     & 1     &145$^a$\\                     
3C~436    &  0.2145   & 2/HEG   & 105 & 172  & 1 & 9318  & 8 &0.3 &  unres    &      &       & part.res.      &    &        & 1     &3  \\                       
3C~452	  &  0.0811   & 2/HEG   & 277 & 79   & 1 & 2195  & 80&2.2 &  ext      & 6.9  & 48    & part.res.      &1.58& 125    & 1     &101$^b$  \\                     
3C~456    &  0.2330   & 2/HEG   & 10  & 20   & 3 & 12746 & 8 &0.3 &  ext      & 6.1  &20     &                &    &        &       &107    \\                                           
3C~458    &  0.2890   &   HEG   &161  & 75   & 1 & 12747 & 8 &0.3 &  non det  &      &       & part.res.(ext) &    &75      & 1     & \\                        
\hline                                                     
3C~017    &  0.2198   & 2/BLO   & 30  & 147 & 1  & 9292  & 8 &0.3 &  unres     &      &       & part.res.      &    &41      &1      &42\\                       
3C~018	  &  0.188    & 2/BLO   &     &     &    & 9293  & 8 &0.3 &  unres     &      &       &                &    &        &       &   \\                                               
3C~033.1  &  0.1809   & 2/BLO   & 216 & 45  & 3  & 9295  & 8 & 0.3&  unres    &      &       & ext            &    &53      &3      &63 \\                                               
3C~111	  &  0.0485   & 2/BLO   & 220 & 62  & 1  & 9279  & 10&0.3 &  unres     &      &       & non det        &    &        & 1     &22$^b$ \\                             
3C~184.1  &  0.1182   & 2/BLO   & 167 & 157 & 1  & 9305  & 8 &0.3 &  unres    &      &       & part.res.      &2.64&24      & 1     &40  \\                    
3C~197.1  &  0.1301   & 2/BLO   & 14  & 2   & 3  & 9360  & 8 &0.3 &  unres    &      &       & unres          &    &        & 2     &    \\                              
3C~219	  &  0.1744   & 2/BLO   & 184 & 40  & 1  & 827   & 20&0.6 &  unres     &      &       & unres          &    &        & 1     &145  \\                          
3C~227	  &  0.0861   & 2/BLO   & 246 & 86  & 1  & 6842  & 30&1.7 &  unres     &      &       & ext(unr)       &    &35,119  &3      &165$^b$ \\                          
3C~273    &  0.1583   &   BLO   &     &     &    & 4879  & 40&1.1 &  unres     &      &       & ext(no bg)     &    &        & 1     & \\                     
3C~287.1  &  0.2159   & 2/BLO   & 112 & 91  & 3  & 9309  & 8 &0.3 &  unres     &      &       &                &    &        &       &142\\                                               
3C~303	  &  0.141    & 2/BLO   & 38  & 97  & 3  & 1623  & 15&0.6 &  unres     &      &       &                &    &        &       &0$^a$ \\                                               
3C~323.1  &  0.264    & 2/BLO   &     &     &    & 9314  & 8 &0.3 &  unres     &      &       & ext(no bg)     &    &        & 1     & \\                      
3C~332    &  0.1517   & 2/BLO   & 81  & 20  & 1  & 9315  & 8 &0.3 &  unres     &      &       & unres          &    &        & 1     &54\\                             
3C~382    &  0.0578   & 2/BLO   & 179 & 50  & 1  & 4910  & 55&0.1 &  unres     &      &       & part.res.(ext) &2.14&112     & 1     &85$^b$ \\             
3C~390.3  &  0.0561   & 2/BLO   & 231 & 145 & 1  & 830   & 35&2.3 &  unres     &      &       & part.res.      &0   &62      & 1     &82$^b$ \\                       
3C~410    &  0.2485   &   BLO   & 13  & 129 & 3  & 12742 & 8 &0.3 &  unres     &      &       &                &    &        & 10    &      \\                                          
3C~445    &  0.0562   & 2/BLO   & 576 & 171 & 1  & 7869  & 50&3.7 &  ext      & 14   &44     & ext(non det)   &    &37      & 3     &80$^b$\\                              
3C~459	  &  0.2199   & 2/BLO   & 8   & 94  & 3  & 12734 & 8 &0.2 &  ext      &      &102    &                &    &        &       &160 \\                                                       
\hline
\hline
\end{longtable}

Column description: 

(col. 1) 3CR name; 

(col. 2) redshift from \citet{spinrad1985}; 

(col. 3) morphological FR type from \citet{buttiglione09}
 and spectroscopic classification into High
Excitation Galaxy (HEG) or Broad Line Object (BLO) from  \citet{buttiglione09}.
Except for 3C~433 and 3C~458, all the unclassified objects could be considered FR~II according to their
morphology and 178 MHz
luminosities.

(col. 4)Angular size of the radio emission in arcsec; 

(col. 5) Position angle of the
radio source, generally measured from hot spot to hot spot. 

(col. 6) References for the radio properties:
1. \citet{privon08} 2. \citet{martel99} 3. \citet{deKoff96} 4. \citet{saripalli09};

(col. 7) Chandra observation identification number;

(col. 8) Approved exposure time for the observation in ks.

(col. 9) Off-axis angle coordinate in arcmin.

(col. 10)  Morphology of the soft (0.5-2 keV) X-ray emission: extended (ext), 
possibly extended (poss. ext), unresolved (unres),
non detected (non det). X-ray jets have been seen in
3C~15: \citet{kataoka2003}, 3C~31: \citet{hardcastle2002}, 3C~66B: \citet{hardcastle2001}, 3C~78: \citet{massaro2008proc}
3C~264: \citet{perlman2010}, 3C~270: \citet{worrall10}, 3C~273: \citet{sambruna2001}, 3C~296: \citet{hardcastle2005}, 
3C~346: \citet{dulwich2009}, 3C~371: \citet{pesce2001}, 3C~465: \citet{Hardcastle2005_2}.

(col. 11) Angular size of the soft X-ray emission in arcsec;

(col. 12) Position angle of the more extended direction in the X-ray image;

(col. 13) Morphology of the optical emission line region as col.9; 

(col. 14) Angular size of the emission line region in arcsec;

(col. 15) Position angle of the more extended direction in the optical line images;

(col. 16) references for the optical line region properties: 1. \citet{privon08} 2.\citet{tremblay09}
3. \citet{mccarthy95} 4. \citet{baum88} 5. \citet{martel04} 6. \citet{edwards2009} 7. \citet{sarzi2006}
 8. \citet{hes1996} 9. \citet{prieto1997} 10. \citet{hippelein96} 11. \citet{morganti92} 12. \citet{husemann08};

(col. 17) Host galaxy major axis position angle from \citet{deKoff96}, $^a$\citet{saripalli09}, $^b$\citet{martel99}.

\end{landscape}
\twocolumn

%% file: fit.tex
\begin{table*}
\begin{tabular}{l|l|l l  |c|c}

\hline\hline
 Name  & Model name   & Model parameters  &   Nuclear component            & $\chi^2_\nu$ (dof) &F(0.5-2 keV)\\
\hline
3C~171 & pha(mekal+zpha(po))   &  kT=1.3$_{-0.2}^{+0.3}$ [keV]  ab=0.03$_{-0.02}^{+0.1}$  &$\Gamma$=1.7fix zNh=8.8e22 [cm$^{-2}$] fix  & 1.25(4)  & 1.19e-14  \\ 
       &                       &  ab=0.3 fix                                            &                                           & 1.83(5)  & \\
       &                       &  ab=0.5 fix                                           &                                           &  2.26(5)  & \\
       & pha(pow+zpha(po))     &  $\Gamma=3.1^{+0.5}_{-0.5}$                            &$\Gamma$=1.7fix zNh=8.8e22[cm$^{-2}$] fix & 1.26(5)  & 1.33e-14  \\
       &                       &  $\Gamma$=1.7 fix                                    &                                           & 4.76(6)  &           \\
       & pha(xstar+zpha(po))   &  N=2.2$^{+0.3}_{-2.2}$e21 [cm$^{-2}$]    log$\xi$=2.5$^{+0.6}_{-0.1}$  &$\Gamma$=1.7fix zNh=8.8e22[cm$^{-2}$] fix  & 0.94(4)  & 1.89e-14  \\
\hline
3C~305 & pha(mekal)          & kT = 1.06$\pm$ 0.1 [keV]   ab=0.3 fix                  & &                                    & 4.17e-14 \\
       & pha(po)               &  $\Gamma$=2.5$\pm$ 0.2                                & &                                    & 4.62e-14 \\
       & pha(xstar)            & N=(2.17$\pm$ 0.09)e21 [cm$^{-2}$] log$\xi$=5.0$\pm$1.0 & &                                    & 4.07e-14 \\   
\hline
3C~403 & pha(mekal+zpha(po)) & kT = 0.38$\pm$ 0.05 [keV] ab=0.00$\pm$0.01       &$\Gamma$=1.7fix zNh=(19.7$\pm$5.2) e22 [cm$^{-2}$]   && 9.12e-15 \\
       & pha(pow+zpha(po))   & $\Gamma$=3.9$\pm$0.3                               &$\Gamma$=1.7fix zNh=(24.1$\pm$ 7.1) e22 [cm$^{-2}$]  && 8.97e-15 \\
       & pha(xstar+zpha(po)) & N=(8.3$\pm$12.0)e23 [cm$^{-2}$] log$\xi$=5.0$\pm$3.3&$\Gamma$=1.7fix  zN=(18.8{$\pm$ 6.8})e22 [cm$^{-2}$] && 1.25e-14  \\
\hline
3C~433 & pha(mekal+zpha(po)) & kT = 0.7$\pm$ 0.1 [keV] ab=0.1$\pm$0.09              &$\Gamma$=1.7fix   zNh=(8.6$\pm$ 2.1)e22 [cm$^{-2}$]    &          & 1.04e-14 \\            
       & pha(pow+zpha(po))   & $\Gamma$=3.5$\pm$ 0.5                                &  $\Gamma$=1.7fix zNh=(10.3$\pm$2.8)e22 [cm$^{-2}$]    &          & 1.03e-14 \\
       & pha(xstar+zpha(po)) & N=(9.9$\pm$ 2.7)e21 [cm$^{-2}$] Log$\xi$=3.9$\pm$ 1.2 &$\Gamma$=1.7fix   zNh=(12.1$\pm$ 3.4)e22 [cm$^{-2}$]   & &1.15e-14\\
\hline
3C~445 & pha(mekal+zpha(po)) & kT = 0.39$\pm$0.05 [keV] ab=0.00$\pm$0.01  &   $\Gamma$=1.7fix zNh=(11.0$\pm$ 6.5)e22[cm$^{-2}$]&0.85(4) &1.46e-14\\ 
       &                       & ab= 0.3 fix                                   &                                                   & 2.84(5) & \\
       &                       & ab= 0.5 fix                                   &                                                   & 2.91(5) & \\
       & pha(pow+zpha(po))   & $\Gamma$=3.7$\pm$ 0.2 &$\Gamma$=1.7 fix  zNh=(14.7$\pm$ 7.9)e22[cm$^{-2}$]       &  1.22(5)      & 1.03e-14 \\
       &                       & $\Gamma$=1.7 fix      &                                                        & 10.89(6)        &          \\ 
       & pha(xstar+zpha(po)) & N=(1.1$\pm$ 0.1)e22 [cm$^{-2}$] Log$\xi$=5.0$\pm 3.4$ & $\Gamma$=1.7fix zNh=(15.9$\pm$ 8.6)e22 [cm$^{-2}$] & 2.45(4)&1.59e-14 \\  
\hline\hline
\end{tabular}

\caption{
Results  of the  spectral  fits for the extended emission  of the 5 sources with the highest
  number of counts. For 3C~305, 3C~403, and 3C~433 the spectrum is grouped to at least 10 counts/bin and we
  apply  a Poissonian  statistic (cstat). The  fluxes reported  in the
  table  are  corrected for Galactic absorption.  
  For 3C~171 and 3C~445 the spectrum is grouped by at least 20 counts/bin
  ($\chi^2$ statistic).
  In three cases we add in the models a power-law component to fit the emission observed in the hard band.
Notes. 3C~171: in all the models, we fixed the nuclear component parameters at $\Gamma$=1.7 
and N$_H$=8.8\,10$^{22}$ cm$^{-2}$ (see \citealt{hardcastle10});
3C~305: the spectrum do not require a nuclear related component.
We do not report in table the flux measurement if the model is not representative of the
spectrum (according to the $\chi^2$ value).}
\label{fit}
\end{table*}

%% file: flux.tex
\begin{table*}
\begin{tabular}{l|c|c|c|ccc|c|c|c|c|c}
\hline\hline
Name & N$_{\rm H}$ & cts & S/N & \multicolumn{3}{|c|}{F(0.5-2 keV)}&L(0.5-2 keV)
&F$_{\rm [O~III]}$  &L$_{\rm [O~III]}$  & F$_{\rm [O~III]}$/F$_{\rm sX}$ & F$_{\rm [O~III]}$/F$_{\rm sX}$\\
     &  10$^{20}$     &        &     &  \multicolumn{3}{|c|}{[10$^{-14}$ erg cm$^2$ s$^{-1}$]} & 10$^{41}$ &10$^{-14}$  & 10$^{41}$  & ext& nuc\\
     & [cm$^{-2}$]    &        &     & pha(po)& pha(mek) & pha(xstar) & [erg s$^{-1}$] & [erg cm$^2$ s$^{-1}$]  &   [erg s$^{-1}$]           & & \\
\hline

3C 33 & 3.23  & 44   & 6.5 & 0.904     &  0.829     & 1.78      &  1.48      & 5.92      & 4.91 &3.3& 2.3\\
3C 63 & 2.47  & 31   & 5.4 & 1.77      &  1.72      & 1.89      &  15.8      &           & &\\
3C 98 & 10.2  & 25   & 4.7 & 0.592     &  0.587     & 0.805     &  0.171     &           & &\\
3C 171*& 5.65  & 134  & 11.3& 1.15      &  1.52      & 1.34      &  30.4      &  3.09     & 51.8     & 2.3& 3.1\\
3C 180& 13.6  & 6    & 2.3 &$<$0.664   &$<$0.736    &$<$0.846   &  $<$11.8   & 1.80      & 698     &$>$2.1  &\\
3C 234& 1.76  & 13   & 3.6 & 0.670     &  0.651     & 0.715     &  6.75      & 4.13      & 39.1    &5.8& 2.5\\
3C 305*& 1.31  & 89   & 9.3 & 4.38      &  3.98      & 4.87      &  1.92      & 11.1      &4.36     & 2.3& 2.3\\
3C 379.1&5.43 & 3    & 1.7 &$<$0.266   & $<$0.280   &$<$0.317   &  $<$6.24   & 0.332     &6.54     &$>$1.04&\\
3C 403*& 12.2  & 51   & 7.0 & 0.494     & 0.467      & 0.641     &  0.520     & 4.79      & 3.89    & 7.5& 2.0\\
3C 433*& 7.77  & 83   & 8.7 & 1.01      & 0.971      & 1.27      &  3.25      &           & &\\
3C 445*& 4.49  & 92   & 9.2 & 0.884     & 0.816      & 1.03      &  0.756     &  3.60     &2.64    & 3.5&3.8\\
3C 452& 9.64  & 12   & 2.3 &$<$0.06    &$<$0.0594   &$<$0.0736  &  $<$0.117  &  0.467    &0.738   &$>$6.3&1.5\\
3C 456& 3.70  & 11   & 3.2 & 0.597     & 0.606      & 0.670     &  14.4      &           & &\\
3C 459& 5.24  & 13   & 3.6 & 0.739     & 0.754e     & 0.858     &  11.9      &           & &\\
\hline\hline
\end{tabular}
\caption{Results for the 14 sources for which it was possible to carry out a spectral analysis. 
We mark with a star symbol the 5 sources with a sufficiently high number of counts 
for which we perform in addition a spectral fit.
We converted the observed number counts in a flux value adopting three different models: a power-law, mekal, 
  or Xstar, absorbed by the Galactic N$_{\rm H}$ column density.
  We report the unabsorbed soft X-ray fluxes for all 3 models and the luminosity for the \xstar\ model in the soft 
X-ray band (0.5-2 keV). 
  We also give the [O~III] flux and luminosity in the same regions used for the extraction of the X-ray spectra. }
\label{flux}
\end{table*}